\documentclass[preprint,tightenlines,preprintnumbers,aps,eqsecnum,nofootinbib,superscriptaddress,showpacs,floatfix]{revtex4-1}
\usepackage{longtable}
\usepackage{bm}
\usepackage{amsmath,amssymb,amsbsy}
\usepackage[pdftex]{graphicx,color}
\usepackage[sort&compress]{natbib}
\usepackage[colorlinks=true, linkcolor=blue, filecolor=blue, urlcolor=blue, citecolor=blue, pdftex=true, plainpages=false]{hyperref}

\newcommand{\ie}{\textit{i.e.}}
\newcommand{\cf}{\textit{cf.}}
\newcommand{\eg}{\textit{e.g.}}
\newcommand{\vs}{\textit{vs.}}

\newcommand{\CM}{{\cal M}}

\newcommand{\MeV}{\textrm{MeV}}

\newcommand{\GeV}{\textrm{GeV}}

\newcommand{\fm}{\textrm{fm}}

\newcommand{\spose}[1]{\hbox to 0pt{#1\hss}}

\newcommand{\inapprox}{\mathrel{\spose{\lower 3pt\hbox{$\mathchar"218$}}
 \raise 2.0pt\hbox{$\mathchar"232$}}}

\newcommand{\chpt}{\raise0.4ex\hbox{$\chi$}PT}

\newcommand{\rschpt}{rS\chpt}

\newcommand{\aCoarse}{0.12}
\newcommand{\aFine}{0.09}
\newcommand{\aMediumCoarse}{0.15}

\newcommand{\figref}[1]{Fig.~\ref{fig:#1}}
\newcommand{\Figref}[1]{Figure~\ref{fig:#1}}
\newcommand{\figrefs}[2]{Figs.~\ref{fig:#1} and \ref{fig:#2}}

\newcommand{\tabref}[1]{Table~\ref{tbl:#1}}
\newcommand{\tabrefs}[2]{Tables~\ref{tbl:#1} and \ref{tbl:#2}}
\def\secref#1{Sec.~\ref{sec:#1}}
\def\secrefs#1#2{Secs.~\ref{sec:#1} and \ref{sec:#2}}

\newcommand{\eqn}[1]{\label{eq:#1}}

\newcommand{\eq}[1]{Eq.~(\ref{eq:#1})}

\renewcommand{\case}[2]{\ensuremath{{\textstyle\frac{#1}{#2}}}}

\def\LP{\left(}		
\def\RP{\right)}	
\newcommand{\BE}{\begin{displaymath}}
\newcommand{\EE}{\end{displaymath}}
\newcommand{\BNE}{\begin{equation}}
\newcommand{\ENE}{\end{equation}}
\newcommand{\BEA}{\begin{eqnarray}}
\newcommand{\EEA}{\nonumber\end{eqnarray}}
\newcommand{\EL}{\nonumber\\}

\newcommand{\rhoA}{\ensuremath{\rho_{A^4}}}

\newcommand{\rhoAQq}{\ensuremath{\rho_{A^4_{Qq}}}}
\newcommand{\rhoAbq}{\ensuremath{\rho_{A^4_{bq}}}}
\newcommand{\rhoAcq}{\ensuremath{\rho_{A^4_{cq}}}}
\newcommand{\rhoVQq}{\ensuremath{\rho_{V^1_{Qq}}}}
\newcommand{\ZAQq}{\ensuremath{Z_{A^4_{Qq}}}}
\newcommand{\ZVQq}{\ensuremath{Z_{V^4_{Qq}}}}
\newcommand{\ZVQQ}{\ensuremath{Z_{V^4_{QQ}}}}

\newcommand{\ZVqq}{\ensuremath{Z_{V^4_{qq}}}}

\newcommand{\Zvqq}{\ZVqq}
\newcommand{\ZvQQ}{\ZVQQ}

\newcommand{\aphiH}{\ensuremath{a^{3/2}\phi_H}}

\newcommand{\cSW}{\ensuremath{c_{\mathrm{SW}}}}
\newcommand{\crit}{\ensuremath{\kappa_{\mathrm{crit}}}}
\newcommand{\kapch}{\ensuremath{\kappa_{\mathrm{c}}}}
\newcommand{\kapbot}{\ensuremath{\kappa_{\mathrm{b}}}}
\newcommand{\kaptuned}{\ensuremath{\kappa_{\mathrm{tuned}}}}
\newcommand{\kapsim}{\ensuremath{\kappa_{\mathrm{sim}}}}

\begin{document}

\preprint{FERMILAB-PUB-11/651-T}

\title{\boldmath $B$- and $D$-meson decay constants from three-flavor lattice QCD}

\author{A.~Bazavov}
\affiliation{Physics Department, Brookhaven National Laboratory, Upton, NY, USA}

\author{C.~Bernard}
\affiliation{Department of Physics, Washington University, St.~Louis, Missouri, USA}

\author{C.M.~Bouchard}
\affiliation{Physics Department, University of Illinois, Urbana, Illinois, USA}
\affiliation{Fermi National Accelerator Laboratory, Batavia, Illinois, USA}
\affiliation{Department of Physics, The Ohio State University, Columbus, OH, USA}

\author{C.~DeTar}
\affiliation{Physics Department, University of Utah, Salt Lake City, Utah, USA}

\author{M.~Di~Pierro}
\affiliation{School of Computing, DePaul University, Chicago, Illinois, USA}

\author{A.X.~El-Khadra}
\affiliation{Physics Department, University of Illinois, Urbana, Illinois, USA}

\author{R.T.~Evans}
\affiliation{Physics Department, University of Illinois, Urbana, Illinois, USA}

\author{E.D.~Freeland}
\affiliation{Physics Department, University of Illinois, Urbana, Illinois, USA}
\affiliation{Department of Physics, Benedictine University, Lisle, Illinois, 60532, USA}

\author{E.~G\'amiz}
\affiliation{Fermi National Accelerator Laboratory, Batavia, Illinois, USA}
\affiliation{CAFPE and Depto. de F\'{\i}sica Te\'orica y del Cosmos, Universidad de Granada, Granada, Spain}

\author{Steven~Gottlieb}
\affiliation{Department of Physics, Indiana University, Bloomington, Indiana, USA}

\author{U.M.~Heller}
\affiliation{American Physical Society, Ridge, New York, USA}

\author{J.E.~Hetrick}
\affiliation{Physics Department, University of the Pacific, Stockton, California, USA}

\author{R.~Jain}
\affiliation{Physics Department, University of Illinois, Urbana, Illinois, USA}

\author{A.S.~Kronfeld}
\affiliation{Fermi National Accelerator Laboratory, Batavia, Illinois, USA}

\author{J.~Laiho}
\affiliation{SUPA, School of Physics and Astronomy, University of Glasgow, Glasgow, UK}

\author{L.~Levkova}
\affiliation{Physics Department, University of Utah, Salt Lake City, Utah, USA}

\author{P.B.~Mackenzie}
\affiliation{Fermi National Accelerator Laboratory, Batavia, Illinois, USA}

\author{E.T.~Neil}
\affiliation{Fermi National Accelerator Laboratory, Batavia, Illinois, USA}

\author{M.B.~Oktay}
\affiliation{Physics Department, University of Utah, Salt Lake City, Utah, USA}

\author{J.N.~Simone}
\affiliation{Fermi National Accelerator Laboratory, Batavia, Illinois, USA}

\author{R.~Sugar}
\affiliation{Department of Physics, University of California, Santa Barbara, California, USA}

\author{D.~Toussaint}
\affiliation{Department of Physics, University of Arizona, Tucson, Arizona, USA}

\author{R.S.~Van~de~Water}
\email[E-mail contact: ]{ruthv@bnl.gov}
\affiliation{Physics Department, Brookhaven National Laboratory, Upton, NY, USA}

\collaboration{Fermilab Lattice and MILC Collaborations}
\noaffiliation

\date{\today}

\begin{abstract}
We calculate the leptonic decay constants of $B_{(s)}$ and $D_{(s)}$ mesons in lattice QCD using staggered 
light quarks and Fermilab bottom and charm quarks.   
We compute the heavy-light meson correlation functions on the MILC asqtad-improved staggered gauge 
configurations which include the effects of three light dynamical sea quarks.
We simulate with several values of the light valence- and sea-quark masses (down to $\sim m_s/10$) and at 
three lattice spacings ($a \approx$ 0.15, 0.12, and 0.09~fm) and extrapolate to the physical up and down 
quark masses and the continuum using expressions derived in heavy-light meson staggered chiral perturbation 
theory.
We renormalize the heavy-light axial current using a mostly nonperturbative method such that only a small 
correction to unity must be computed in lattice perturbation theory and higher-order terms are expected 
to be small.
We obtain 
    $f_{B^+} = 196.9(8.9)$~MeV, 
    $f_{B_s} = 242.0(9.5)$~MeV, 
    $f_{D^+} = 218.9(11.3)$~MeV, 
    $f_{D_s} = 260.1(10.8)$~MeV,
and the $\mathrm{SU}(3)$ flavor-breaking ratios 
    $f_{B_s}/f_{B} = 1.229(26)$ and 
    $f_{D_s}/f_{D} = 1.188(25)$, 
where the numbers in parentheses are the total statistical and systematic uncertainties added in quadrature.
\end{abstract}

\maketitle

\section{Introduction}
\label{sec:Intro}

Leptonic decays of $B$ and $D$ mesons, in which the hadron annihilates weakly to a $W$ boson, are important probes of heavy-to-light quark flavor-changing interactions.
When combined with a nonperturbative lattice QCD calculation of the 
heavy-light pseudoscalar meson decay constant, $f_B$ or $f_D$, a precise experimental measurement 
of the leptonic decay width allows the determination of the Cabibbo-Kobayashi-Maskawa (CKM) quark-mixing matrix element $|V_{ub}|$ or $|V_{cd}|$.
Conversely, if the relevant CKM matrix element is known from an independent process such as semileptonic decay or from CKM-unitarity 
constraints, a comparison of the decay constant from lattice QCD simulations with that measured by experiment provides a straightforward test of the Standard Model.
As the lattice and experimental determinations become more precise, this test will become more sensitive and may ultimately reveal, through the appearance of a discrepancy, the presence of new physics in the quark flavor sector.

Improved determinations of the $B$ meson decay constant $f_B$ are of particular importance given the 
current, approximately 3-$\sigma$ tension in the CKM unitarity triangle that may indicate the presence of 
new physics in $B_d$-mixing or  $B \to \tau \nu$ decay~\cite{Lenz:2010gu,Lunghi:2010gv,Laiho:2011nz}.
The experimental uncertainty in the branching fraction ${\mathcal{B}}(B \to \tau \nu)$ is at present 
$\sim30\%$~\cite{Hara:2010dk,:2010rt}, but this error is expected to be reduced to $\sim 10\%$ by Belle II 
at KEK-$B$ in as little as five or six years~\cite{Masuzawa:2010zz,Iijima:HINTS09}, at which point even 
modest improvements in the determination of $f_B$ will significantly help constrain the apex of the CKM 
unitarity triangle and isolate the source of new physics~\cite{Lunghi:2009ke}.
 
Because leptonic decays are ``gold-plated" processes in numerical 
lattice QCD simulations (they have a single stable hadron in the 
initial state and no hadrons in the final state~\cite{Davies:2003ik}), they can be determined accurately using present lattice methods.  
Currently all realistic lattice calculations of  $f_{D_{(s)}}$ and $f_{B_{(s)}}$ that include the effects of three light dynamical quarks use staggered lattice 
fermions~\cite{Susskind:1976jm,Sharatchandra:1981si} for the up, down, and strange quarks.  Because staggered fermions are computationally cheaper than other lattice fermion formulations, 
they allow for QCD simulations with dynamical quarks as light as $0.05m_s$, several lattice spacings, down to $a \approx 0.045$~fm, large physical volumes, and high statistics.  
This enables lattice determinations of many light-light and heavy-light meson quantities with controlled systematic uncertainties.  
The results of staggered lattice calculations are largely in excellent numerical agreement with experimental results \cite{Davies:2003ik}.  
This includes both postdictions, such as the pion and kaon decay constants \cite{Aubin:2004fs}, and predictions, as in the case of the $B_c$ meson mass \cite{Allison:2004be}.  
Such successes give confidence that further calculations using the same methods are reliable.   
This is essential if lattice QCD calculations of hadronic weak matrix elements are to be used to test the Standard Model and search for new physics.  

The staggered dynamical quark simulations used here employ the fourth-root procedure 
(``rooting'') for eliminating unwanted extra quark degrees of freedom that
arise from lattice fermion doubling. The rooting method is not standard 
quantum field theory, and at nonzero lattice spacing it
leads to violations of unitarity \cite{Prelovsek:2005rf,Bernard:2006zw,Bernard:2007qf,Aubin:2008wk}
that can be considered nonlocal~\cite{Bernard:2006ee}.
Nevertheless, there are strong arguments \cite{Shamir:2004zc,Shamir:2006nj}
that the desired local, unitary theory of QCD is reproduced by the rooted staggered
lattice theory in the continuum limit.
Further, one can show \cite{Bernard:2006zw,Bernard:2007ma} that the  
unitarity-violating lattice artifacts in the pseudo-Goldstone boson sector can be
described, and hence removed, using  rooted staggered chiral
perturbation theory (\rschpt), which is a  low-energy effective description of the rooted
staggered lattice theory~\cite{Lee:1999zxa,Aubin:2003mg,Sharpe:2004is}.
When coupled with other analytical and numerical
evidence (see Refs.~\cite{Sharpe:2006re,Kronfeld:2007ek,Golterman:2008gt,Bazavov:2009bb} for reviews
and Ref.~\cite{Donald:2011if} for a recent study),
this gives us confidence that the rooting procedure is valid.
Indeed, the validity of the rooted staggered lattice simulations is 
of critical importance to flavor
physics phenomenology,  since a majority of the unquenched, 
three-flavor lattice results for hadronic weak matrix elements used to determine CKM matrix 
elements and as inputs to constraints on the CKM unitarity triangle come from 
such simulations~\cite{Laiho:2009eu}. 

\bigskip

In this paper, we present new results for the leptonic decay constants of heavy-light mesons containing bottom and charm quarks.  
We use the ``2+1" flavor asqtad-improved gauge configurations made publicly-available by the MILC Collaboration~\cite{Bernard:2001av}.  
These ensembles include the effects of three light, dynamical sea-quark flavors: one with mass~$m_h$ near $m_s$ (the physical strange-quark mass) and the other two with mass $m_l$ as small as~$0.1m_h$.
We generate light valence quarks for the $B$ and $D$ mesons using the same staggered action as in the sea sector, and generate heavy bottom and charm quarks using the clover action~\cite{Sheikholeslami:1985ij} 
with the Fermilab interpretation~\cite{ElKhadra:1996mp}. 
Because the Fermilab method uses knowledge of the heavy-quark limit of QCD to systematically eliminate heavy-quark discretization 
errors, exploiting ideas of Symanzik~\cite{Symanzik:1983dc,Symanzik:1983gh} and of heavy-quark   
effective theory (HQET)~\cite{Kronfeld:2000ck,Harada:2001fi,Harada:2001fj},
it is well-suited for both bottom and charm quarks.
We simulate with many values for the light up/down quark mass (the mass of our lightest pion in both the sea and valence sectors is $\approx 250$ MeV), 
and at three lattice spacings ranging from $a\approx 0.09$~fm to $a\approx 0.15$~fm.  We then extrapolate our numerical lattice data to the physical up and down quark masses and continuum guided by expressions 
derived in staggered chiral perturbation theory for heavy-light mesons  (HMS$\chi$PT)~\cite{Aubin:2005aq,Laiho:2005ue,Aubin:2007mc}.  

We renormalize the heavy-light axial current with a mostly 
nonperturbative approach, computing the flavor-diagonal (heavy-heavy 
and light-light) renormalization factors nonperturbatively and then 
calculate the remaining flavor off-diagonal correction factor 
($\rhoAQq$) in lattice perturbation theory \cite{ElKhadra:2001rv,Harada:2001fi,ElKhadra:2007qe}.  
This procedure has the advantage that $\rhoAQq$ is close to unity.  
Furthermore, tadpole diagrams cancel in the ratio needed to obtain $\rhoAQq$, thereby improving the convergence of the perturbative series.  
Empirically, the size of the 1-loop contribution to $\rhoAQq$ is found to be small.

Our results for the charmed-meson decay constants improve upon our 
published results for $f_D$ and $f_{D_s}$ in Ref.~\cite{Aubin:2005ar} 
in several ways.  The coarsest lattices used in this work have a 
smaller lattice spacing ($a\approx 0.15$~fm) than those used in our 
previous work ($a\approx0.18$~fm).  The number of configurations in the 
two  most chiral ensembles with $a\approx0.12$~fm has been
increased, approximately by factors of 1.4 (sea $m_l=0.1m_h$) and 1.7 (sea $m_l=0.14m_h$).  We 
have added new data on a new $a\approx0.09$~fm sea-quark ensemble with 
a light quark mass of $0.1m_h$. 
We now obtain our 
results from a combined analysis of our entire data set (all 
partially-quenched mass combinations and lattice spacings).  
Furthermore, we 
now compute the bottom meson decay constants $f_B$ and $f_{B_s}$.  
We have presented reports on this project at several conferences~\cite{Bernard:2006zz,Bernard:2007zz,Bernard:2009wr,Bazavov:2009ii,Simone:2010zz}; in our final analysis of this data set we also improve upon 
bottom and charm quark mass-tuning, with increased statistics and a more sophisticated analysis of heavy-quark discretization effects. 

\bigskip

This paper is organized as follows.  In Sec.~\ref{sec:Method}, we present an overview of the calculation, including the gluon and light-quark actions used in generating 
the gauge configurations and the light- and heavy-quark actions used 
in constructing the heavy-light meson correlators.
We also introduce the mostly nonperturbative method for matching the 
lattice heavy-light current to the continuum, and the treatment of heavy-quark discretization errors from the Fermilab action within our chiral-continuum extrapolation.
Next, in 
Sec.~\ref{sec:SimDetails}, we describe the details of our numerical 
simulations and we present the parameters used, such as the light-quark 
masses and lattice spacings.  We also describe the procedure for tuning 
the hopping parameter in the clover action so that it corresponds to 
$b$ and $c$ quarks.  
In Sec.~\ref{sec:2point}, we define the two-point correlation functions used to extract the decay constant at each value of the light-quark mass and lattice spacing.
We use two different fitting procedures to obtain the decay constants that differ in their treatment of the statistical errors, choice of fit ranges and number of states, and choice of input correlators.
We include the difference between the two in our estimate of the fitting systematic uncertainty.
Next, we present the numerical details of the calculation of the heavy-light axial-current renormalization factor in Section~\ref{sec:HLCurrents}.
Putting the results of the two previous sections together, in Sec.~\ref{sec:ChPT}, we extrapolate the renormalized decay constant data at unphysical quark masses and nonzero lattice spacing to the physical light quark masses and zero lattice spacing using HMS$\chi$PT.  
In Sec.~\ref{sec:Errors}, we estimate the contributions of the various 
systematic uncertainties to the decay constants, discussing each item 
in our error budget separately.  We present the final results for the 
decay constants in Sec.~\ref{sec:Results}, and compare them to other 
lattice QCD calculations and to experiment.  We describe the impact of 
our results for current flavor physics phenomenology and then conclude 
by discussing the ongoing improvements to our calculations, and their 
future impact on searches for new physics in the quark flavor 
sector.

Appendix~\ref{app:HQcutoff} applies HQET to the Fermilab action to obtain explicit expressions for heavy-quark discretization effects.
Appendix~\ref{apdx:twoPointFitsJackknife} contains the complete set of fit results for the heavy-light pseudoscalar meson mass and renormalized decay constant for all combinations of sea-quark mass, light valence-quark mass, and heavy-quark mass used in the chiral-continuum extrapolation.  These results will be included as an EPAPS attachment upon publication.

\section{Methodology}
\label{sec:Method}

The decay rate for a charged pseudoscalar meson $H$ (with flavor content $Q$ and $\bar{q}$) to leptons is, 
in the Standard Model,
\begin{equation}
	\Gamma(H\to\ell\nu) = \frac{M_H}{8\pi} f_H^2 \left| G_F V^*_{Qq}m_\ell\right|^2 
        \left(1-\frac{m_\ell^2}{M_H^2}\right)^2,
	\label{eq:methods:rate}
\end{equation}
where
$M_H$ is the mass of the meson~$H$, 
$G_F$~is the Fermi constant, and
$V_{Qq}$ is the pertinent element of the CKM matrix.
The decay constant $f_H$ parameterizes the pseudoscalar-to-vacuum matrix element of the axial 
vector current,
\begin{equation}
	\left\langle0|\mathcal{A}^\mu|H(p)\right\rangle = ip^\mu f_H,
	\label{eq:0AD}
\end{equation}
where $p^\mu$ is the 4-momentum of the pseudoscalar meson.
The flavor contents of the associated vector current and CKM matrix element are 
given in Table~\ref{tab:methods:flavor}.
\begin{table}[t]
	\centering
	\caption[tab:methods:flavor]{Flavor content of the axial vector 
	current and associated CKM matrix element.}
	\label{tab:methods:flavor}
	\begin{tabular}{c@{\quad}c@{\quad}c}
		\hline\hline
		$H$ & $\mathcal{A}^\mu$ & $V$  \\
		\hline
		$D$   & $\bar{d}\gamma^\mu\gamma^5 c$ & $V^*_{cd}$ \\
		$D_s$ & $\bar{s}\gamma^\mu\gamma^5 c$ & $V^*_{cs}$ \\
		$B$   & $\bar{b}\gamma^\mu\gamma^5 u$ &  $V_{ub}$  \\
		$B_s$ & $\bar{b}\gamma^\mu\gamma^5 s$ &     ---     \\
		\hline\hline
	\end{tabular}
\end{table}
Note that the neutral $B_s$ decays to a charged lepton pair with an amplitude proportional to~$f_{B_s}$; 
hence the CKM factor in the decay rate involves more than one CKM matrix element.
Because this process is loop-suppressed in the Standard Model, it is potentially sensitive to new physics 
effects.
These formulas hold for all pseudoscalar mesons; in the normalization convention used here, 
$f_\pi(|V_{ud}|/0.97425)=130.41\pm0.20~\textrm{MeV}$ \cite{Rosner:2010ak}.

In Eq.~\eqref{eq:0AD}, the 1-particle state assumes the relativistic 
normalization convention.
For mesons containing a heavy quark, however, it is more convenient to 
pull out factors of $M_H$ to ensure a smooth $M_H\to\infty$ limit:
\begin{equation}
 	\left\langle0|\mathcal{A}^\mu|H(p)\right\rangle (M_H)^{-1/2} = 
		i(p^\mu/M_H) \phi_H.
	\label{eq:0AH}
\end{equation}
In lattice QCD, the normalization of states on the left-hand side falls 
out of correlation functions more naturally.
Thus, most of our analysis, including error analysis, focuses on~$\phi_H$.
We then obtain $f_H=\phi_H/\sqrt{M_H}$ using the experimentally measured 
value of the meson mass~\cite{Nakamura:2010zzi}.

\bigskip

To compute the decay constants with lattice gauge theory, we must 
choose a discretization for the heavy quark, the light quark, and the 
gluons.
As in previous work \cite{Aubin:2004ej, 
Aubin:2005ar,    
Bernard:2008dn,  
Bailey:2008wp,   
Bernard:2010fr}, 
we choose the Fermilab method for heavy quarks~\cite{ElKhadra:1996mp}
and staggered quarks with the asqtad action~\cite{Lepage:1998vj} 
for the light (valence) quark.  The gauge action is Symanzik improved, with couplings chosen to remove 
order $\alpha_s a^2$ errors from gluon loops~\cite{Luscher:1985zq}, but 
not those from quark loops~\cite{Hao:2007iz} (which became available 
only after the gauge-field generation was well underway).

For heavy bottom and charm quarks, we use the Sheikholeslami-Wohlert (SW) ÒcloverÓ 
action~\cite{Sheikholeslami:1985ij} with the Fermilab interpretation~\cite{ElKhadra:1996mp}, which connects to the continuum limit as $am_Q\to0$.
This is an extension of the Wilson action~\cite{Wilson:1975id}, which retains the Wilson action's smooth 
limit as $am_Q\to\infty$ and also remains well behaved for $m_Qa\approx1$.
Because this lattice action respects heavy-quark spin-flavor symmetry, one can apply HQET to organize the 
discretization effects. 
In essence, one uses HQET to develop the $1/m_Q$ expansion both for continuum QCD and for lattice gauge 
theory (LGT)~\cite{Kronfeld:2000ck,Harada:2001fi,Harada:2001fj}. 
Discretization effects are then captured order-by-order in the heavy-quark expansion by the difference of 
the short-distance coefficients in the descriptions of QCD and LGT. 
Thus, in principle, the lattice heavy-quark action can be improved to arbitrarily high orders in $1/m_Q$ by 
adjusting a sufficiently large number of parameters in the lattice action. 
(See Ref.~\cite{Oktay:2008ex} for details at dimension 6 and~7. 
In principle, the adjustment can be done nonperturbatively, such as in the scheme of 
Ref.~\cite{Lin:2006ur}.)
In practice, we tune the hopping parameter $\kappa$ and the clover coefficient \cSW\ of the SW action, to 
remove discretization effects through order $1/m_Q$ in the heavy-quark expansion. 

The HQET analysis of cutoff effects could be applied to any lattice action with heavy-quark symmetry, such 
as the action of lattice NRQCD~\cite{Lepage:1992tx}. 
In the latter case, it is simply a different perspective on the usual approach to lattice NRQCD, which 
derives the heavy-quark Lagrangian formally, and then replaces derivatives with difference operators.
A key feature of the Wilson, SW, Fermilab and OK~\cite{Oktay:2008ex} actions is their well-behaved continuum limit, which is especially important for charm.
For $m_Qa<1$, one can analyze the cutoff effects in a complementary way with the Symanzik effective action~\cite{Symanzik:1983dc,Symanzik:1983gh}. 
This two-pronged attack shows that the difference of short-distance coefficients, mentioned above, vanishes as a suitable power of lattice spacing~$a$.
In this paper, we shall use our knowledge of this behavior to constrain heavy-quark discretization effects in several steps of our analysis.
See Secs.~\ref{sec:tuning}, \ref{sec:ChPT}, and Appendix~\ref{app:HQcutoff} for details.

\bigskip

The lattice and continuum currents are related by a matching factor~$Z_{A^\mu}$~\cite{Harada:2001fi}:
\begin{equation}
    Z_{A^\mu} A^\mu \doteq \mathcal{A}^\mu +
        \mathrm{O}\left(\alpha_s a\Lambda f_i(m_Qa)\right) +
        \mathrm{O}\left(a^2\Lambda^2 f_j(m_Qa)\right),
    \label{eq:methods:ZA=A}
\end{equation}
where $\doteq$ denotes equality of matrix elements, and the functions $f_{i,j}$ that depend on $m_Qa$ stem from the difference in the HQET short-distance coefficients.
In the Fermilab method, they remain of order~1 for all values of $m_Qa$~\cite{ElKhadra:1996mp,Oktay:2008ex}, and they are given explicitly in Appendix~\ref{app:HQcutoff}.
In this work, we compute $Z_{A^\mu}$ mostly nonperturbatively~\cite{ElKhadra:2001rv} and partly in one-loop perturbation theory. 
As shown in the analysis of Ref.~\cite{Harada:2001fi}, many of the Feynman diagrams in the perturbative expansion of \ZAQq\ are common or 
similar to those in the flavor-conserving renormalization factors \ZVQQ\ and \ZVqq, which can be computed nonperturbatively.
Therefore, we define \rhoAQq\ by
\begin{equation}
    \ZAQq = \rhoAQq \sqrt{\ZVqq\ZVQQ},
    \label{eq:rho}
\end{equation}
evaluating only \rhoAQq\ in lattice perturbation theory.

The flavor-conserving factors account for most of the value of the heavy-light renormalization factor \ZAQq.  They are obtained by enforcing the 
normalization condition, at zero momentum transfer,
\begin{equation}
    1 = \ZVqq \langle H_q|V^4_{qq}|H_q\rangle,
    \label{eq:lightZvNorm}
\end{equation}
where $H_q$ is a hadron containing a single quark of flavor~$q$, and $V^\mu_{qq}$ is the lattice version of the degenerate vector current.
This condition holds for all discretizations and quark masses and, hence, the heavy quark (\ie, \ZVQQ) as well.
The remaining correction factor \rhoAQq\ is close to unity due to the cancellation of most of the radiative corrections including tadpole graphs.
Although such cancellations have only been explicitly shown at 1-loop in lattice perturbation theory~\cite{Harada:2001fi,ElKhadra:2007qe}, we expect similar cancellations to persist at higher orders.
Therefore, the perturbative truncation error in the heavy-light renormalization factor is subdominant.

\section{Lattice Simulation Details}
\label{sec:SimDetails}

\subsection{Parameters}
\label{sec:Params}

\begin{table}[tp]
\centering
\caption{The MILC three-flavor lattices and valence asqtad quark masses
used in this work.
All of the valence masses were used in version II of the correlator fits (Sec~\ref{sec:bootstrap}),
while only the ones in bold print were used in version I (Sec~\ref{sec:jackknife}).}
\label{tbl:latticeAndValenceDetails}
\begin{tabular*}{\textwidth}{l@{\extracolsep{\fill}}ll@{\extracolsep{\fill}}c@{\extracolsep{\fill}}c@{\extracolsep{\fill}}r@{$\times\kern-0.4em$}l@{\extracolsep{\fill}}l}
	\hline\hline
	$\approx a\;[\fm]$ &$am_h$ & $~am_l$ & ~~$u_0$ & $r_1/a$& $n_{\rm conf}$ & $n_{\rm src}$ &\quad valence $am_q$ \\
	\hline	   	          
	$0.09$     &$0.031$  &$0.0031$   &$0.8779$   &$3.69$  &$435$ & $4$   &${\bf 0.0031},0.0037,0.0042,{\bf 0.0044},0.0052,{\bf 0.0062},$ \\
&&&&&\multicolumn{2}{c}{ }&${\bf 0.0087},{\bf 0.0124},{\bf 0.0186},{\bf 0.0272},{\bf 0.031}$ \\
	           &         &$0.0062$    &$0.8782$ &$3.70$  &$557$ & $4$   &${\bf 0.0031},0.0037,{\bf 0.0044},0.0052,{\bf 0.0062},$ \\
&&&&&\multicolumn{2}{c}{ }&${\bf 0.0087},{\bf 0.0124},{\bf 0.0186},{\bf 0.0272},{\bf 0.031}$ \\
		   &         &$0.0124$  &$0.8788$  &$3.72$  &$518$ & $4$   &${\bf 0.0031},{\bf 0.0042},{\bf 0.0062},{\bf 0.0087},{\bf 0.0124},$ \\
&&&&&\multicolumn{2}{c}{ }&${\bf 0.0186},{\bf 0.0272},{\bf 0.031}$ \\
	$0.12$     &$0.05$   &$0.005$    &$0.8678$  &$2.64$  &$678$ & $4$   &${\bf 0.005},0.006,{\bf 0.007},0.0084,{\bf 0.01},0.012,{\bf 0.014},$ \\
&&&&&\multicolumn{2}{c}{ }&$0.017,{\bf 0.02},0.024,{\bf 0.03},{\bf 0.0415}$ \\
	           &         &$0.007$    &$0.8678$  &$2.63$  &$833$ & $4$   &${\bf 0.005},0.006,{\bf 0.007},0.0084,{\bf 0.01},0.012,{\bf 0.014},$ \\
&&&&&\multicolumn{2}{c}{ }&$0.017,{\bf 0.02},0.024,{\bf 0.03},{\bf 0.0415}$ \\
	           &         &$0.01$     &$0.8677$ &$2.62$  &$592$ & $4$   &${\bf 0.005},0.006,{\bf 0.007},0.0084,{\bf 0.01},0.012,{\bf 0.014},$ \\
&&&&&\multicolumn{2}{c}{ }&$0.017,{\bf 0.02},0.024,{\bf 0.03},{\bf 0.0415}$ \\
	           &         &$0.02$    &$0.8688$ &$2.65$  &$460$ & $4$   &${\bf 0.005},0.006,{\bf 0.007},0.0084,{\bf 0.01},0.012,{\bf 0.014},$ \\
&&&&&\multicolumn{2}{c}{ }&$0.017,{\bf 0.02},0.024,{\bf 0.03},{\bf 0.0415}$ \\
	           &         &$0.03$     &$0.8696$  &$2.66$  &$549$ & $4$   &${\bf 0.005},0.006,{\bf 0.007},0.0084,{\bf 0.01},0.012,{\bf 0.014},$ \\
&&&&&\multicolumn{2}{c}{ }&$0.017,{\bf 0.02},0.024,{\bf 0.03},{\bf 0.0415}$ \\
	$0.15$     &$0.0484$ &$0.0097$  &$0.8604$ &$2.13$  &$631$ & $4$   &${\bf 0.0048},{\bf 0.007},{\bf 0.0097},0.013,{\bf 0.0194},0.0242,$ \\
&&&&&\multicolumn{2}{c}{ }&${\bf 0.029},0.0387,{\bf 0.0484}$ \\
	           &         &$0.0194$   &$0.8609$ &$2.13$  &$631$ & $4$   &${\bf 0.0048},{\bf 0.007},{\bf 0.0097},0.013,{\bf 0.0194},0.0242,$ \\
&&&&&\multicolumn{2}{c}{ }&${\bf 0.029},0.0387,{\bf 0.0484}$ \\
	           &         &$0.029$  &$0.8614$  &$2.13$  &$576$ & $4$   &${\bf 0.0048},{\bf 0.007},{\bf 0.0097},0.013,{\bf 0.0194},0.0242,$ \\
&&&&&\multicolumn{2}{c}{ }&${\bf 0.029},0.0387,{\bf 0.0484}$ \\
	\hline\hline
\end{tabular*}
\end{table}

\tabref{latticeAndValenceDetails} lists the subset of the ensembles of lattice gauge 
fields generated by the MILC Collaboration~\cite{Bazavov:2009bb} used in this analysis.  
We now describe each entry in the table.

\bigskip

We analyze data at three lattice spacings: $a\approx 0.15$~fm, $a\approx 0.12$~fm, and $a\approx 0.09$~fm.
The ensembles contain 2+1 flavors of sea quarks, using the asqtad-improved staggered 
action~\cite{Lepage:1998vj}, and the square (fourth) root of the 
staggered determinant for the two~degenerate light sea quarks (one strange 
sea quark).
The sea contains one flavor with mass $m_h$ close to the physical strange quark mass and two
degenerate lighter flavors of mass $m_l$.  The tadpole improvement factor
$u_0$ is a parameter of the gauge and asqtad staggered (sea) quark
action and is determined from the fourth root of the average
plaquette.
We calculate the two-point correlation functions on each ensemble from an average over four different time 
sources. 

The relative lattice scale is determined by calculating $r_1/a$ on each ensemble, where $r_1$ is related to the force between static quarks, $r_1^2F(r_1)=1.0$ \cite{Sommer:1993ce,Bernard:2000gd}.
\tabref{latticeAndValenceDetails} lists $r_1/a$ values for each of the ensembles that result from fitting the calculated $r_1/a$ to a smooth function~\cite{Allton:1996kr}, as explained in Eqs.~(115) and (116) of 
Ref.~\cite{Bazavov:2009bb}. 

In order to
fix the absolute lattice scale, one must compute a physical
quantity which can be compared directly to experiment.
The combination of the PDG's value of $f_\pi$ with MILC's 2009 determination of $r_1f_\pi$ \cite{Bazavov:2009fk} yields $r_1=0.3117(6)({}^{+12}_{-31})$~fm.
From an average of three methods for scale setting, including one based on $\Upsilon$ splittings, the HPQCD collaboration obtains $r_1=0.3133(23)(3)$~fm \cite{Davies:2009tsa}, consistent with MILC. 
Symmetrizing MILC's error range gives $r_1=0.3108(21)$~fm, and a straightforward
average with the HPQCD result then yields $r_1=0.3120(16)$~fm.  
This average omits likely correlations, due to the use of MILC sea-quark configurations by
both groups.  Conservatively assuming a 100\% correlation, we inflate the error
to $0.0022$~fm.
Finally, for convenience, we also shift the central value slightly, back to the 2009 MILC central value.
We thus take $r_1=0.3117(22)$~fm in this paper.

The complete list of light (asqtad) valence quark masses $m_q$ simulated in this analysis is also given in 
\tabref{latticeAndValenceDetails}. 
The mass values are selected to be roughly logarithmically spaced, but to
also include the set of light sea quark masses simulated at each
lattice spacing. We use a multimass solver to compute the valence
quark propagators. The marginal numerical cost of including masses
heavier than our lightest $m_q\sim0.1m_s$ is small and logarithmic
spacing is designed to constrain the chiral logarithms.

\begin{table}
\caption{Table of clover coefficients and $\kappa$ values for charm and bottom used
in heavy-light two-point simulations.}
\label{tbl:runKappaValues}
\begin{tabular}{llccc}
\hline\hline
&&&\multicolumn{2}{c}{$\kappa_{sim}$}\\
$\approx a$~[fm]~ &$am_l/am_h$  & ~~$\cSW$~~ & charm    & bottom \\ \hline
0.09                 &0.0031/0.031  &1.478          & 0.127          & 0.0923 \\
                     &0.0062/0.031  &1.476          &               &       \\
                     &0.0124/0.031  &1.473          &               &       \\
0.12                &0.005/0.05    &1.72           & 0.122          & 0.086  \\
                     &0.007/0.05    &1.72           &               &       \\
                     &0.01/0.05     &1.72           &               &       \\
                     &0.02/0.05     &1.72           &               &       \\
                     &0.03/0.05     &1.72           &               &       \\
0.15                 &0.0097/0.0484 &1.570          & 0.122          & 0.076  \\
                     &0.0194/0.0484 &1.567          &               &       \\
                     &0.0290/0.0484 &1.565          &               &       \\
\hline\hline
\end{tabular}
\end{table}

\bigskip

In \tabref{runKappaValues}, we show the coefficient of the
Sheikholeslami-Wohlert term $\cSW$ of the clover action and
the $\kappa$ values used to compute heavy-light two-point functions. 
The coefficient of the clover term is set to the tadpole-improved tree-level value
$\cSW=u_0^{-3}$. For the $a\approx 0.09$ and $0.15$ ensembles the tadpole 
coefficient is taken from the average plaquette.
We note, however, that at lattice spacing $a\approx 0.12\,\fm$ the tadpole coefficient
$u_0$ appearing in both the valence asqtad action and the heavy quark clover action
is taken from the average of the Landau link evaluated on the $am_l/am_h=0.01/0.05$ 
ensemble. Hence,
in our $a\approx 0.12~\fm$ lattice data there is a mismatch between light valence and
sea quark mass definitions.
As discussed in Sec.~\ref{sec:Errors}, this (inadvertent) choice leads to a small error in the decay constants.
We have remedied this mismatch by using the plaquette $u_0$ everywhere in new runs started while this analysis was underway.

The charm and bottom kappa values listed in \tabref{runKappaValues} are based on our initial kappa tuning 
analysis using about one fourth of our final statistics. 
We then used a larger data set to refine our determination of the $\kappa$ values corresponding to bottom 
and charm as described in the next subsection. 
We adjust our data post-facto to correspond to tuned values of $\kapch$ and $\kapbot$ using the measured 
value of the derivative $\delta \phi/\delta \kappa$.

\subsection{\boldmath Input quark masses $m_c$ and $m_b$}
\label{sec:tuning}

Our method for tuning $\kappa$ for charm and bottom quarks closely
follows that of Ref.~\cite{Bernard:2010fr}, where further details can be found.
We start with the dispersion relation for a
heavy particle on the lattice~\cite{ElKhadra:1996mp}
\begin{equation}
    E^2(\bm{p})= M_1^2 + \frac{M_1}{M_2}\bm{p}^2 + 
        \frac{1}{4} A_4\,(a\bm{p}^2)^2 + 
        \frac{1}{3} A_{4'} a^2 \sum_{j=1}^3 |p_j|^4 + \ldots ,  
    \label{eq:disprel}
\end{equation}
where  
\begin{equation}
    M_1 \equiv E(\bm{0})
\end{equation}
is called the rest mass, and the kinetic mass is given by
\begin{equation}
    M_2^{-1} \equiv \left.\frac{\partial E(\bm{p})}{\partial p_j^2} 
        \right|_{\bm{p}=\bm{0}} .
\end{equation}
These meson masses differ from corresponding quark masses, $m_1$ and 
$m_2$, by binding-energy effects.
The bare mass or, equivalently, the hopping parameter~$\kappa$ must be 
adjusted so that these masses reproduce an experimental charmed or 
$b$-flavored meson mass.
When they differ, as they do when $m_Qa\not\ll1$, one must choose.
Decay constants are unaffected by the heavy-quark rest mass 
$m_1$~\cite{Kronfeld:2000ck}, so it does not make sense to adjust the 
bare mass to $M_1$.  
We therefore focus on $M_2$,  adjusting $\kappa$ to the strange 
pseudoscalars $D_s$ and $B_s$, both because the signal degrades for lighter 
spectator masses and because this avoids introducing an unnecessary systematic uncertainty due to a chiral extrapolation.  

The first step is to compute the correlator 
$C_2^{(S_1S_2)}(t,\bm{p})$ in Eq.~\eqref{eq:methods:corr4norm} (below) for several 
3-momenta $\bm{p}$ and several values of $\kappa$  and light quark mass, 
bracketing charm and bottom, and strange, respectively.
We use all momenta such that $|\bm{p}|\le4\pi/L$.
Second, we fit the time dependence of the multichannel 
correlation matrix $C_2^{(S_1S_2)}$ to a sum
of exponentials---including the usual staggered-fermion oscillating
terms---and extract the ground state energy $aE(\bm{p})$ by minimization 
of an augmented $\chi^2$~\cite{Bernard:2010fr,Lepage:2001ym,Morningstar:2001je}.
Third, we fit the energies to the dispersion relation given
in Eq.~\eqref{eq:disprel}, through $\mathrm{O}(p_i^4)$.
The output of this fit is $aM_1$, $M_1/M_2$, $A_4$, 
and~$A_{4'}$, all as functions of $\kappa$.
Fourth, we form $M_2(\kappa)$ from the first two fit outputs and $r_1/a$,
propagating the error with a single-elimination jackknife. 
Finally, we obtain our tuned $\kapch$ and $\kapbot$ by interpolating in $\kappa$ so that
$M_2(\kappa)$ matches the experimentally known $D_s$ and $B_s$ masses.
The $\kappa$ values used to compute $M_2$ are listed in
Table~\ref{tbl:kappaInputsMethodA}.
For each of the lattice spacings listed, we used the ensemble with
light-to-strange sea-quark mass ratio $am_l/am_h=0.2$.
The resulting tuned values of $\kapch$ and \kapbot\ are shown with errors in
Table~\ref{tbl:kappaFromMesonDispersion}.

\begin{table}[tp]
\centering
    \caption{Hopping-parameter values used to compute the dispersion relation.}
  \begin{tabular}{l@{\quad}c@{\quad}l@{\quad}l} \hline\hline
  	& & \multicolumn{2}{c}{$\kappa_Q$} \\
    $\approx a$~[fm]  & $n_\textrm{conf} \times n_\textrm{src}$   & charm & bottom \\ \hline
    0.09    &$1912\times 4$   &0.1240, 0.1255, 0.1270 &0.090, 0.092, 0.094 \\
    0.12    &$592\times 4$   &0.114, 0.117, 0.119, 0.122, 0.124 &0.074, 0.086, 0.093, 0.106  \\
    0.15    &$631\times 8$   &0.100, 0.115, 0.122, 0.125 &0.070, 0.076, 0.080, 0.090 \\
  \hline\hline
  \end{tabular}
  \label{tbl:kappaInputsMethodA}
\end{table}

\begin{table}[tp]
    \centering
    \caption{Hopping parameter values \kapch\ and \kapbot\ corresponding to 
    charm and bottom.
    The outputs of the tuning are labeled \kaptuned, where the first 
    error is from statistics and the second is from $r_1$, which enters 
    through matching to the experimentally-measured $D_s$ and $B_s$ meson masses.
    The derivative $d\phi/d\kappa$ is used to correct the values of $\phi$ obtained with the simulated values $\kapsim$ listed in Table~\ref{tbl:runKappaValues} to the tuned values given below. }
    \begin{tabular}{l@{\quad}l@{\quad}r@{.}l@{\qquad}l@{\quad}r@{.}l} \hline\hline
    	&  \multicolumn{3}{c}{charm} & \multicolumn{3}{c}{bottom} \\
        $\approx a$~[fm]  & \kaptuned   & \multicolumn{2}{l}{$d\phi/d\kappa$} &  \kaptuned   & \multicolumn{2}{l}{$d\phi/d\kappa$}\\ \hline
        0.09     & 0.12691(18)(13)    & $-21$&66 &  0.0959(13)(3)     & $-7$&41 \\
        0.12     & 0.12136(37)(19)   & $-18$&23 &  0.0856(19)(3)     &  $-6$&82 \\
        0.15    & 0.12093(36)(24)    & $-15$&40 &  0.0788(11)(3)      &  $-6$&07 \\
    \hline\hline
    \end{tabular}
    \label{tbl:kappaFromMesonDispersion}
\end{table}

We constrain the coefficients $A_4$ and $A_{4'}$ with Gaussian priors derived from the HQET theory of cutoff effects,
adding the contribution of the priors to the $\chi^2$ in the minimization procedure~\cite{Lepage:2001ym,Morningstar:2001je}.
(In principle, we could include such priors for $M_1$ and $M_1/M_2$ too,
but in practice we take priors so wide that these fit parameters are 
solely data-driven.)
Neglecting binding energies, we have exact tree-level expressions for 
$a_4$ and $a_{4'}$, the quark analogs of $A_4$ and $A_{4'}$.
The differences $A_4-a_4^{[0]}$ and $A_{4'}-a_{4'}^{[0]}$ stem from both perturbative and nonperturbative effects.
The asymptotics of the former can be estimated along the lines of Appendix~\ref{app:disprel}, and the latter can be deduced following the methods of 
Refs.~\cite{Kronfeld:2000ck,Kronfeld:1996uy}.
Briefly, we constrain $A_n(\kappa)$, $n\in\{4,4'\}$, to a Gaussian with central value
\begin{equation}
    a^{[0]}_n(m_0a) + \alpha_s a^{[1]}_n(m_0a) + \bar{\Lambda}a\,A'_n(m_0a).
    \label{eq:HQM:disprelprior}
\end{equation}
Here $a^{[0]}_n$ is the exact tree-level contribution,
$a^{[1]}_n$ is an estimate of the one-loop contribution, and
$A'_n$ is an expression for the binding-energy contribution.
The width of the Gaussian is determined by combining in quadrature the 
chosen widths of the separate contributions, 
as outlined in Appendix~\ref{app:disprel}.

The details of the $\kapch$ and \kapbot\ determination differ from that of Ref.~\cite{Bernard:2010fr} in two respects.
First, we use the pseudoscalar meson masses rather than the
spin average of pseudoscalar and vector meson masses, leading to a
modest reduction of the statistical error. 
Second, we include the quartic terms in Eq.~(\ref{eq:disprel}), allowing us to fold 
discretization effects directly into the dispersion-relation fit.
Although we consider these two changes improvements, the change in the
tuned $\kappa$ values as compared to Ref.~\cite{Bernard:2010fr} stems 
primarily from the substantial increase in statistics on key ensembles.

The computations of the correlation functions needed 
to extract $\phi_D$ and $\phi_B$ have been carried out using the fiducial values listed in 
Table~\ref{tbl:runKappaValues}. 
These simulation $\kappa$'s were obtained
near the beginning of the project, but while the runs were in progress, we redetermined the hopping parameters utilizing increased statistics and reflecting an updated value of $r_1$~\cite{Bazavov:2009fk}.  
The resulting improved determinations of $\kapch$ and $\kapbot$ differ slightly from the simulation values.
In order to adjust $\phi$ from the simulated value $\kapsim$ to the 
tuned value \kaptuned, we write 
\begin{equation}
    \phi_{\rm tuned} = \phi_{\rm sim} +  \frac{d\phi}{d\kappa} (\kaptuned - \kapsim),
    \eqn{phi-tune-adj}
\end{equation}
where the derivatives $d\phi/d\kappa$ listed in 
\tabref{kappaFromMesonDispersion} are obtained from tuning runs with nearby 
$\kappa$ values.
As explained in \secref{Errors}, these derivatives 
are also used to propagate to the decay constants the statistical and scale uncertainties on 
\kaptuned\ listed in \tabref{kappaFromMesonDispersion}. 

\section{Two-point correlator fits}
\label{sec:2point}

We obtain the unrenormalized decay amplitude for every combination of heavy-quark mass, light-quark mass, and sea-quark ensemble by fitting the heavy-light meson two-point correlation functions, 
described in Sec.~\ref{sec:LatCorrs}.   
We use two independent fitting procedures, which we refer to as ``Analysis~I''  and ``Analysis~II''.  
These procedures differ in several respects.
In Analysis~I, we use a jackknife procedure for estimating errors, while
in Analysis~II, we use a bootstrap procedure.
The two analyses also differ in their methods for handling autocorrelations in the data and in their choices of
fit ranges, priors for masses and amplitudes, and numbers of states included.
In the end, we use Analysis~I (Sec.~\ref{sec:jackknife}) to obtain central values,
and use differences from fits with different distance ranges and/or number of 
states included, and from Analysis~II (Sec.~\ref{sec:bootstrap}) to estimate the systematic error due to choices made in the 
fit procedure.

\subsection{Lattice correlators}
\label{sec:LatCorrs}

The lattice axial-vector current is given by
\begin{equation}
    A^4_a(x) = [\bar{\Psi}(x)\gamma^4\gamma^5\Omega(x)]_a\chi(x),
\end{equation}
where $\chi(x)$ is the one-component field appearing in the staggered action, and 
$\Omega(x)=\gamma_1^{x_1/a}\gamma_2^{x_2/a}\gamma_3^{x_3/a}\gamma_4^{x_4/a}$ is the transformation 
connecting naive and staggered fields~\cite{Kawamoto:1981hw}.
The heavy-quark field $\Psi$ is a four-component (Dirac) spinor field, 
and the remaining free Dirac index is interpreted as a taste label.

To remove tree-level discretization errors in the lattice axial current, the heavy-quark field $\Psi$ is 
``rotated'':
\begin{equation}
    \Psi = [1 + a d_1 \bm{\gamma}\cdot \bm{D}]\psi,
    \label{eq:rotation}
\end{equation}
where $\psi$ is the field appearing in the clover action.
Tree-level improvement is obtained when
\begin{equation}
    d_1 = \frac{1}{2 + m_0a} - \frac{1}{2(1 + m_0 a)},
    \label{eq:methods:d1}
\end{equation}
where
\begin{equation}
    m_0a = \frac{1}{u_0}\left(\frac{1}{2\kappa} - \frac{1}{2\crit} \right)
    \label{eq:methods:m0}
\end{equation}
is the tapdole-improved bare mass.
The critical hopping parameter \crit\ is the one for which the clover-clover 
pion mass vanishes.

As usual in lattice gauge theory, we obtain the matrix element in 
\eqref{eq:0AH} from two-point correlation functions.
We introduce pseudoscalar operators
\begin{equation}
    \mathcal{O}_a^{(S)}(x) = \sum_y [\bar{\psi}(y)S(y,x)\gamma^5\Omega(x)]_a\chi(x),
    \label{eq:methods:source}
\end{equation}
depending on a ``smearing'' function $S$.
In this work, we use two functions, 
the local (or unsmeared) source $S(x,y)=\delta_{xy}$, and the smeared 
source (in Coulomb gauge)
\begin{equation}
    S(x,y) = \delta_{x_4y_4}S(\bm{x}-\bm{y}),
    \label{eq:methods:smeared}
\end{equation}
where $S(\bm{r})$ is the 1$S$ solution of the Richardson potential
for the quarkonium systems~\cite{Richardson:1978bt}. 
We obtain $S(\bm{x}-\bm{y})$  by scaling the radial Richardson wavefunction to lattice units, interpolating it to lattice sites,
and then using it as the spatial source for the heavy-quark propagators~\cite{Menscher:2005kj}.

We introduce two-point correlation functions
\begin{eqnarray}
    \Phi_2^{(S)}(t) & = & \sum_{a=1}^4 \sum_{\bm{x}} \left\langle 
        {A^4_a}^\dagger(t,\bm{x}) \mathcal{O}_a^{(S)}(0) \right\rangle,
    \label{eq:methods:corr4phi} \\
    C_2^{(S_1S_2)}(t,\bm{p}) & = & \sum_{a=1}^4 \sum_{\bm{x}} 
        e^{i\bm{p}\cdot\bm{x}} \left\langle 
        {\mathcal{O}_a^{(S_1)}}^\dagger(t,\bm{x}) \mathcal{O}_a^{(S_2)}(0)
        \right\rangle,
    \label{eq:methods:corr4norm}
\end{eqnarray}
where $\langle\bullet\rangle$ now represents the ensemble average.
For large time separations, $\Phi_2^{(S)}$ is proportional to the matrix element $\phi_H$, 
and the proportionality is determined from $C_2^{(SS)}(t,\bm{0})$.
Each two-point function is constructed from a staggered quark
propagator with local ($\delta$) sources and sinks.
We compute $C_2$ functions for all (four) combinations $S_1,S_2=\delta$ and 1S,
requiring heavy clover quark propagators with
all combinations of 1S smeared and local sources and sinks.
Only the local sink clover propagators are needed to compute the $\Phi_2$ functions.
With the sum over tastes in Eqs.~(\ref{eq:methods:corr4phi}) 
and~(\ref{eq:methods:corr4norm}), the correlation functions $\Phi_2$ 
and $C_2$ can also be cast in a heavy-naive formalism~\cite{Wingate:2002fh}.

The staggered light quarks in the axial-current and pseudoscalar two-point correlation functions lead to the presence of opposite-parity states that oscillate in time as $(-1)^t$.
Hence the two-point functions take the following forms:
\begin{eqnarray}
    \Phi_2^{(S)}(t) &=& 
        A_{\Phi}^{(S)}  \LP e^{-Mt} + e^{-M(T-t)}  \RP  +  \widetilde{A}_{\Phi}^{(S)} 
            \LP-1\RP^t \LP e^{-\widetilde{M}t} + e^{-\widetilde{M}(T-t)}  \RP \EL
    &+& A_{\Phi}^{\prime\,(S)} \LP e^{-M^\prime t} + e^{-M^\prime(T-t)}  \RP  +  \ldots ,
\label{eq:Phi_exp} \\
    C_2^{(S_1 S_2)}(t,\vec{p}=0) &=& 
        A^{(S_1)} A^{(S_2)}  \LP e^{-Mt} + e^{-M(T-t)}  \RP  +  \widetilde{A}^{(S_1)} \widetilde{A}^{(S_2)} 
            \LP-1\RP^t \LP e^{-\widetilde{M}t} + e^{-\widetilde{M}(T-t)}  \RP \EL
    &+& A^{\prime\,(S_1)} A^{\prime\,(S_2)} \LP e^{-M^\prime t} + e^{-M^\prime(T-t)}  \RP  +  \ldots ,
\label{eq:C2_exp}
\end{eqnarray}
where a prime denotes a standard excited state of the same parity and a tilde denotes the mass or amplitude of an opposite-parity state.
The oscillating behavior is visible throughout the entire lattice temporal extent, and must be included in fits to extract the ground-state mass and amplitudes.

We then obtain the renormalized decay amplitude in lattice units from the ratio
\BNE
    \aphiH = \sqrt{2}\frac{\ZAQq A_{\Phi}^{(S)}}{A^{(S)}},
    \label{eq:aphiH}
\ENE
where $A_{\Phi}^{(S)}$ and $A^{(S)}$ are the amplitudes of the ground state exponentials defined in Eqs.~(\ref{eq:Phi_exp}) and~(\ref{eq:C2_exp}),
and the renormalization factor \ZAQq\ is discussed in Sec.~\ref{sec:HLCurrents}.

\subsection{Analysis I}
\label{sec:jackknife}

Our primary analysis of two-point correlation functions $\Phi_2^{(S)}$ and $C_2^{(S_1S_2)}$---``Analysis~I''---proceeds as follows.
The amplitudes  $A_{\Phi}^{(S)}$ and $A^{(S)}$ in Eq.~(\ref{eq:aphiH})
are determined from fits to multiple correlators using the full data correlation matrix.
In Analysis I, we start by fitting combinations A, B, C and D in Table~\ref{tbl:fitCombinations}.
\begin{table}
\caption{Combinations of two-point functions that can be used to extract $\aphiH$.
All combinations of two and three correlators are shown.
Additional combinations of four or more correlators are not enumerated.}
\begin{tabular}{l|ccccccc}
\hline\hline
two-point         &\multicolumn{6}{c}{fit combination} \\
function          &A          &B           &C          &D          &E         &F           \\ \hline
$\Phi_2^{(1S)}(t)$    &$\bullet$  &            &$\bullet$  &$\bullet$  &$\bullet$ &$\bullet$   \\[0.5mm]
$\Phi_2^{(\delta)}(t)$    &           &$\bullet$   &$\bullet$  &$\bullet$  &$\bullet$ &$\bullet$   \\[0.5mm]
$C_2^{(1S,1S)}(t)$      &$\bullet$  &            &           &           &$\bullet$ &$\bullet$   \\[0.5mm]
$C_2^{(\delta,\delta)}(t)$      &           &$\bullet$   &           &           &$\bullet$ &            \\[0.5mm]
$C_2^{(\delta,1S)}(t)$      &           &            &$\bullet$  &           &          &$\bullet$   \\[0.5mm]
$C_2^{(1S,\delta)}(t)$      &           &            &           &$\bullet$  &          &            \\
\hline\hline
\end{tabular}
\label{tbl:fitCombinations}
\end{table}
We find combination A, which uses the axial-current correlator with a $1S$
smeared source and the pseudoscalar
correlator with a $1S$ smeared source and sink, to be suitable.
The extra complexity of combinations of three correlators (C and D) give
little benefit, and the errors from
combination A are somewhat smaller than those from combination~B.

For fits to charm-light meson correlators, we include just one simple exponential
(the desired state) and one oscillating exponential, which we call a ``1+1 state fit''.
We choose the minimum distance, $t_\textit{min}$, such that contributions from excited states are small compared to our statistical errors. 
Because we fit two propagators simultaneously while imposing the constraint that the masses be equal, 
this is a six parameter fit: two amplitudes for each propagator and a common mass for each of the simple
and oscillating exponentials.
To help stabilize the fit, the amplitudes and mass of the oscillating state are weakly constrained by 
Gaussian priors, which are incorporated
as additional terms in $\chi^2$ in the fitting procedure~\cite{Lepage:2001ym,Morningstar:2001je}.
The central values for these priors are determined by a trial fit where the prior
for the opposite parity
mass is set to $500 \pm 250$ MeV above the ground state mass and the amplitudes are
unconstrained.
Then the jackknife fits use central values for the opposite parity state amplitudes
and mass determined by the trial fit, with widths (Gaussian) that are typically
three to ten times the error estimates on these parameters, so in the end the priors
make a negligible contribution to $\chi^2$.
(Although 500 MeV is a reasonable guess for the mass gap to the first excited state of
the meson, we actually expect that this excited state in the fit approximates the contributions
of a number of physical states, likely including both single and multiparticle channels.)
Empirically, the width of the prior is made narrow enough to insure that the fits converge
to sensible values.
We propagate the uncertainties in the correlator fits to the subsequent chiral-continuum extrapolation with a jackknife procedure.
In the jackknife resamples, we center the priors at
the values found in the fit to the full ensemble, again with widths that are typically
three to ten times the error estimates on these parameters.  

The bottom meson correlators fall off much more rapidly with $t$, so it is difficult
to take a large enough minimum distance to insure that excited state contributions
are negligible.  Therefore we use a fit with two simple exponentials and one oscillating
exponential or a ``2+1 state fit''.  The mass of this excited state is also weakly
constrained by priors in the same way that the opposite parity mass is, except that
the width of the prior on the excited state mass is set to 200 MeV.

\begin{figure}[p]
\centering
\includegraphics[width=1.0\textwidth]{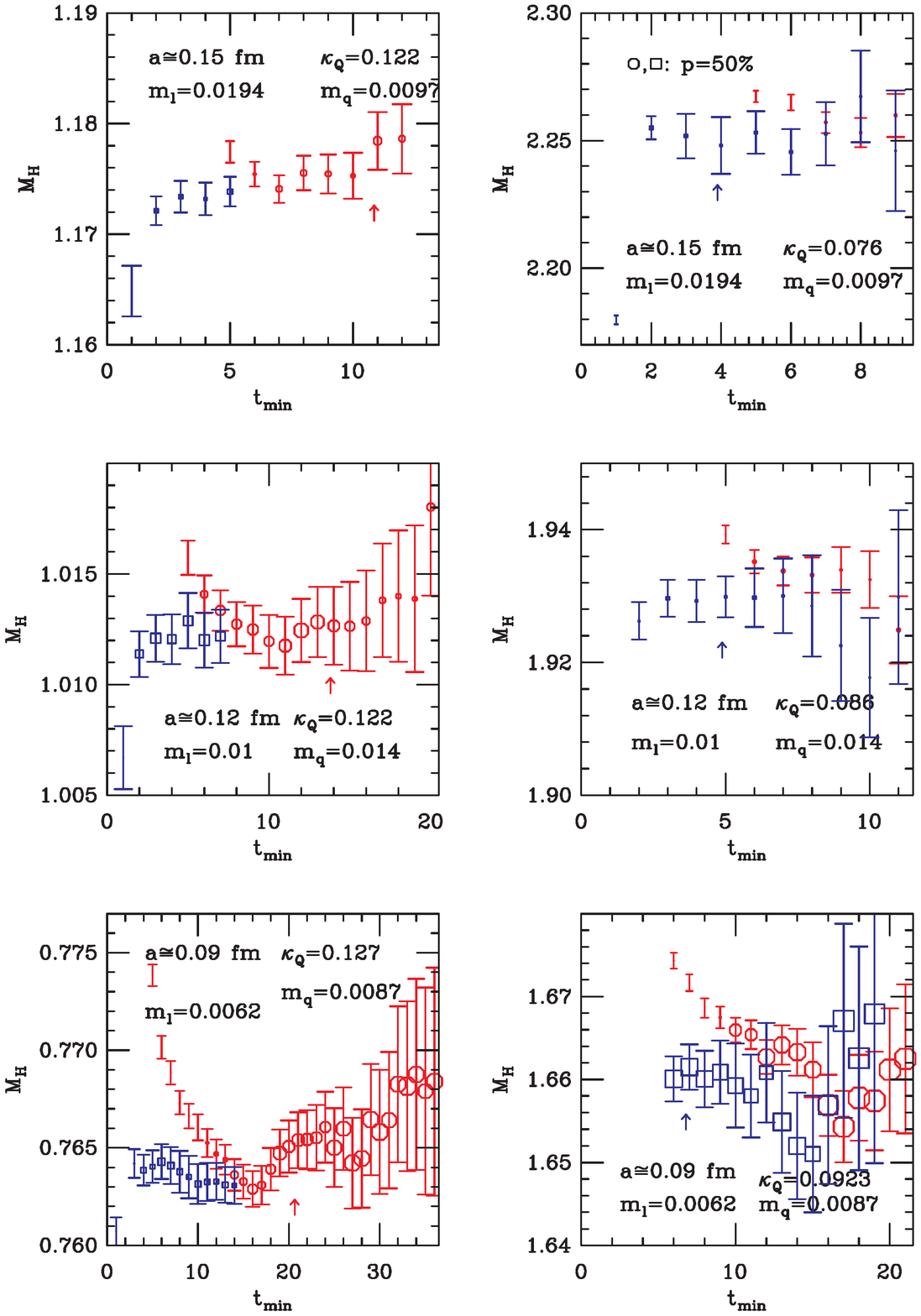}
\vspace{-1.5in}
\caption{Ground-state rest mass $M_H$ versus minimum distance $t_\textit{min}$ included in the fit.
For each lattice spacing, we show an ensemble with the dynamical light-quark
mass $m_l$ in the middle of the range.  Similarly, we show correlators with 
a valence quark mass $m_V$ in the center of the ranges used.
The top two panels are at $a\approx 0.15$~fm, the middle two at $a\approx 0.12$~fm and
the bottom two at $a\approx 0.09$~fm.   In each row the
left panel shows results for charm and the right-panel shows results for bottom.
The size of each plot symbol is proportional to the $p$ value (confidence level) of the fit,
with the symbol size in the legends of the upper right panel corresponding to $p=50\%$.
The red octagons are for fits including one state of each parity (``1+1 fits'') and
the blue squares are for fits including an excited state of the same parity as
the ground state (``2+1 fits'').
In each panel, the arrow indicates the fit that is used in Sec.~\ref{sec:ChPT}.}
\label{fig:mass_vs_dmin}
\end{figure}

Figure~\ref{fig:mass_vs_dmin} shows the heavy-light pseudoscalar mass as a function
of the minimum time used in the fit.  The left-hand plots show sample fits to bottom correlators,
while the right-hand plots show sample fits to charm correlators.
We select fitting ranges to give reasonable fits for all sea-quark ensembles and all valence-quark masses.
We quantify the goodness-of-fit with the ``$p$ value''~\cite{Nakamura:2010zzi}, which is the probability that a
fit with this number of degrees of freedom would have a $\chi^2$ larger than this value.
Table~\ref{tab:fit_ranges} gives the fit ranges for charm-light and bottom-light
correlators on the three lattice spacings, 
both for the fits used for the central values and for alternate fits used in estimating systematic
errors from choices of fit parameters.
The meson masses, $\aphiH$ values, and $p$ values for the data set used in Analysis I are tabulated
in Appendix~\ref{apdx:twoPointFitsJackknife}.

\begin{table}
\caption{Numbers of states and time ranges used in two-point Analysis~I.
    In the number of states, ``1+1'' means one simple exponential and one oscillating
    state (opposite parity).
    The fits in columns two through five were used for the central values, while
    the fits in columns six through nine were used in estimating systematic errors
    from the choice of fit ranges (see Sec.~\ref{subsec:fitErrors}).}
\label{tab:fit_ranges}
\begin{tabular}{l@{\qquad}rr@{\quad}rr@{\qquad}rr@{\quad}rr}
\hline\hline
    	&  \multicolumn{4}{c}{central fits} & \multicolumn{4}{c}{alternate fits} \\
 & \multicolumn{2}{c}{charm} & \multicolumn{2}{c}{bottom}  & \multicolumn{2}{c}{charm}  & \multicolumn{2}{c}{bottom}  \\
 $\approx a$ [fm] & $n_\textrm{states}$ & t range & $n_\textrm{states}$ & t range & $n_\textrm{states}$ & t range & $n_\textrm{states}$ & t range \\
\hline
0.15 & 1+1 & 11--23 & 2+1& 4--20 & 1+1 & 12--23 & 1+1 & 9--20 \\
0.12 & 1+1 & 14--31 & 2+1& 5--22 & 1+1 & 16--31 & 1+1& 12--22 \\
0.09 & 1+1 & 21--47 & 2+1& 7--30 & 1+1& 23--47 & 1+1& 16--30 \\
\hline\hline
\end{tabular}
\end{table}

The decay amplitude $\aphiH$ is highly correlated among different light valence-quark
masses on the same ensemble.  To propagate the correlations among the different
valence masses to the subsequent chiral-continuum extrapolation, in Sec.~\ref{sec:ChPT},
we use a single-elimination jackknife procedure to estimate the covariance matrix of $\aphiH$
for the selected valence quark masses.   This is done by computing the covariance
matrix of the single elimination jackknife samples, and multiplying by
$\LP N-1 \RP^2$, where $N$ is the number of configurations in the ensemble.
In fact, when all valence quark masses
are kept, the covariance matrices are close enough to singular to be
unmanageable.  This reflects the fact that the correlators for intermediate
valence masses can be very accurately predicted from the correlators for nearby
masses, so some of the correlators provide very little new information.  
Therefore, we omit some valence quark masses, using only those set in bold in 
Table~\ref{tbl:latticeAndValenceDetails}.

We use a single elimination jackknife rather than an omit-$J$ jackknife because a
large number of samples are needed to compute a reliable covariance matrix.
Successive configurations in the ensemble are not independent, however, so we must
take autocorrelations into account.
We do so by repeating the calculation after first blocking the data by a factor of four.  
(This block size of four is determined from tests on the $a\approx 0.12$ fm lattices using
fit Analysis~II, for which it gives a reasonable compromise between
suppressing autocorrelations and leaving enough data points for the statistical analysis.)
We then compute, for each valence-quark mass $i$, the ratio $R_i$ of the diagonal
element of the covariance matrix with a block size of four to the same element of
the unblocked covariance matrix:
\BNE
	R_i = \sigma_{ii}^{(4)} / \sigma_{ii},
\ENE
where the superscript denotes the block size.
The rescaled covariance matrix for $\aphiH$ is given by
\BNE
    C_{ij}^{(4)} = C_{ij} \sqrt{R_i R_j},
    \label{eq:autocor_rescale}
\ENE
which preserves the eigenvalue structure of the covariance
matrix, whereas simply using the covariance matrix of the blocked data
would be more likely to produce spurious small eigenvalues.
The rescaling factors $R_i$ themselves have errors, and in many cases turn out to be
less than one.  In such cases, we do not
replace the $R_i$ by one, despite the fact that this would likely be a better estimate of
the individual $R_i$. Doing so would yield a covariance
matrix with a bias toward larger errors, and could produce misleading
estimates of goodness-of-fit in the later analysis.

Finally, we combine the covariance matrices from all of the individual ensembles
into larger covariance matrices, one each for the charm and bottom $\aphiH$.
Since different ensembles are statistically independent, these large covariance
matrices are block diagonal, with each block containing the correlations among different light
valence-quark masses on a single sea-quark ensemble.

\subsection{Analysis II}
\label{sec:bootstrap}

Analysis~II is a second, independent analysis of the two-point correlators that uses the bootstrap method to propagate correlated errors from the two-point analysis through to the chiral extrapolations.
In Analysis~II, we block average the two-point correlator data over four sequential
configurations (which themselves are spaced by more than four trajectories) before the analysis.
In the bootstrap procedure, we resample the data (with replacement), taking the number of sampled configurations to be equal to the number of blocked configurations in each bootstrap ensemble.
For each bootstrap ensemble, we recompute the covariance matrix.
During the bootstrap process, we randomly draw from a gaussian distribution new prior mean values of each constrained parameter
belonging to an excited state while keeping the widths fixed.
The ground state parameters are given loose priors so that the fitted values are determined by the data.
To help stabilize the fits, the ground state prior means are not randomized in the bootstrap.
Prior values for the energy splittings are taken
from a chiral quark model calculation for the $D$ and $B$ meson systems~\cite{Pierro:2001uq}.
Prior widths are taken to be about $200\,\MeV$ for excited states.
Excited state amplitudes $\log(A^{(S)})$ have a prior width $\sigma_{\log A}=2$.

On each gauge ensemble, the same sequence of gauge resamplings is taken
for all valence $m_q$ to preserve correlations among $\aphiH$ values. 
Our final results are based upon 4,000 bootstrap replications
of the data.  We use the central values of $\aphiH$ from the fits to the entire ensemble in the chiral-continuum extrapolation, and use the bootstrap values to obtain the covariance matrix. 

To optimize the determination of $\aphiH$, we compare simultaneous fits of up to six two-point functions; the various combinations of up to four functions are listed in \tabref{fitCombinations}.
At a minimum, one axial-current correlator
must be paired with one propagator (combinations A or B in \tabref{fitCombinations}) to extract~$\aphiH$.
Combination A, using smeared operators, is used in Analysis~I, described above.
Because fits of four or more two-point functions over a wide time range can lead
to a poorly determined data covariance matrix having large rank relative to the number of available configurations, 
we focus on combinations having two or three correlators.
Unlike combination A,
combination B does not take advantage of smeared sources and the ratio does not show convincing plateaus
over the range of times with decent signal to noise.
Comparing combination C to D, the smeared source in C is less noisy than the smeared sink in D.

Given these considerations, for fits to charm correlators, we use two-point function combination C to obtain $\aphiH$
which uses both of the axial current functions. 
We look for stability of the ground-state mass and amplitude
when varying $t_\textit{min}$, $t_\textit{max}$, and the number of excited
states included in the fit. We also compare fit results from other combinations
of correlators to check that we have isolated the correct ground-state energy and matrix element. 
Our final results come from fits accounting for two pseudoscalar states and two (oscillating)
opposite-parity states. 

For fits to bottom correlators, we use combination B for our final results;  this is the same set used in Analysis~I. Combination C gives fits with rather low confidence levels for the $B$ meson and tends
to result in larger errors for $\aphiH$.  Again, we examine fits varying the time range; we also try fits with up to three pseudoscalar states plus three oscillating opposite parity states.
We use these fits and fits to alternate combinations of two-point correlators as a consistency check.

\bigskip

The fit results from the two different analyses are consistent with each other for most cases, but there are 
a few cases where they differ by a standard deviation or more (see \Figref{Bscatter}).
The $\aphiH$ results from the two analyses are propagated through the chiral-continuum extrapolations in 
\secrefs{Dresults}{Bresults}. The resulting differences in the extrapolated results in turn provide the 
basis for our systematic error analysis due to fit choices given in Sec.~\ref{subsec:fitErrors}. 

\section{Heavy-light current matching}
\label{sec:HLCurrents}

In this section, we discuss in more detail the ingredients of Eq.~(\ref{eq:rho}), which allow a ``mostly nonperturbative'' matching procedure~\cite{ElKhadra:2001rv}.

\subsection{Perturbative calculation of \boldmath\rhoAQq}

The perturbative expansion of \rhoAQq\ can be written as
\begin{equation}
    \rhoAQq = 1 + \alpha_s(q^*) \rhoA^{[1]}(m_Qa,m_qa) + \ldots.
\label{eq:rho-PT}
\end{equation}
where $\alpha_s$ is the strong coupling and $\rhoA^{[1]}$ is the one-loop coefficient.
The one-loop coefficient is calculated in Ref.~\cite{ElKhadra:2007qe} using lattice perturbation theory, where we see explicitly that
$\rhoA^{[1]}$ is small because most of the one-loop corrections cancel.  The self-energy contributions cancel exactly (to all orders, in fact), and, in practice, we are in a region where $\rhoA^{[1]}(m_Qa,m_qa)$, viewed as a function of $m_Qa$, has two zeroes.  Therefore the renormalization factor $\rhoAQq$ is close to unity for both bottom and charm.

The perturbative calculation of $\rhoAQq$ in Eq.~(\ref{eq:rho-PT}) proceeds as follows.
We use $\alpha_s(q^*)$ defined in the $V$~scheme \cite{Lepage:1992xa} as determined in Ref.~\cite{Mason:2005zx}, and take $q^*=2/a$, which is close to the optimal choice of 
Refs.~\cite{Lepage:1992xa,Hornbostel:2002af} for a wide range of quark masses.
The one-loop coefficients $\rhoA^{[1]}$ are computed for light-quark masses
$am_q = 0.001, 0.01, 0.04$ to cover the range used in this analysis. 
From these we obtain \rhoAQq\ at other light-quark masses by linear interpolation in $am_q$. 
For illustration, Table~\ref{tab:rho} lists \rhoAbq\ and \rhoAcq\ evaluated at the light 
valence mass $am_q = 0.01$ for the eleven sea-quark ensembles used in this work.
Note that the sea-quark mass dependence is indirect, via the plaquette used to determine $\alpha_s(q^*)$.
The dependence on the light-quark mass in the current is very mild:
for bottom, \rhoAbq\ changes with $am_q$ by $0.07$--$0.2$\%, depending 
on lattice spacing, and for charm, \rhoAcq\ changes by around $0.1$\%. 
On the fine ensembles, the $am_q$ dependence is almost as large as the total one-loop correction because the overall cancellation, especially in \rhoAcq, is so fortuitously good.

\begin{table}
    \centering
    \caption{The perturbative correction factor \rhoAQq\ for the heavy-light current $A^4$ at the simulated charm and 
        bottom heavy quark $\kappa$ values given in Table~\ref{tbl:runKappaValues} and at $am_q = 0.01$ for the different sea-quark ensembles.
        The statistical errors associated with the numerical integration are negligible.}
    \label{tab:rho}
    \begin{tabular}{l@{\quad}r@{/}l@{\quad}c@{\quad}c}
        \hline\hline
    $\approx a$~[fm] & $am_l$&$am_h$ & \rhoAbq & \rhoAcq \\
    \hline
    0.09 & 0.0031&0.031 & 1.0026 & 1.0000 \\
         & 0.0062&0.031 & 1.0026 & 1.0000 \\
         & 0.0124&0.031 & 1.0026 & 1.0000 \\
    \hline
    0.12 & 0.005&0.05 & 1.0081 & 0.9959 \\
         & 0.007&0.05 & 1.0081 & 0.9959 \\
         & 0.010&0.05 & 1.0081 & 0.9959 \\
         & 0.020&0.05 & 1.0080 & 0.9960 \\
         & 0.030&0.05 & 1.0079 & 0.9961 \\
    \hline
    0.15 & 0.0097&0.0484 & 1.0270 & 0.9937 \\ 
         & 0.0194&0.0484 & 1.0267 & 0.9938 \\
         & 0.0290&0.0484 & 1.0265 & 0.9938 \\
    \hline\hline
    \end{tabular}
\end{table}

\subsection{Nonperturbative computation of \boldmath\ZVqq and \ZVQQ}

The nonperturbative part of the matching factor $\ZAQq$ is obtained from the temporal components of the clover-clover and staggered-staggered vector currents.
At zero-momentum transfer, the (correctly normalized) vector current simply counts flavor-number, so it is possible to obtain $Z_{V^4}$ nonperturbatively for any discretization and any mass~\cite{ElKhadra:2001rv}.

For the staggered-staggered current, we compute
\begin{equation}
    C_3^{(S_1S_2)}(t_2,0,t_1) =
        \sum_{ab} \sum_{\bm{x},\bm{y}} \left\langle
        \mathcal{O}_a^{(S_1)}(t_2,\bm{y}) V^4_{ab}(0)
        {\mathcal{O}_b^{(S_2)}}^\dagger(t_1,\bm{x}) \right\rangle,
    \label{eq:methods:ZVll}
\end{equation}
where, as in Eq.~(\ref{eq:methods:source}), $\mathcal{O}_a^{(S)}$ is a smeared or local clover-staggered 
meson operator with mass chosen to optimize the signal, and
\begin{equation}
    V^4_{ab}(x) = 
        \bar{\chi}(x)[\Omega^\dagger(x)\gamma^4\Omega(x)]_{ab}\chi(x)
\end{equation}
is the temporal component of the staggered-staggered vector current.
The three-point functions $C_3$ are computed from the same staggered quarks used
for the clover-staggered two point functions.
The staggered quark is transformed into an improved naive quark
by applying the $\Omega$ matrix; this naive quark at time $t_1$
is then used as the source term when computing the charm propagator.
We smear the source at $t_1$ so that $S_1=S_2$.

We compute \ZVqq using a $D_q$ meson [\cf\ \eq{methods:source}],
which provides a good signal. 
The three-point function $C_3^{(S_1S_2)}(t_2,0,t_1)$ contains states 
of both the desired and the opposite parity, with the latter carrying 
oscillating $(-1)^t$ dependence. 
We construct $C_3^{(S_1S_2)}(t_2,0,t_1)$ with local sources $S_1 = S_2=\delta$ and compute it
at multiple even and odd values 
of $t_1$ and $t_2$ in order to disentangle the ground-state amplitude from the other contributions. 

Within the time range $t_1<0<t_2$ and in the limit of large separations, $|t_1|, t_2\gg a$, \linebreak
\begin{eqnarray}
    C_3^{(\delta,\delta)}(t_2,0,t_1) & = & \Zvqq^{-1} A^2 \exp\left(-E(t_2-t_1)\right) \nonumber \\
    &+& Z^{\prime} A B \left[ (-1)^{t_1} \exp\left(E^\prime t_1 - E t_2\right) + (-1)^{t_2}\exp\left(E t_1 - E^\prime t_2\right) \right] \nonumber \\
    &+& Z^{\prime\prime}  B^2 (-1)^{(t_1+t_2)} \exp\left(-E^\prime(t_2-t_1)\right) + \ldots, \label{eq:3pt4ZVqq}
\end{eqnarray}
neglecting contributions from excited states. 
We extract $\Zvqq$ from a minimum $\chi^2$ fit to the three-point function using the right-hand side of Eq.~(\ref{eq:3pt4ZVqq}) as the model function.
The fit is linear in the free parameters $\ZVqq^{-1}$, $Z^{\prime\prime}$ and $Z^{\prime}$,
while we fix the ground-state energies $E$ and $E^\prime$, and
the operator overlaps $A$ and $B$ to the values determined by fitting the 
two-point function $C_2^{(\delta)}(t,\bm{0})$.
We use a single-elimination jackknife procedure to compute the data covariance matrix.

\tabref{ZvqqResults} presents our results for $\Zvqq$ on the ensembles used in this work.
\begin{table}[tp]
\caption{Light-light vector current renormalization factor $\protect\Zvqq$.  Values in bold are used in computing
the heavy-light current renormalizations. With our conventions, the tree-level value
of $\protect\Zvqq$ is 2.  A colon is used to represent the range of time values
included in the fit.}
\label{tbl:ZvqqResults}
\begin{tabular}{lr@{/}lcccll} \hline\hline
$\approx a$ [fm]  & $am_l$&$am_h$ & $n_\textrm{conf}$    & $-t_1$     &$t_2$    &$am_q$   &\Zvqq    \\ \hline
0.09      & 0.0124&0.031       & 518        &23:12  &11:13  &0.0272   & \bf{1.868(49)} \\ 
          &\multicolumn{2}{c}{}&           &23:12  &11:13  &0.0124   &1.883(69) \\ \hline
0.12     &   0.01&0.05   & 592        &15:9   &7:11   &0.03     &\bf{1.853(45)} \\
          &  0.007&0.05   & 523        &20:7   &7:12   &0.03     &1.882(56) \\ \hline
0.15      & 0.0097&0.0484 & 631        &20:5   &4:12   &0.0484   &\bf{1.704(34)} \\
          &\multicolumn{2}{c}{}&           &20:5   &4:12   &0.029    &1.709(40) \\
          &\multicolumn{2}{c}{}&           &20:5   &4:12   &0.0242   &1.711(42) \\
          &\multicolumn{2}{c}{}&           &20:5   &4:12   &0.0194   &1.707(45) \\
          &\multicolumn{2}{c}{}&           &20:5   &4:12   &0.0097   &1.662(55) \\
\hline\hline
\end{tabular}
\end{table}
The three-point functions for the $\Zvqq$ calculation are generated at a single source
time, $t_{\mathrm{src}}=0$ (instead of the four used for two-point functions $\Phi_2^{(S)}$ and  $C_2^{(S_1S_2)}$).
At $a\approx 0.12~\fm$ we have results at two values of the sea quark masses 
which are consistent within errors. At $a\approx 0.09$ and $0.15~\fm$ we have
results for several values of $m_q$. We do not see evidence for a dependence
upon $m_q$ with current statistics. The errors, however, increase at smaller
quark mass.
Hence, we use the $\Zvqq$ corresponding to $m_q\sim m_s$ in Eq.~(\ref{eq:aphiH}).
In the table, they are set in \textbf{bold}.

For the clover-clover current, we compute
\begin{equation}
    \widetilde{C}_3^{(S_1S_2)}(t_2,t_1,0) =  \sum_{\bm{x},\bm{y}} \left\langle
        {\widetilde{\mathcal{O}}^{(S_1)\dagger}} (t_2,\bm{y}) V^4_{QQ}(t_1,\bm{x})
        \widetilde{\mathcal{O}}^{(S_2)}(0) \right\rangle,
    \label{eq:methods:ZVhh}
\end{equation}
where
\begin{equation}
    V^4_{QQ}(x) = \bar{\Psi}(x)\gamma^4\Psi(x)
\end{equation}
is the temporal component of the (rotated) clover-clover vector current.
The clover-clover bilinear
\begin{equation}
    \widetilde{\mathcal{O}}^{(S)}(x) = \sum_y \bar{\psi}(y)S(x,y)\gamma^5s(x)
    \label{eq:methods:tildesource}
\end{equation}
consists of a heavy-quark field corresponding to charm or bottom, as the case 
may be, and a light clover-quark field $s$ with mass chosen to 
provide a good signal.
At large time separations, these three-point functions are proportional to $\ZVQQ^{-1}$, with the proportionality coming from
\begin{equation}
    \widetilde{C}_2^{(S_1S_2)}(t) = \sum_{\bm{x}} \left\langle
        \widetilde{\mathcal{O}}^{(S_1)\dagger}(t,\bm{x}) 
        \widetilde{\mathcal{O}}^{(S_2)}(0) \right\rangle.
\end{equation}

We compute \ZVQQ
using a $\bar{Q}s$ meson, where the  strange quark is simulated with the 
clover action to circumvent oscillating opposite-parity states [\cf\ \eq{methods:tildesource}]. 
We restrict our calculation of $\widetilde{C}_{2,3}$ to $S=S_1=S_2$
using both local and 1S smearing functions.  The function
$\widetilde{C}_2$ combines a local-local clover quark with mass around
$m_s$ and a heavy clover quark propagator with source and sink $S$.
The function $\widetilde{C}_3$ requires the same heavy- and light-quark
propagators as needed in $\widetilde{C}_2$. An additional heavy-quark
propagator originating from $t_2$ has as its source the light quark
propagator restricted to $t_2$, multiplied by $\gamma_5$ and convolved
with smearing function $S$.

\begin{table}
\caption{Heavy-heavy vector current renormalization factor \ZvQQ 
computed at several $\kappa$ values, covering the charm and
bottom quark masses, for three lattice spacings.}
\label{tbl:ZvQQresults}
\begin{tabular}{lllll}
\hline\hline
$\approx a$ [fm]   & $am_l/am_h$    & $n_\textrm{conf} \times n_\textrm{src}$     &$\kappa_Q$   &$\ZvQQ$  \\ \hline
0.09       &$0.0062/0.031$   &1912$\,\times\,$2                 &0.1283       &0.2749(4) \\
           &                 &                                  &0.127        &0.2830(4) \\
           &                 &                                  &0.110        &0.3856(6) \\
           &                 &                                  &0.0950       &0.4730(8) \\
           &                 &                                  &0.0931       &0.4840(9) \\ \hline
0.12      &$0.007/0.05$     &2110$\,\times\,$2                 &0.124        &0.2899(4) \\
           &                 &                                  &0.122        &0.3028(4) \\
           &                 &                                  &0.116        &0.3410(5) \\
           &                 &                                  &0.098        &0.4507(7) \\
           &                 &                                  &0.086        &0.5209(10) \\
           &                 &                                  &0.074        &0.5894(15) \\ \hline
0.15       &$0.0194/0.0484$  &631$\,\times\,$2                  &0.122        &0.3195(14) \\
           &                 &                                  &0.118        &0.3440(16) \\
           &                 &                                  &0.088        &0.5195(48) \\
           &                 &                                  &0.076        &0.5898(81) \\
\hline\hline
\end{tabular}
\end{table}

In Eq.~(\ref{eq:methods:tildesource}), we use a random color wall source with three
dilutions for both the heavy and light spectator quarks that
originate from $t=0$. We generate two- and three-point functions for
both local-local and smeared-smeared source-sink combinations where
the smearing is applied to the heavy quark.
We compute the 2- and 3-point functions at several values of $\kappa$ spanning
a range from around the charm quark to the bottom quark. 
We determine $\ZvQQ$ from a fit to the plateaus in the
jackknifed ratio of the three-point and two-point functions.
Our results are summarized in \tabref{ZvQQresults}.

\begin{figure}
    \includegraphics[width=0.90\textwidth]{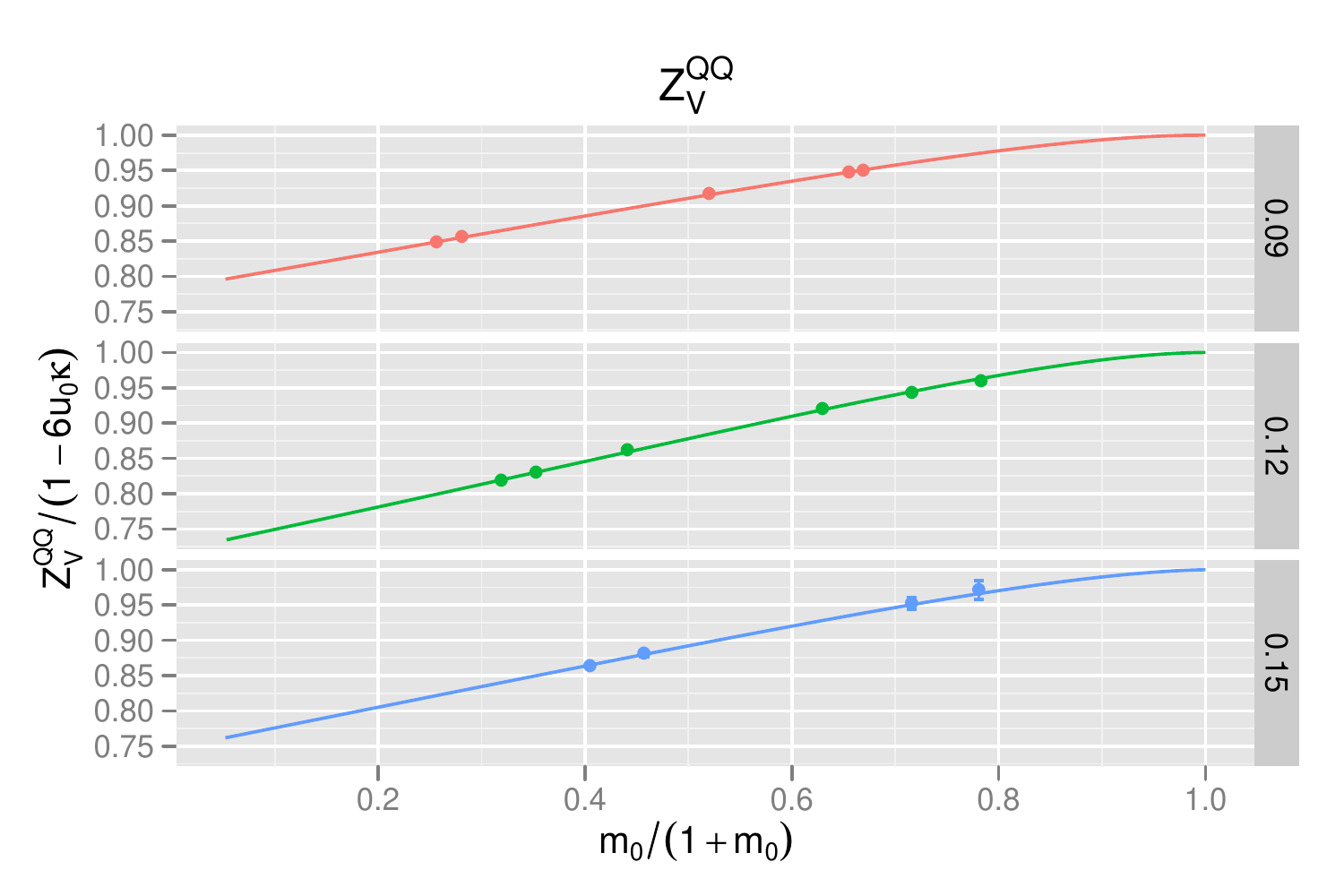}
    \caption{Plot of $\ZvQQ/(1-6 u_0 \kappa)$  \vs\ $m_0a/(1+m_0a)$ for the three lattice spacings.}
    \label{fig:rhoZ:ZVQQ}
\end{figure}

In order to properly normalize the derivative $d\phi/d\kappa_Q$ (see \eq{phi-tune-adj}),
we need values of $\ZvQQ$ at $\kappa$ values other than those used in the $\ZvQQ$
simulations. 
We therefore fit the simulation results to the interpolating quartic polynomial 
\begin{equation}
    \ZvQQ(\kappa) = 1 + \sum_{j=1}^4 c_j \kappa^j
    \label{eq:interpZVQQ}
\end{equation}
which reproduces the infinite mass limit $\ZvQQ\to1$.
Our codes employ the hopping parameter version of the action; so, at tree level $c_1=-6u_0$ and for $j>1$, $c_{j}=0$.
We constrain the interpolation parameters to the tree-level values taking $\sigma_{j}=4$ as the widths.
\tabref{ZvQQruns} shows values for $\ZvQQ$ interpolated to the nominal charm and bottom $\kapsim$
used in our decay constant runs.
\begin{table}
\caption{Heavy-heavy vector current renormalization factor $\ZvQQ$ corresponding to the charm and bottom 
$\kappa_{\rm sim}$ values used in the decay constant simulations.}
\label{tbl:ZvQQruns}
\begin{tabular}{l@{\quad}ll@{\quad}ll}
\hline\hline
           &\multicolumn{2}{c}{charm} &\multicolumn{2}{c}{bottom} \\
$\approx a$ [fm]   &$\kappa_Q$        &$\ZvQQ$      &$\kappa_Q$     &$\ZvQQ$ \\ \hline
0.09       &0.127             &0.2829(4)    &0.0923        &0.4891(9) \\
0.12      &0.122             &0.3029(4)    &0.086         &0.5216(10) \\
0.15       &0.122             &0.3199(14)   &0.076         &0.5868(81) \\
\hline\hline
\end{tabular}
\end{table}
Figure~\ref{fig:rhoZ:ZVQQ} plots the data in Table~\ref{tbl:ZvQQresults} together with 
the interpolation of Eq.~(\ref{eq:interpZVQQ}).
To aid perturbative intuition, the values of $\ZvQQ$ in the figure are scaled by the tree-level expression $1-6u_0\kappa$;
the relation between $\kappa$ and $m_0a/(1+m_0a)$ can be inferred from Eq.~(\ref{eq:methods:m0}).

\section{Chiral and continuum extrapolation}
\label{sec:ChPT}

In this section, we present the combined chiral and continuum extrapolations used to obtain the physical values of the $B_{(s)}$ and $D_{(s)}$ meson decay constants.
We first discuss the use of $SU(3)$ chiral perturbation theory for heavy-light mesons in Sec.~\ref{sec:ChPT_framework}, giving the formulas used for the chiral fits and 
describing our method for incorporating heavy-quark and light-quark discretization effects into the extrapolation.  We then show the chiral fits for the $D$ system in 
Sec.~\ref{sec:Dresults}, and for the $B$ system in Sec.~\ref{sec:Bresults}.

\subsection{Chiral Perturbation Theory framework}
\label{sec:ChPT_framework}

The errors introduced by the chiral and (light-quark) continuum
extrapolations are controlled with rooted staggered chiral
perturbation theory (\rschpt)~\cite{Lee:1999zxa,Aubin:2003mg} applied to heavy-light mesons.
In Ref.~\cite{Aubin:2005aq}, the heavy-light decay constant was calculated to one-loop in rS$\chi$PT at 
leading order in the heavy-quark expansion [$(1/M_H)^0$], where $M_H$ is a generic heavy-light meson mass. 
A replica trick is used in \rschpt\ to take
into account the effect of the fourth root of the staggered
determinant~\cite{Bernard:2006zw,Bernard:2007ma}.

In addition to using the form calculated in Ref.~\cite{Aubin:2005aq}, we
also use a chiral fit form that includes, in the loops, the effects of hyperfine splittings
(\eg, $M_B^*-M_B$) and flavor splittings (\eg, $M_{B_s}-M_B$). 
These splittings are $\sim\! 100$ MeV, and so not much smaller than $M_\pi$, despite the fact that
they are formally of order $1/M_H$.
Since the lightest pseudoscalar meson masses in our simulations are $\sim 225$ MeV, it is not 
immediately obvious that including the splittings is necessary or useful.
Their inclusion is motivated, first of all, by the observation of Arndt and Lin
\cite{Arndt:2004bg} that finite-volume  effects in the one-loop diagrams can be substantially
larger with the splittings present.  This is mainly due to the fact that accidental
cancellations in finite volume effects between different diagrams 
at $(1/M_H)^0$ disappear once splittings are included. As described below,
it is not difficult to include the splitting effects into the 
calculation of Ref.~\cite{Aubin:2005aq}. We also discuss the extent to which
including the splittings, but not other effects that could occur at
order $1/M_H$, is a systematic approximation. In practice, we do fits both including and
omitting the splittings, and use the difference as one estimate of the chiral extrapolation
error.  For central values, we include the splittings, because this yields
a more conservative estimate of finite-volume effects.

With staggered quarks, the (squared) pseudoscalar meson masses are
\begin{equation}
	M^2_{ab,\xi} = B_0(m_a+m_b) + a^2\Delta_\xi,
	\label{eq:mesonmass}
\end{equation}
where $m_a$ and $m_b$ are quark masses, $B_0$ is a parameter of \chpt,
and the representation of the meson under the taste symmetry group is
labeled by $\xi=P, A, T, V, I$ \cite{Lee:1999zxa}.
The exact non-singlet chiral symmetry of staggered
quarks as $m_a,m_b\to0$ ensures that $\Delta_P=0$.
All of these pseudoscalars appear in the
``pion'' cloud around the heavy-light meson in the simulation,
and all of them therefore affect the decay constant.

Working at leading order [$(1/M_H)^0$] in the heavy-quark expansion and at one loop, or
next-to-leading order (NLO), in the chiral expansion, the  \rschpt\ expression for the decay constant 
with light valence quark $q$ takes 
the form \cite{Aubin:2005aq}
\begin{eqnarray}
	\phi_{H_q} = \phi_H^0 \Bigg[ &&
1 + \frac{1}{16\pi^2f^2}
  \frac{1+3g_\pi^2}{2}
  \Biggl\{-\frac{1}{16}\sum_{e,\xi} \ell(M_{eq,\xi}^2)
  \nonumber \\*
&&{}- \frac{1}{3}
\sum_{j\in \CM_I^{(2,q)}} 
\frac{\partial}{\partial M^2_{Q,I}}\left[ 
      R^{[2,2]}_{j}(
    \CM_I^{(2,q)};  \mu^{(2)}_I) \ell(M_{j}^2) \right]
    \nonumber \\*&&{} 
     -   \biggl( a^2\delta'_V \sum_{j\in \hat\CM_V^{(3,q)}}
     \frac{\partial}{\partial M^2_{Q,V}}\left[ 
       R^{[3,2]}_{j}( \hat\CM_V^{(3,q)}; \mu^{(2)}_V)
    \ell(M_{j}^2)\right]
        + [V\to A]\biggr)  
   \Biggr\} \nonumber \\*&&{}+
 p(m_q,m_l,m_h,a^2) \Bigg] ,
	\label{eq:phiq}
\end{eqnarray}
where $m_q$ is the light valence-quark mass, 
 $e$ runs over the sea quarks, the lighter two of which  have masses $m_l$, and the heavier,
$m_h$.%
\footnote{The physical values of the average up-down quark mass and of the strange-quark mass are denoted by 
$\hat m=(m_u+m_d)/2$ and $m_s$, respectively.} 
The parameter $\phi_H^0$ is independent of the light masses, and $p$ is an analytic function.
We fit the charm and bottom systems separately, so $\phi_H^0$ depends, in practice, on the
heavy-quark mass.
The meson mass $M_{Q,\xi}$ is similar to $M_{ab,\xi}$ in \eq{mesonmass}, but constructed from
a valence quark-antiquark, $q\bar q$.
The light-meson decay constant $f\approx f_\pi\cong 130.4$~MeV and
  the $H$-$H^*$-$\pi$  coupling $g_\pi$ controls the size of the one-loop effects.
Taste-violating hairpin diagrams, which arise only at non-zero lattice spacing,
are parameterized by $\delta'_A$ and $\delta'_V$.
The residue functions $ R^{[n,k]}_{j}(\{M\},\{\mu\})$ are defined in Ref.~\cite{Aubin:2003mg}.
Chiral logarithms are written in terms of  the functions $ \ell(M^2)$ \cite{Bernard:2001yj}:
\begin{eqnarray}
        \ell(M^2) &=&  M^2 \ln \frac{M^2}{\Lambda_\chi^2} \qquad{\rm [infinite\ volume]}, \label{eq:ell}\\
        \ell( M^2) &=&  M^2 \left(\ln \frac{M^2}{\Lambda_\chi^2} + \delta_1(ML)\right) \qquad{\rm [spatial\ volume}\ L^3 {\rm ]},
\eqn{ell-FV} \\
        \delta_1(ML)  &\equiv&  \frac{4}{ML} \sum_{\bm{r}\ne\bm{0}} \frac{K_1(|\bm{r}|ML)}{|\bm{r}|} \eqn{delta1} .
\end{eqnarray}
Here $\Lambda_\chi$ is the chiral scale, 
$K_1$ the Bessel function of imaginary argument, and $\bm{r}$ any non-zero three-vector with integer components.
The mass sets in the residue functions of \eq{phiq} are
\begin{eqnarray}
  \mu^{(2)} & = & \{M^2_U,M^2_S\} ,\\*
  \CM^{(2,q)} & = & \{M_Q^2, M_{\eta}^2\},\\*
  \hat\CM^{(3,q)} & = & \{M_Q^2, M_{\eta}^2, M_{\eta'}^2\} ,
\end{eqnarray}
where $M_U$ ($M_S$) is the mass of the pseudoscalar $l\bar l$ ($h\bar h$) meson.

The salient feature of the chiral extrapolation
of $\phi_{H_q}$ is that the chiral logs 
have a characteristic curvature as $m_q\to 0$~\cite{Kronfeld:2002ab}.  At non-zero lattice spacing,
the presence of the additive splittings $a^2\Delta_\xi$ in the meson
masses reduces the curvature of the chiral logarithms.
The characteristic curvature returns, however, as the continuum limit is approached.

To combine data from several lattice spacings into one chiral extrapolation, it is necessary to convert lattice units to (some sort of) physical units.
As mentioned in Sec.~\ref{sec:Params}, we convert in two steps, first by canceling lattice units with the appropriate power of $r_1/a$.
In particular, pseudoscalar meson masses [\cf\ Eq.~(\ref{eq:mesonmass})] become $r_1^2M^2_{ab,\xi}=(r_1/a)^2(aM_{ab,\xi})^2$,
and the decay constant [\cf\ Eq.~(\ref{eq:phiq})] becomes $r_1^{3/2}\phi_H=(r_1/a)^{3/2}(a^{3/2}\phi_H)$, with $a^{3/2}\phi_{H}$ determined from Analyses~I or~II (\cf\ Sec.~\ref{sec:2point}).
Strictly speaking, one must take the quark mass dependence
of $r_1$ into account, either separately or by modifying the right-hand side of Eq.~(\ref{eq:phiq}) accordingly.
At the present level of accuracy, we ignore this subtlety, canceling units ensemble-by-ensemble with the computed $r_1/a$.
Since $r_1$ is expected to depend smoothly on $m_l$ and $m_h$, we are unlikely to introduce an uncontrolled error into the extrapolated decay constants.
(After completing the chiral-continuum extrapolation in $r_1$ units, we then use $r_1=0.3117(22)$~fm (\cf\ Sec.~\ref{sec:Params}) to convert to~MeV.)

To quantify the size of NLO (and higher) corrections to \chpt, it is useful to define dimensionless 
parameters $x_q$, $x_l$ and $x_h$ proportional to the quark masses $m_q$, $m_l$ and $m_h$: 
$$x_{q,l,h} \equiv \frac {(r_1 B_0)(r_1/a)(2am_{q,l,h})}{8 \pi^2 f_\pi^2 r_1^2}\ .  $$
Since the splittings $a^2\Delta_\xi$ are added to the quark mass terms in
\eq{mesonmass}, it is similarly useful to define
\begin{eqnarray}
    x_{\Delta_\xi} & \equiv & \frac {r_1^2a^2 \Delta_\xi}{8 \pi^2 f_\pi^2 r_1^2}, \\[0.7em]
    x_{\bar \Delta} & \equiv & \frac {r_1^2a^2 \bar\Delta}{8 \pi^2 f_\pi^2 r_1^2},
\end{eqnarray}
where $\bar \Delta$ is the average pion splitting
\begin{equation}
    \bar\Delta = \case{1}{16}(\Delta_P + 4\Delta_A + 6\Delta_T + 4\Delta_V  + \Delta_I).
    \label{eq:avgDelta}
\end{equation}
The $x_i$ are in ``natural'' units for \chpt,  in the sense that one expects that 
chiral corrections, when written as
series in the $x_i$, have coefficients [or low-energy constants (LECs)] that are of order~1.

We then take the analytic function $p$ in
\eq{phiq} to have the following form at NLO
\begin{equation}
	 L_{\rm val} (x_q + x_{\Delta_{\rm val}}) + 
 L_{\rm sea} (2x_l+x_h + 3x_{\Delta_{\rm sea}} )
 + L_a \frac{a^2}{16 \pi^2 f_\pi^2 r_1^4} ,
	\label{eq:pq}
\end{equation}
where $L_{\rm val}$, $L_{\rm sea}$ and $L_a$ are quark-mass-independent LECs that
we fit from our data,
and we define
\begin{eqnarray}
x_{\Delta_{\rm val}} &\equiv& \frac{9}{5}x_{\bar\Delta} - \frac{4}{5}x_{\Delta_I}, \label{eq:Delta-val} \\
x_{\Delta_{\rm sea}} &\equiv& \frac{9}{11}x_{\bar\Delta} + \frac{2}{11}x_{\Delta_I}, \label{eq:Delta-sea}
\end{eqnarray}
The low-energy constants $L_{\rm val}$, $L_{\rm sea}$ and $L_a$ depend
implicitly on the chiral scale $\Lambda_\chi$,
so that the complete expression, \eq{phiq}, is independent of $\Lambda_\chi$.
As in Ref.~\cite{Aubin:2005aq}, we choose to include 
the $a^2$ dependent terms $x_{\Delta_{\rm sea}}$ and  $x_{\Delta_{\rm val}}$ 
in the coefficients of
$L_{\rm val}$ and $L_{\rm sea}$ so that these coefficients represent those combinations of meson masses
that arise naturally under a change of $\Lambda_\chi$ in the chiral logarithms.

The LEC $L_a$ arises from analytic taste-violating effects; it serves as a counterterm to absorb changes proportional to
the taste-violating hairpins $\delta'_A$ and $\delta'_V$ under a change in chiral scale.  As such, we take the $a^2$
coefficient of $L_a$ in \eq{pq} to vary with lattice spacing like $x_{\Delta_{\rm val}}$.
As long as $L_a$ then appears as an independent fit parameter, the introduction of
the $x_{\Delta_{\rm sea}}$ and  $x_{\Delta_{\rm val}}$  terms in the coefficients of $L_{\rm val}$ and $L_{\rm sea}$ in \eq{pq}
has a negligible effect on the results from the chiral fits.
However, we find that the introduction of
these terms significantly reduces the magnitude
of $L_a$; in other words, most of the discretization error from
the light quarks appears to be due to the $a^2$ dependence of the light meson masses 
in the chiral loops. 
We leave  $L_{\rm val}$ ,  $L_{\rm sea}$ and $L_a$
unconstrained in the fits that determine central values; their size
is of $\mathrm{O}(1)$ as expected (and is in fact $\le0.6$).

In the region of the strange-quark mass, the data for the decay constants show 
some curvature, and at least some quadratic terms 
in the quark masses (NNLO effects) must in general be added in order to obtain acceptable 
($p>0.01$) fits.
There are four such LECs, giving a NNLO contribution to $p$ of the form
\begin{equation}
 Q_1 x_q^2 + Q_2(2x_l+x_h)^2 + Q_3x_q(2x_l+x_h)+ Q_4(2x_l^2 + x_h^2).
\eqn{Qs}
\end{equation}
Fits omitting the $Q_1$ and $Q_3$ terms give poor confidence levels and are rejected; adding the $Q_2$ and
$Q_4$ terms does not change the fit results much, but increases over-all errors by up to 30\%.
To be conservative, we include all four terms in fits for central values; other acceptable fits (for 
example, fixing $Q_2$ or $Q_4$ or both to zero) are included among the alternatives used to estimate the 
systematic error of the chiral extrapolation.

For the central-value fits, the $Q_i$ are mildly constrained by Gaussian priors with central value~0 and width~0.5,
since that is roughly the expected size in natural units.
After fitting, the posterior values satisfy $|Q_i|\le 0.5$, and $Q_1$ and $Q_3$
have errors $\approx 0.05$ (much less than the prior width), indicating that they are constrained by the data. 
$Q_2$ and $Q_4$ have errors $\sim 0.5$, indicating that they are largely constrained by the
priors.
Changing the prior widths for the $Q_i$ to 1.0 has a negligible effect on central values and errors of the decay constants,
although the posterior $Q_2$ and $Q_4$ typically increase in size and error, as expected.

While the chiral form introduced so far gives acceptable simultaneous fits to our data from all available lattice spacings,
we still need to estimate the size of heavy-quark and generic light-quark discretization errors. Following 
the Bayesian approach advocated in Refs.~\cite{Lepage:2001ym,Morningstar:2001je}, we add constrained lattice-spacing-dependent terms
to the fit function until the statistical errors of the results cease to increase appreciably.
For the heavy quark, we take up to six such terms, 
$f_E(m_0a)$, $f_X(m_0a)$, $f_Y(m_0a)$, $f_B(m_0a)$, $f_3(m_0a)$, and 
$f_2(m_0a)$, where $m_0$ is the
heavy quark bare mass. 
Details about the origin and form of these six functions are given 
in Appendix~\ref{app:HQcutoff}.  These functions estimate fractional (not absolute) errors, and as such are included within the square brackets in Eq.~(\ref{eq:phiq}) (or its equivalent, Eq.~(\ref{eq:phiq-Delta}) below).
The first three are $\mathrm{O}(a^2)$ corrections and are added to the fit with 
coefficients $z_i\,(a \Lambda)^2$, $i\in\{E,X,Y\}$, where $\Lambda$ is a scale characteristic of the heavy-quark expansion, and the $z_i$
are parameters with prior value 0 and prior width 1 (for $f_Y$) or $\sqrt{2}$ (for
$f_E$ and $f_X$, since they each appear twice in the analysis of  Appendix~\ref{app:HQcutoff}).
The next two terms are  $\mathrm{O}(\alpha_s a)$ corrections,
added with coefficients $z_i\,\alpha_s a \Lambda$, $i\in\{B,3\}$, with 
$z_i$ taken to have prior value 0 and prior width 1 (for $f_B$) or $\sqrt{2}$ (for $f_3$, again because it appears twice).
The final term arises from the propagation to the decay constants
of heavy-quark errors in the tuning of the heavy-quark hopping 
parameter, $\kappa$.
It comes in with coefficient $z_2\,(a \Lambda)^3$, with $z_2$ having prior value 0 and prior width 1. 
We take a large value $\Lambda=700$~MeV, which provides 
conservatively wide priors, especially for the first five terms. 
Once one of each of the first two types of terms is added, the errors 
already reach $\sim\!80\%$ of their values with all six added.

Similar terms representing generic light-quark errors, which are not automatically included in the fit function
(unlike taste-violating terms), may also be added.  With the asqtad staggered action, generic discretization effects
are of $\mathrm{O}(\alpha_s a^2)$.  
We allow the physical LECs $\phi_H^0$,
$L_{\rm val}$, $L_{\rm sea}$,
$Q_1$, $Q_2$, $Q_3$, and $Q_4$, to have small relative
variations with lattice spacing with coefficients $C_i \alpha_s (a \Lambda)^2$, where $i$ stands for any of the seven physical LECs,
$\Lambda$ is again taken to be 700 MeV, and the $C_i$ have prior value 0 with prior width 1.  This corresponds to a maximum
of about a $3\%$ difference for a given LEC between the $a\approx\aCoarse$~fm and the $a\approx\aFine$~fm ensembles.
Once several heavy-quark discretization terms have been introduced, these light-quark terms further increase the 
total error of individual decay constants by less than $10\%$. However, the errors on
the decay constant ratios $f_{D_s}/f_{D^+}$ and $f_{B_s}/f_{B^+}$ are significantly
increased by light-quark discretization effects, because the heavy-quark effects
on the ratios cancel to first approximation. For our central values, we include all six heavy-quark and all seven light-quark terms,
so the total error from a given fit should estimate all (taste-conserving) discretization errors, 
as well as normal statistical effects.
To estimate ``heavy-quark'' and ``light-quark'' discretization effects separately, 
we set to zero the light- or
heavy-quark discretization terms, respectively, and then subtract the statistical errors in quadrature.  
Such separate errors are not relevant 
to any final results quoted below, but are
included as separate lines in the error budget for informational purposes.

As mentioned above, our preferred fit form modifies \eq{phiq} by including the effects of hyperfine and flavor splittings
of the heavy-light mesons in one-loop diagrams.  
We now briefly describe how one may adjust the results of Ref.~\cite{Aubin:2005aq}
to include these splittings. In \eq{phiq}, the contributions proportional to $g_\pi^2$ come from diagrams with
internal $H^*$ propagators, namely the left-hand diagrams in Fig.~5 of
Ref.~\cite{Aubin:2005aq}.  Contributions with no factor of  $g_\pi^2$ come from
diagrams with light-meson tadpoles,
namely the right-hand diagrams in Fig.~5 of Ref.~\cite{Aubin:2005aq}.  The latter have
no internal heavy-light propagators, so are unaffected by any heavy-light splittings. The
splittings in the former diagrams depend on whether the light-meson line is connected
(Fig.~5a, left, of Ref.~\cite{Aubin:2005aq}), or disconnected
(Fig.~5b, left). In the disconnected case, the $H^*$ in the loop always has the same
flavor ($q$) as the external $H_q$, so there is no flavor splitting between the two,
only a hyperfine splitting.  In the connected case, the $H^*$ in the loop has the flavor of
the virtual sea quark loop (which we labeled by $e$ in \eq{phiq}), 
so there is flavor splitting with the external $H_q$, in addition
to the  hyperfine splitting.

Let  $\Delta^*$ be the (lowest-order) hyperfine splitting, and $\delta_{eq}$  be the
flavor splitting between a heavy-light meson with  light quark of flavor $e$ and one
of flavor $q$. At lowest order,  $\delta_{eq}$ is  proportional to the quark-mass
difference (or light-meson squared mass difference),
which can be written in
terms of a parameter
$\lambda_1$:
\begin{equation}\eqn{deltaeq}
    \delta_{eq} \cong 2 \lambda_1 B_0(m_e-m_q)  \cong  \lambda_1 (M^2_{E} - M^2_Q) ,
\end{equation}
where $M_{E}$ is the mass of an $e \bar e$ light meson. Here we have used the notation
of Arndt and Lin \cite{Arndt:2004bg} and included a factor of $B_0$ in the middle expression;
$B_0$ is omitted in the notation of Ref.~\cite{Boyd:1994pa}, Eq.~(16), and of
Ref.~\cite{Aubin:2005aq}, Eq.~(45).

By convention, the mass of the external $H$ is removed in the heavy quark effective 
theory, so
the mass shell is at $\bm{k}=\bm{0}$, where $\bm{k}$ is the external three-momentum. 
When there is no splitting, the internal $H^*$ has its pole at the same place, which makes
the integrals particularly simple, giving the chiral log function
$\ell(M^2)$, \eq{ell}.
If a splitting $\Delta$
is present, the integrals involve a significantly more complicated function,
which we denote 
\begin{equation}
    J(M,\Delta) = (M^2-2\Delta^2)\log(M^2/\Lambda^2) +2\Delta^2 -4\Delta^2 F(M/\Delta)  \qquad \textrm{[infinite volume]}.
\eqn{Jdef}
\end{equation}
Here the function $F$ is most simply expressed \cite{Stewart:1998ke,Becirevic:2003ad}
\begin{equation}
F(1/x) = 
    \begin{cases} 
     -\frac{\sqrt{1-x^2}}{x_{\phantom{g}}}\left[\frac{\pi}{2} - \tan^{-1}\frac{x}
    {\sqrt{1-x^2}}\right], & \text{if $ |x|\le 1$,} \\
    \frac{\sqrt{x^2-1}}{x}\ln(x + \sqrt{x^2-1}), & \text {if $ |x|\ge 1$,}
    \end{cases}
    \eqn{Fdef}
\end{equation}
which is valid for all~$x$.

It is then straightforward to write down the generalization of \eq{phiq} to include splittings.
The basic rule is to replace
\begin{equation}
    \ell(M^2) \to  J(M,\Delta)
    \eqn{replacement}
\end{equation}
in the terms proportional to~$g_\pi^2$.
It is not hard to show that $J(M,0)=\ell(M^2)$, so this replacement is consistent with
the original result neglecting the splittings.
In making the replacements, one must choose the correct value of the splitting $\Delta$ in each term.
As mentioned above, in terms that come from the
diagram with a disconnected light-meson propagator, one must put $\Delta=\Delta^*$.  But
in terms that come from the
diagram with a connected light-meson propagator, one must put $\Delta=\Delta^* + \delta_{eq}$,
because the internal heavy-light meson is
a $H^*_e$, while the external meson  is an $H_q$.
The result for the heavy-light meson decay amplitude including the splittings is then
\begin{eqnarray}\label{eq:phiq-Delta}
	\phi_{H_q} = \phi_H^0 \Bigg[ &&
1 + \frac{1}{16\pi^2f^2}
  \frac{1}{2}
  \Biggl\{-\frac{1}{16}\sum_{e,\Xi} \ell(M_{eq,\Xi}^2)
  \nonumber \\*&&{}- \frac{1}{3}
\sum_{j\in \CM_I^{(2,x)}} 
\frac{\partial}{\partial M^2_{X,I}}\left[ 
      R^{[2,2]}_{j}(
    \CM_I^{(2,x)};  \mu^{(2)}_I) \ell(M_{j}^2) \right]
    \nonumber \\*&&{} 
     -   \biggl( a^2\delta'_V \sum_{j\in \hat\CM_V^{(3,x)}}
     \frac{\partial}{\partial M^2_{X,V}}\left[ 
       R^{[3,2]}_{j}( \hat\CM_V^{(3,x)}; \mu^{(2)}_V)
    \ell(M_{j}^2)\right]
        + [V\to A]\biggr) \nonumber \\*&&{}
  -3g_\pi^2\frac{1}{16}\sum_{e,\Xi} J(M_{eq,\Xi},\Delta^*+\delta_{eq})
  \nonumber \\*&&{}- g_\pi^2
\sum_{j\in \CM_I^{(2,x)}} 
\frac{\partial}{\partial M^2_{X,I}}\left[ 
      R^{[2,2]}_{j}(
    \CM_I^{(2,x)};  \mu^{(2)}_I) J(M_{j},\Delta^*) \right]
    \nonumber \\*&&{} 
    \hspace{-1.2cm} -3g_\pi^2 \biggl( a^2\delta'_V \sum_{j\in \hat\CM_V^{(3,x)}}
     \frac{\partial}{\partial M^2_{X,V}}\left[ 
       R^{[3,2]}_{j}( \hat\CM_V^{(3,x)}; \mu^{(2)}_V)
    J(M_{j},\Delta^*)\right]
        + [V\to A]\biggr)  
   \Biggr\} \nonumber \\*&&{}+
 p(m_q,m_l,m_h,a^2) \Bigg] .
\end{eqnarray}

It is also straightforward to include finite-volume effects into \eq{phiq-Delta}.  
One simply replaces
\begin{equation}
    J(M,\Delta) \to J(M,\Delta) + \delta J(M,\Delta,L) ,
\eqn{J-FV}
\end{equation}
where $\delta J(M,\Delta,L)$ is the finite-volume correction in a spatial volume~$L^3$.
The correction can be written in terms of functions defined in
Refs.~\cite{Arndt:2004bg,Aubin:2007mc}:
\begin{equation}
\delta J(M,\Delta,L)  = \frac{M^2}{3}\delta_1(ML) - 16\pi^2\left[\frac{2\Delta}{3}J_{FV}(M,\Delta,L)
+\frac{\Delta^2-M^2}{3} K_{FV}(M,\Delta,L)\right] \ ,
\end{equation}
with
\begin{equation}
    K_{FV}(M,\Delta,L) \equiv \frac{\partial}{\partial \Delta} J_{FV}(M,\Delta,L) ,
\end{equation}
and $\delta_1(ML)$ as given in \eq{delta1}. 

Before turning to the fit details and results, we briefly discuss the extent to 
which including the splittings as in \eq{phiq-Delta}, and not other possible $1/M_H$
effects, is a systematic improvement on \eq{phiq}.   In fact, 
in a parametric sense within the power counting introduced
by Boyd and Grinstein \cite{Boyd:1994pa}, this is a systematic
improvement, as long as we make some further specifications as to how
\eq{phiq-Delta} should be applied. As we detail below, however, the power counting
of Ref.~\cite{Boyd:1994pa} is only marginally applicable to our data. For that reason
we ultimately fit to both \eq{phiq-Delta}  and \eq{phiq} and take the difference as one
measure of the chiral extrapolation error.

For the following discussion, let $\Delta$ be a generic splitting ($\Delta^*$ or $\delta_{eq}$ or a linear combination of the two), and
$M$ be a generic light pseudoscalar mass. The power counting introduced in  Ref.~\cite{Boyd:1994pa} takes
\begin{equation}
    \frac{\Delta^2,\; \Delta M,\; M^2}{M_{H}} \ll \Delta\sim M.
    \eqn{power-counting}
\end{equation}
For our data, treating $\Delta$ and $M$ as the same size is not dangerous, even though $\Delta$ is significantly smaller than our simulation $M$ values---at worst this means
that we include some terms unnecessarily.
The condition $M^2/M_H\ll\Delta$, which is necessary to drop other $1/M_H$ contributions as still higher order, is marginally valid, however.
For the $D$ system,
$M_K^2/M_D\approx130~\MeV$, which is roughly of the same
size as $\Delta^*$ and $\delta_{sd}$. For the $B$ system,
$M_K^2/M_B\approx47~\MeV$, of the same size as $\Delta^*$ but somewhat less than
$\delta_{sd}$.   For the purposes of the chiral extrapolation, however, what matters 
is the applicability of the power counting at the lowest simulated light meson masses, not
its applicability at $M_K$.%
\footnote{We assume here that the fit to the data is good over the full mass range simulated.
It is not important for the chiral extrapolation that the fit be systematic in the region around $M_K$, but 
it must describe the data in that range so that we can correctly interpolate to the physical kaon mass.}
For our lightest simulated pions with mass $\sim M_K/2$, we can reduce the left hand side of the inequality
in \eq{power-counting} by a factor of four, at which point it becomes reasonably
applicable. 

Having tentatively accepted the power counting of \eq{power-counting}, it is clear that
$F(M/\Delta)$ in \eq{Jdef} should be treated as $\mathrm{O}(1)$. Then the difference between
$J(M,\Delta)$ and the chiral logarithm it replaces, $\ell(M^2)$ is of the same order
as $\ell(M^2)$ itself, so including the splittings becomes mandatory at the one-loop order
to which we are working.  The next question is whether \eq{phiq} includes \emph{all}
effects to this order.  As discussed by Boyd and Grinstein, the key issue is whether operators
with two or more derivatives (two or more powers of residual momentum $\bm{k}$)
on the heavy fields can contribute.  Such operators are suppressed
by $1/M_H$ relative to the leading-order heavy-light Lagrangian, which has a single
derivative.  Since we are keeping $\Delta^*$, which is also in principle a $1/M_H$
effect, one might worry that such operators could contribute at the same order. 
The power counting implies, however, that the relevant diagrams pick up
a factor of $(\Delta,M)/M_H$ relative to the terms being kept in \eq{phiq-Delta}.  The
reason for the difference is that the explicit extra factor of $\bm{k}$ 
turns into $\Delta$ or $M$---the only
dimensional constants available---after integration.  In the term that 
generates the
hyperfine splitting itself, in contrast, the dimensional quantity balanced against $1/M_H$ 
is $\Lambda$---a heavy-quark QCD scale---rather than $M$. The power counting in \eq{power-counting}
effectively treats $\Lambda$ as larger than $M$ (so that $\Delta \sim \Lambda^2/M_{H}\sim M$). 
Similarly, the term that generates the flavor splittings has a single
factor of $m_q$ and no residual momentum, and \eq{power-counting} effectively
takes $m_q\sim \bm{k}$ in such terms.

Boyd and Grinstein do find some other contributions at the same order as \eq{phiq-Delta},
but most come from terms that are simply $\Lambda/M_H$ times terms in
the leading-order heavy-light Lagrangian or current, and
thus give simply an overall  factor times the result without them.
The exceptions are the terms multiplied by $g_2$ in Eq.~(15) of Ref.~\cite{Boyd:1994pa}
and by $\rho_2$ in Eq.~(18) of Ref.~\cite{Boyd:1994pa}.  These are operators that have the same
dimension as the original Lagrangian current operators, but that violate heavy-quark spin
symmetry, and therefore give different contributions to the pseudoscalar
and vector  meson decay constants at this order. Since we are only looking
at pseudoscalar meson  decay constants here, however, and since these effects are flavor-independent,
we can also absorb all of the $1/M_H$ effects into (1) the effects of the splittings in the
loop, described by \eq{phiq-Delta}, and (2) an overall
factor in front of the full one-loop result. 

The overall factor  in  \eq{phiq-Delta} is $1/(16 \pi^2 f^2)$. Since $f$
is not fixed at one loop, one should in any case allow it to vary over a reasonable range, 
which we take to be $f_\pi$ to $f_K$. 
We allow such variations 
even when we fit to the form without
splittings, \eq{phiq}. The difference between using  $f_\pi$ and $f_K$
corresponds to a 45\% change in the size of the one-loop coefficient, but produces only a 1 to 3 MeV change
in the decay constants.%
\footnote{Most of the change in the size of the overall coefficient is compensated by a change in the LECs 
that come from the fit to our data.}
We therefore assume that any further 
$1/M_H$ uncertainty in $1/(16 \pi^2 f^2)$ has negligible effects on our results.

Finally, there is a question of whether terms coming from taste violations contribute something
new at the same order in which we include splittings.  Since taste-violating terms in the
Lagrangian can enter just like light-quark masses, this is a possibility in principle.
Corresponding to the terms in the quark masses that generate flavor splittings of
heavy-light mesons (\cf\ Eq.~(45) of
Ref.~\cite{Aubin:2005aq}), there are taste-violating terms given in Eq.~(51) of that paper.
Just as for the quark-mass terms, however, we are only interested here in contributions
that change the heavy-light meson mass, not ones coupling the mesons to pion fields.  
When the pion fields are set to zero, all the terms in  Eq.~(51) of Ref.~\cite{Aubin:2005aq} 
just give a constant heavy-light meson mass term proportional to $a^2$ that
contributes equally to the $H$ and $H^*$ masses of all valence flavors. 
Terms that produce a hyperfine splitting would have to also violate heavy quark spin symmetry, and hence be of order $a^2\Lambda/M_H$. Similarly, terms that produce
flavor splitting would need to violate flavor symmetry, and hence be of
order $a^2m_q/\Lambda_\chi$.
Both such contributions are higher order in our power-counting.
Since there is no splitting, there is no contribution to the decay constants because the effect
will vanish when we put the external $B$ or $D$ meson on mass shell.

In our chiral fits, we take the physical light-quark masses,
as well as the parameters $B_0$, $a^2\Delta_\xi$,  $\delta'_A$, and $\delta'_V$, 
from the MILC Collaboration's results of \rschpt\ fits to light pseudoscalars masses
and decay constants \cite{Aubin:2004fs,Bernard:2007ps} on ensembles that include lattice spacing
$a\approx \aMediumCoarse$~fm through $a\approx 0.06$~fm.
Table~\ref{tbl:params} shows the values used. 
\begin{table}[tp]
    \caption{Inputs to our heavy-light chiral fits taken from the MILC Collaboration's light-meson chiral fits \cite{Aubin:2004fs,Bernard:2007ps}. 
        The physical bare-quark masses $m_u$, $m_d$, $\hat m \equiv (m_u +m_d)/2$, and $m_s$ are determined by demanding that the charged pion and kaons take their physical masses after the removal of 
        electromagnetic effects.
        Errors in the masses are due to statistics, chiral extrapolation systematics, scale determination, 
        and (for $m_d$ and $m_u$) the estimate of electromagnetic effects, respectively.
        ``Continuum'' values are found from chiral fits that have been extrapolated to the continuum, but 
        masses are still in units of the ``fine'' ($a\approx0.09$~fm) lattice spacing, and with the 
        fine-lattice value of the mass renormalization.  
        Values for $r_1^2 a^2 \delta'_A$ and $r_1^2 a^2 \delta'_V$ take into account newer MILC analyses \cite{Bazavov:2009fk} as noted in the text.
        The light-meson analysis determining these quantities assumes that they scale like the taste-violating splittings $\Delta_\xi$ and are larger by a factor of 1.68 on the $\aMediumCoarse~$fm lattices than on 
        the $\aCoarse~$fm lattices, and smaller by a factor 0.35 on the $\aFine~$fm lattices than on the $\aCoarse~$fm lattices. 
        The statistical and systematic errors on $r_1  B_0$ and $r_1^2 a^2 \Delta_{\xi}$  are not given here;
        such errors have negligible effect on the heavy-light decay constants.}
    \label{tbl:params}
    \begin{tabular*}{\textwidth}{c@{\extracolsep{\fill}}l@{\extracolsep{\fill}}l@{\extracolsep{\fill}}l@{\extracolsep{\fill}}l}
        \hline\hline
        Quantity & \multicolumn{4}{c}{Lattice spacing} \\
        & \quad$a\approx \aMediumCoarse~$fm  &  \quad$a\approx \aCoarse~$fm & \quad$a\approx \aFine~$fm & \quad ``continuum'' \\
        \hline                            
        $a m_s\times10^2$ &  $4.29(1)(8)(6)$ &  $3.46(1)(10)(5)$ &  $2.53(0)(6)(4)$ & $2.72(1)(7)(4)$\\
        $a \hat m\times10^3$ & $1.55(0)(3)(2)$ & $1.25(0)(4)(2)$ & $0.927(2)(27)(13)$  & $0.997(2)(32)(14)  $\\
        $a m_d\times10^3$&  $2.20(0)(4)(3)(5)$ & $1.78(0)(6)(3)(4)$ & $1.31(0)(4)(2)(3)$ & $1.40(0)(5)(2)(3)$ \\
        $a  m_u\times10^4$&  $8.96(2)(17)(13)(49)$ & $7.31(2)(23)(10)(40)$ & $5.47(1)(16)(8)(30)$ & $5.90(1)(19)(9)(32)$ \\
        \hline
        $r_1 B_0 $  & \quad 6.43 & \quad 6.23 & \quad 6.38 & \quad 6.29 \\
        \hline
        $r_1^2 a^2 \Delta_{A}$ & $\hphantom{-}0.351$ & $\hphantom{-}0.205$ & $\hphantom{-}0.0706$ & \quad 0 \\
        $r_1^2 a^2 \Delta_{T}$ & $\hphantom{-}0.555$ & $\hphantom{-}0.327$ & $\hphantom{-}0.115$  & \quad 0 \\
        $r_1^2 a^2 \Delta_{V}$ & $\hphantom{-}0.721$ & $\hphantom{-}0.439$ & $\hphantom{-}0.152$  & \quad 0 \\
        $r_1^2 a^2 \Delta_{I}$ & $\hphantom{-}0.897$ & $\hphantom{-}0.537$ & $\hphantom{-}0.206$  & \quad 0 \\
        \hline
        $r_1^2 a^2 \delta'_A$ &\quad--- &  $-0.28(6)$ & \quad--- & \quad 0 \\
        $r_1^2 a^2 \delta'_V$ &\quad---&  $\hphantom{-}0.00(7)$ & \quad---& \quad 0 \\
        \hline\hline
    \end{tabular*}
\end{table}
In general, we use older MILC determinations since newer versions, {\it e.g.}, those in Ref.~\cite{Bazavov:2009fk}, do not cover the full range of lattice
spacings employed here (but are consistent where they overlap).  
The exceptions are the values of the taste-violating hairpin
parameters $r_1^2 a^2 \delta'_A$ and $r_1^2 a^2 \delta'_V$.  For them, the newer
analysis including two-loop chiral logarithms gives larger systematic errors and a changed
sign of the central value of $r_1^2 a^2 \delta'_V$, which has always been consistent with zero.
For these parameters, we therefore use the wider ranges listed in Table~\ref{tbl:params}, 
which encompasses both types of analyses. 
For comparison, the results of the
analysis of Ref.~\cite{Bernard:2007ps} were
 $r_1^2 a^2 \delta'_A = -0.30(1)(4)$ and $r_1^2 a^2 \delta'_V = -0.05(2)(4)$.

In order to fit \eq{phiq-Delta} to our lattice data, it is also necessary to input values for the
hyperfine splitting $\Delta^*$ and for $\lambda_1$ in \eq{deltaeq}.
For $B$ mesons, we have~\cite{Nakamura:2010zzi}
\begin{eqnarray}
    \Delta^* = M_{B^*}-M_B & \approx &  45.8~\MeV,  \eqn{Bvalues1} \\
    \delta_{sd} = M_{B_s}-M_{B} & \approx& 87.0~\MeV, \eqn{Bvalues2} \\
    \lambda_1 &\approx& 0.192~\GeV^{-1}, \eqn{Bvalues3}
\end{eqnarray}
where we use $M_E= M_S=0.6858(40)~\GeV$ \cite{Davies:2009tsa} and $M_Q=M_{\pi^0}\approx135.0~\MeV$ to obtain $\lambda_1$ from the experimental data.
Similarly, for $D$ mesons, we have
\begin{eqnarray}
    \Delta^* = M_{D_0^*}-M_{D_0} & \approx & 142.1~\MeV, \eqn{Dvalues1} \\
    \delta_{sd} = M_{D_s} - M_{D_\pm} & \approx & 98.9~\MeV,  \eqn{Dvalues2}\\
    \lambda_1 &\approx& 0.219~\GeV^{-1}. \eqn{Dvalues3}
\end{eqnarray}

In the chiral fit, we input the relevant physical
 $\Delta^*$ and $\lambda_1$ from either Eqs.~(\ref{eq:Bvalues1})--(\ref{eq:Bvalues3}) or~(\ref{eq:Dvalues1})--(\ref{eq:Dvalues3}), and then use \eq{deltaeq}
with the actual $m_e$ and $m_q$ from each data point, and $B_0$ the slope for a given 
ensemble, from \tabref{params}. We emphasize here that $B_0$ comes from a simple
tree-level chiral fit 
of light meson masses to \eq{mesonmass}. This is adequate for our purposes, since the
resulting meson masses are only used within the one-loop chiral logarithms.

\bigskip 

We can now present the actual chiral fits and show how we extract results and systematic 
errors from them. 
Recall that we compute $\phi_{H_q}$ for many combinations of the valence and
light sea-quark masses, and at three lattice spacings: $a \approx$~\aMediumCoarse, \aCoarse, and \aFine~fm.
We fit all the decay constant data to the
form given either by \eq{phiq-Delta} or by \eq{phiq}. One-loop finite-volume 
effects are included through \eq{J-FV} or \eq{ell-FV}. There are four unconstrained
free parameters in our fits: the LO parameter $\phi^0_H$, and
the one-loop LECs $L_{\rm val}$, $L_{\rm sea}$, $L_a$ [\eq{pq}].
The central fit 
fixes the chiral coupling $f$ at $f_\pi$, but a range of couplings are considered
in alternative fits, as described in more detail in \secref{Errors}.
Similarly, 
the $H$-$H^*$-$\pi$  coupling $g_\pi$, which is poorly constrained by our data, is taken
in the range $0.51\pm0.20$. This encompasses a range of
phenomenological and lattice determinations~\cite{Casalbuoni:1996pg,Stewart:1998ke,Anastassov:2001cw,Abada:2002vj,Arnesen:2005ez,Ohki:2008py,Bulava:2010ej}, as discussed in Ref.~\cite{Bernard:2008dn}. In
the central fit, $g_\pi$ is held fixed at 0.51, while it is varied in alternative fits described
in \secref{Errors}.
Although changing $g_\pi$ is equivalent to
changing $f$ when splittings are omitted [\cf\ \eq{phiq}], the effects are inequivalent
when splittings are included [\cf\ \eq{phiq-Delta}].  This 
is especially true of the finite-volume effects, for which the splittings have the potential
to produce significant changes \cite{Arndt:2004bg}.

Some additional parameters constrained by Bayesian priors are also included in the chiral fits,
as discussed above. The taste-violating hairpin parameters $\delta'_V$ and $\delta'_A$
are given by the ranges in \tabref{params}. In addition,
up to six heavy-quark and up to seven light-quark lattice-spacing dependent
terms, are added for investigation of discretization effects.
Except where otherwise noted, all twelve such terms are included
in the fits plotted below: this gives errors that include true statistical errors
plus our estimate of 
discretization effects from the heavy quarks and generic (taste non-violating) discretization
errors from the light quarks.  In addition, some or all of the (mildly) constrained 
NNLO LECs, $Q_1,\dots,Q_4$, are included.  Again, unless otherwise noted, the fits below include
all four such parameters; such fits tend to give larger (and hence more conservative)
errors than fits that restrict the number of these parameters.  In total, there are 23
fit parameters in the central fits: the 19 constrained parameters listed in this paragraph, and the 4 unconstrained parameters listed in the previous paragraph.

\subsection{Chiral fits and extrapolations for the $D$ system}
\label{sec:Dresults}

\begin{figure}[b]
\centering
    \includegraphics[width=11.0cm]{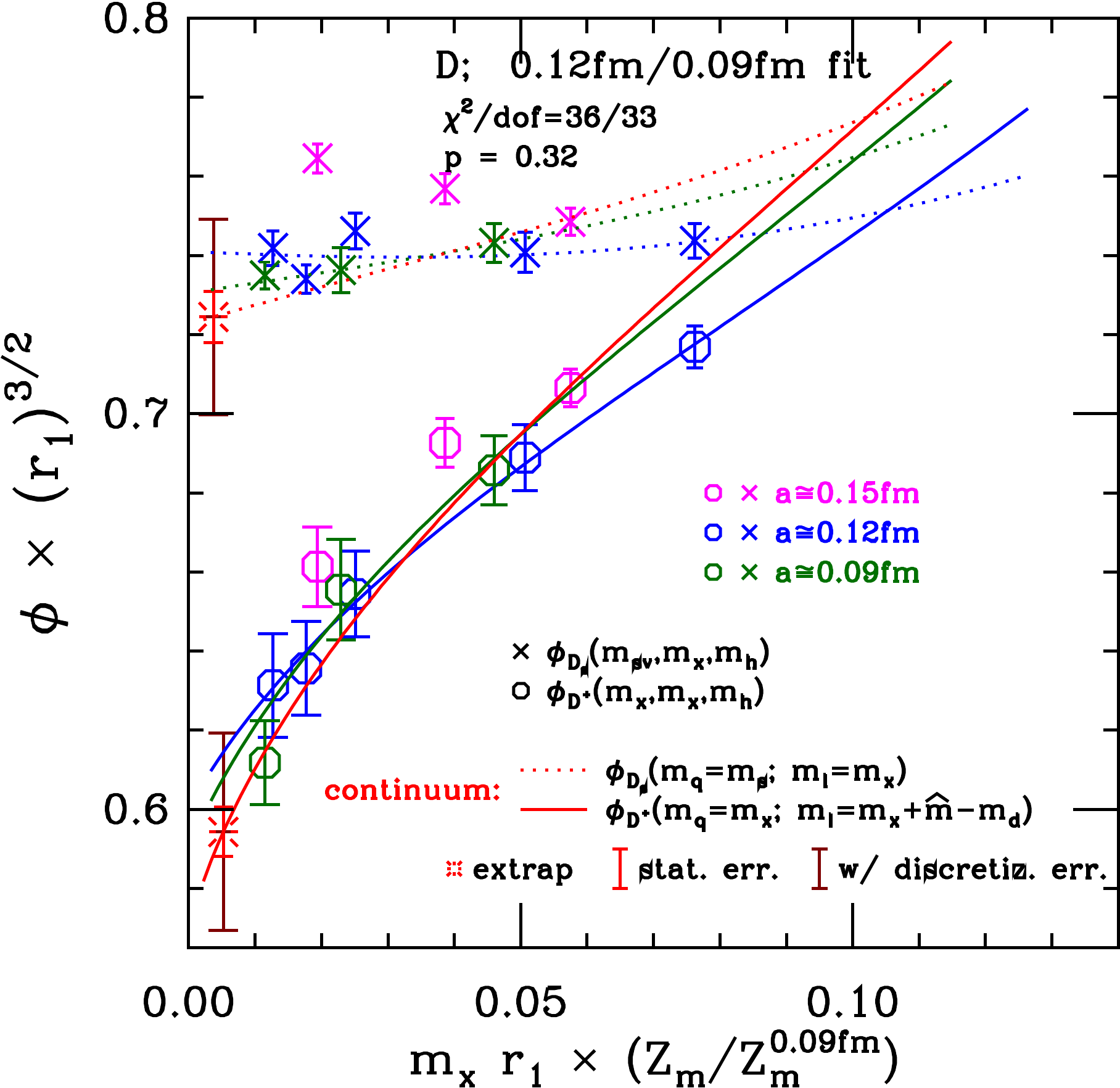}
    \caption{Central chiral fit for the $D$ system, based on Analysis~I of the fits to 2-point correlators. 
        Only (approximately) unitary points are shown.
        Data from ensembles at $a\approx\aMediumCoarse$~fm, $a\approx\aCoarse$~fm and $a\approx\aFine$~fm are shown, but the $a\approx\aMediumCoarse$~fm ensembles are not included in the fit.  
        The bursts show extrapolated values for $\phi_{D_s}$ and $\phi_{D^+}$, with the purely statistical errors in bright red and the statistical plus discretization errors in darker red. 
        The physical strange-quark mass corresponds to an abscissa value of $m_x\approx 0.1$.}
    \label{fig:fDcentral}
\end{figure}

\Figref{fDcentral} shows our central chiral fit to $r_1^{3/2}\phi_{D^+}$ and $r_1^{3/2}\phi_{D_s}$. 
Data from ensembles at $a\approx\aMediumCoarse$~fm,
$a\approx\aCoarse$~fm and $a\approx\aFine$~fm are shown, but the $a\approx\aMediumCoarse$~fm ensembles are not included
in the fit.  
The points and covariance matrix are obtained from Analysis~I (Sec.~\ref{sec:jackknife}) of the two-point functions. 
For clarity, only the unitary (full QCD) points
are shown for $\phi_D$ (and approximately unitary
for $\phi_{D_s}$), but the fit is to all the partially-quenched 
data on the $a\approx\aCoarse$~fm and $a\approx\aFine$~fm ensembles.
The fit properly takes into account the covariance 
of the data; $\chi^2/{\rm dof}$ and the $p$ value (goodness of fit) are reasonable, as shown.
The points in \figref{fDcentral} are plotted as a function of mass $m_x$, where, for $\phi_{D^+}$,
the light valence mass $m_q$ and the light sea mass $m_l$ are given by $m_q=m_l=m_x$.
For $\phi_{D_s}$, only $m_l=m_x$ varies, while $m_q$ is held fixed at the value
$m_{s_v}$ near the physical strange mass $m_s$.%
\footnote{On the $a\approx0.15$~fm ensembles, $m_{s_v}$ is equal to the 
    value of the strange sea quark mass $m_h$ ($am_{s_v}=0.0484$), but on the
    other two ensembles we take it lower than $m_h$, because $m_h$ has been 
    chosen somewhat larger than the physical strange mass.  In the figure,
    $a m_{s_v}=0.415$ for the $a\approx0.12$~fm ensembles and $a 
    m_{s_v}=0.272$ for the $a\approx0.09$~fm ensembles.}
In order to be able to compare ensembles at different lattice spacings, we have adjusted
the bare quark masses by the ratio $Z_m/Z^{\mathrm{0.09\,fm}}_m$, where  $Z_m$ is
the (one-loop) mass renormalization constant \cite{Aubin:2004ck}, and 
$Z^{\mathrm{0.09\,fm}}_m$ is its value on the $a\approx0.09$ fm ensembles.

The continuum extrapolation is carried out by taking the fitted parameters and
setting $a^2=0$ in all taste-violating terms
(parameterized by $\Delta_\xi$, $\delta'_A$, $\delta'_V$, and $L_a$), all heavy-quark
discretization effects (parameterized by $z_E$, $z_X$, $z_Y$, $z_B$, $z_3$, and $z_2$) and all generic light-quark
discretization effects (parameterized by $C_i$). 
The red lines
(solid for  $\phi_{D^+}$, dotted for  $\phi_{D_s}$) show the effect of extrapolating
to the continuum and setting the strange quark mass (both sea, $m_h$, and valence, $m_{s_v}$)
to the physical value $m_s$. 

Finally, the bursts give the result after the chiral extrapolation in the continuum,
\ie, setting $m_x=m_d$ for $\phi_{D^+}$, and $m_x=\hat m$ for $\phi_{D_s}$.
The larger, dark red, error bars on the bursts show the total error from the fit,
which includes heavy-quark and generic light-quark discretization errors using Bayesian priors, as
described above.
The smaller, bright red error bars, show purely statistical errors, which are computed
by a fit with all the discretization prior functions turned off.
In plotting the red line for $\phi_{D^+}$, the light sea mass is shifted slightly 
($m_l=m_x+\hat m-m_d$) so that it takes its proper mass when $m_x=m_d$. (We neglect
isospin violations in the sea.)
The small mass differences between $\hat m$ and $m_d$ (and the corresponding
difference between $\hat m$ and $m_u$ for the $B^+$) produce changes in $\phi$ that are
much smaller than our current errors, but we include them here with an eye to future
work, where the precision will improve.

\begin{figure}[b]
\centering
    \includegraphics[width=11.0cm]{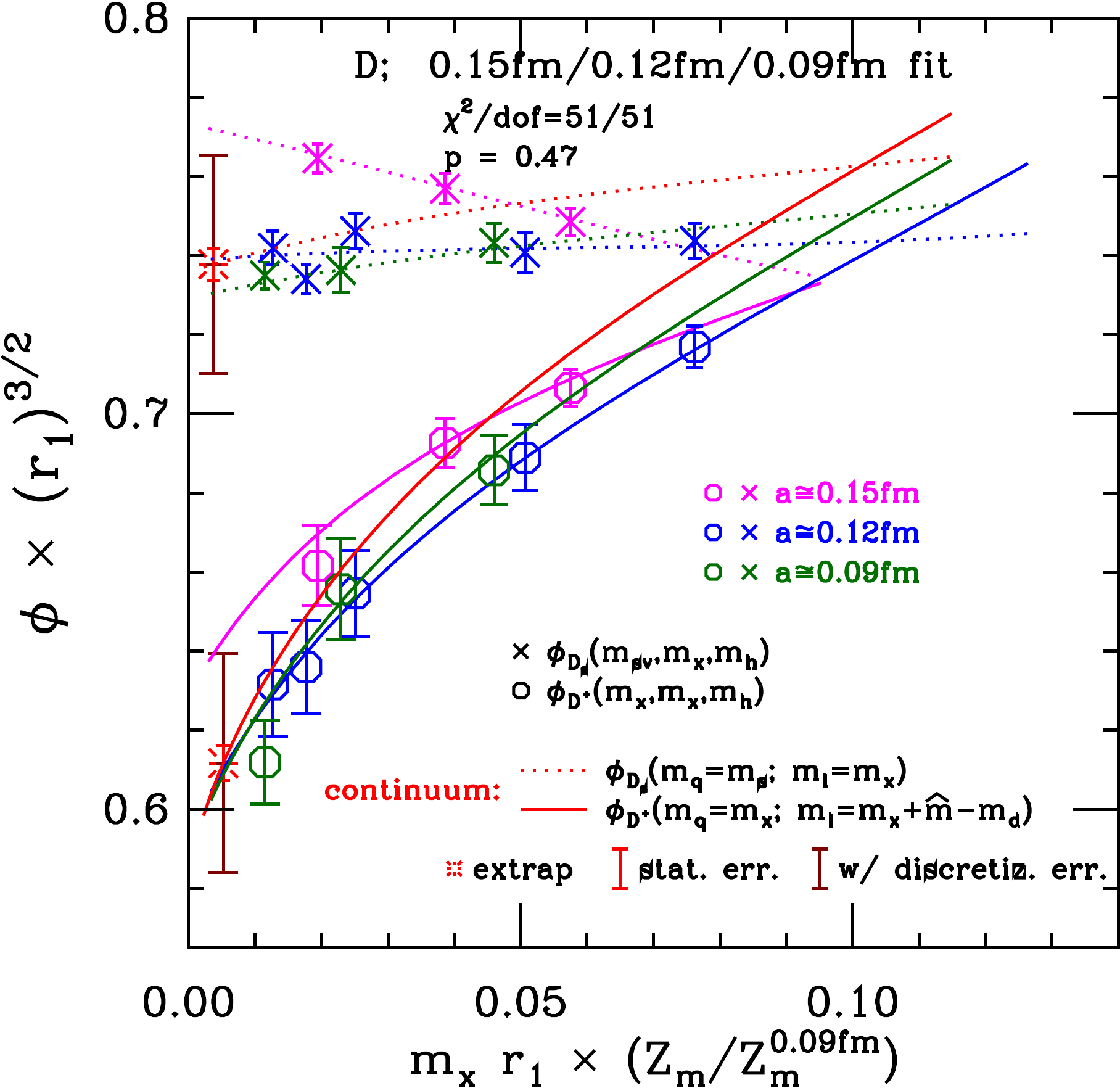}
    \caption{Same as \figref{fDcentral}, but including points at $a\approx\aMediumCoarse$~fm in the chiral-continuum fit.}
    \label{fig:fDmcf}
\end{figure}

The trend of the data for the coarsest lattice spacing ($a\approx\aMediumCoarse$~fm, the magenta points in \figref{fDcentral})
tends to be rather different than for the finer lattice spacings, especially
for the $D_s$, which is why we
exclude  the $a\approx\aMediumCoarse$~fm data from the central fit.
This trend is even more exaggerated for the $B$ system, but with particularly large statistical errors;
see \figref{fBcentral} below.  Nevertheless, the effect of
including the $a\approx\aMediumCoarse$~fm points in the fit is a rough indication  
of the size of discretization errors.
\Figref{fDmcf} shows what happens to the fit when these points are included: $\phi_{D^+}$ and $\phi_{D_s}$ each move up an 
amount comparable to (but less than)
the size of the larger (dark red) error bars, which represent heavy and generic light
quark discretization errors (as well as statistical errors, which are smaller).  
The consistency is reassuring. 

\begin{figure}[b]
\centering
    \includegraphics[width=11.0cm]{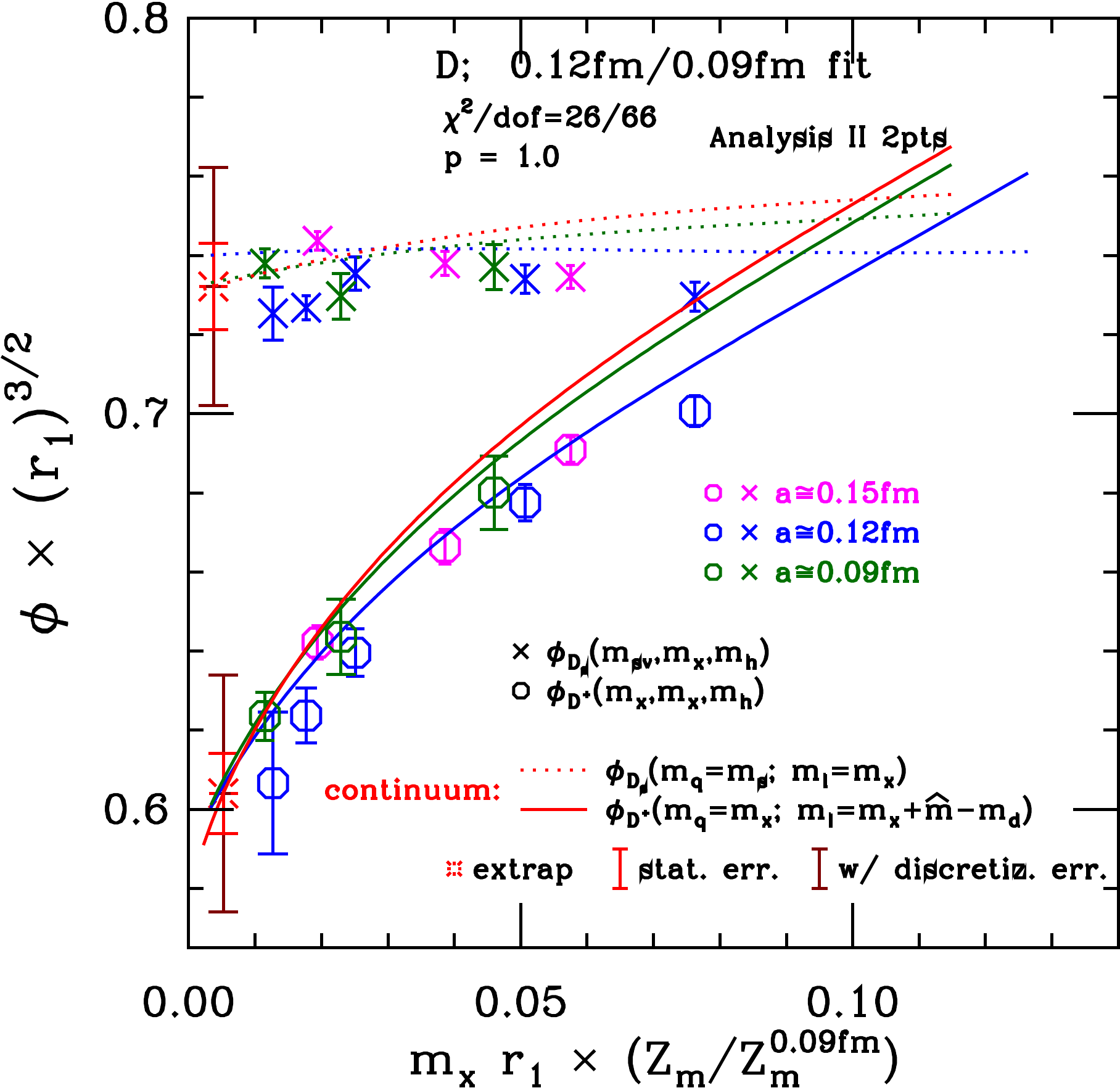}
    \caption{Same as \figref{fDcentral}, but using Analysis~II of the 2-point function.}
    \label{fig:fDjim}
\end{figure}

As discussed in \secref{2point}, we also examine Analysis~II of the 2-point functions.
\Figref{fDjim} shows the effect of using Analysis~II in the chiral fits. 
The differences in the decay constant results between \figref{fDcentral} and \figref{fDjim}
are included in the decay-constant error budgets as a ``fitting error".  Note that the covariance
matrix calculation in Analysis~II results in an apparent underestimate of $\chi^2$ (and, consequently, a high apparent $p$ value).  We believe that this stems from
binning of the data to remove autocorrelation effects, which has the disadvantage
of reducing the number of samples used to compute the covariance matrix.
It is then difficult to determine small eigenvalues accurately. Indeed the eigenvalues of the (normalized)
correlation matrix tend to have a lower bound of $\sim10^{-4}$ to $10^{-3}$ with this approach,
whereas they typically go down to $10^{-5}$ in Analysis~I.
[Recall that in Analysis~I we keep all samples, and deal with autocorrelation effects by
Eq.~(\ref{eq:autocor_rescale}).]  Nevertheless, the difficulty
with  small eigenvalues explains only a small fraction of the difference between the results from Analyses~I and~II.
For example, $f_D$ is changed by only 0.2~MeV when we smooth eigenvalues from Analysis~I
that are less than $10^{-3}$,
following the method of Ref.~\cite{Bernard:2002pc}.  This may be compared to
the total difference between $f_D$ in Analyses~I and~II,  which is 1.7 MeV.  

\subsection{Chiral fits and extrapolations for the $B$ system}
\label{sec:Bresults}

Results for the $B$ system closely resemble those for the $D$ system in most respects.  
One important difference is that the signal-to-noise ratio is worse for the $B$ system because the mass difference 
that controls the noise, $2m_B-m_{\eta_b}-m_\pi$, increases
with the mass of the heavy quark~\cite{Lepage:TASI}. 
Therefore, the preferred fit in Analysis~I for the charm
case (1 simple exponential + 1 oscillating exponential at large $t_\textit{min}$) is too noisy
here, and we must use fits with an extra excited state and smaller $t_\textit{min}$ (see \secref{jackknife}).
Consequently, our $B$-system results have larger statistical errors.
On the other hand, heavy-quark discretization
errors are smaller in the $B$ system. In the HQET analysis of discretization effects they appear in the heavy-quark expansion, which works better for $B$'s to begin 
with~\cite{Oktay:2008ex}.

\begin{figure}[b]
\centering
\includegraphics[width=11.0cm]{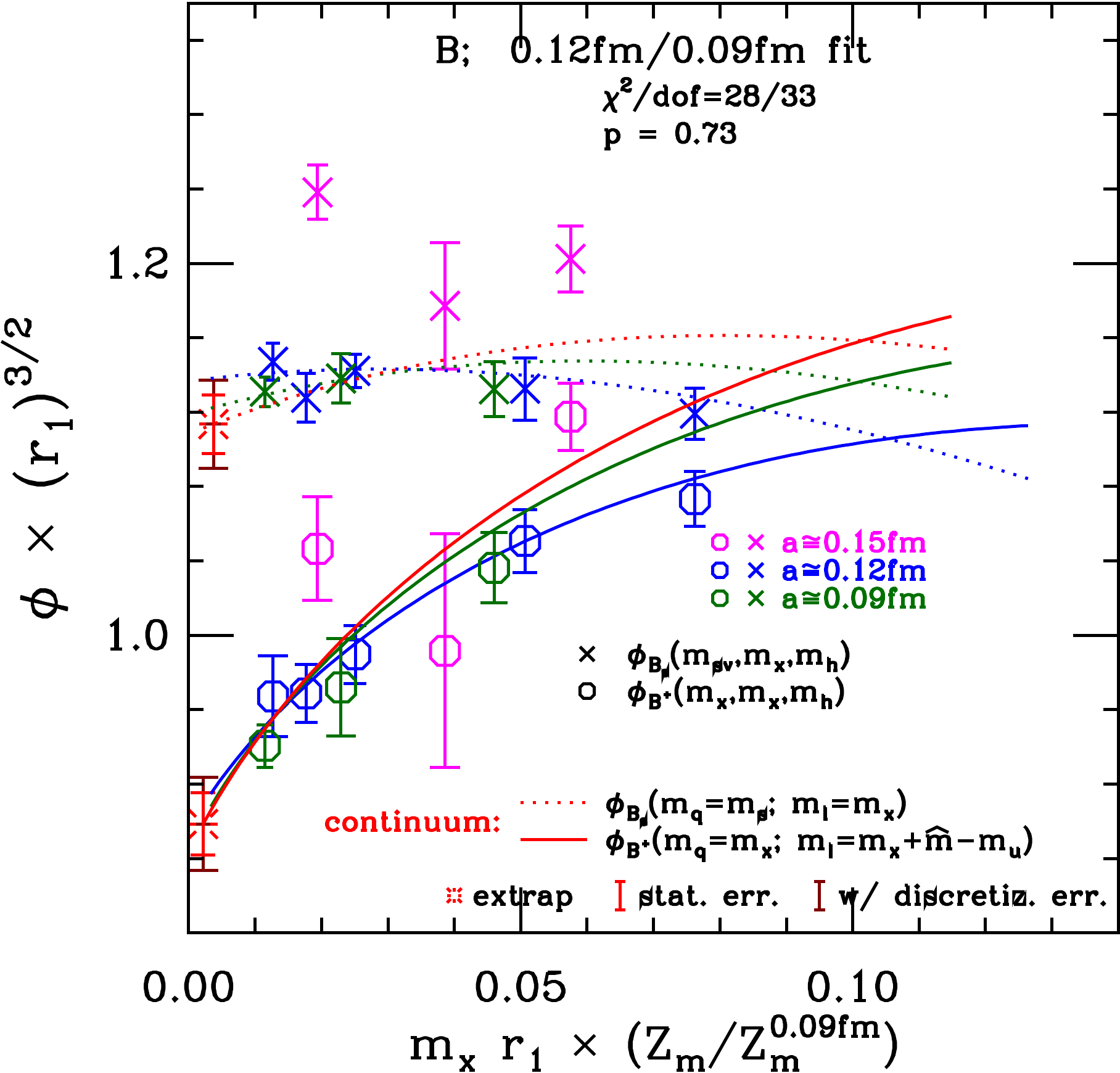}
    \caption{Central chiral fit for the $B$ system, with data from Analysis~I of the 2-point functions.
    Only (approximately) unitary points are shown.
    Data from ensembles at $a\approx\aMediumCoarse$~fm, $a\approx\aCoarse$~fm, and $a\approx\aFine$~fm are shown, but the $a\approx\aMediumCoarse$~fm ensembles are not included in the fit.
    The bursts show extrapolated values for $\phi_{B_s}$ and $\phi_{B^+}$, with the purely statistical errors in bright red and the statistical plus discretization errors in darker red. 
    The physical strange-quark mass corresponds to an abscissa value of $m_x\approx 0.1$.}
    \label{fig:fBcentral}
\end{figure}

\Figref{fBcentral} shows, for unitary points only, our central chiral fit for the $B$ system.
This is based on Analysis~I of the 2-point functions.
As in \figref{fDcentral}, 
the red lines (solid for $\phi_{B^+}$, dotted for $\phi_{B_s}$) show
the effect of extrapolation to the continuum and setting the strange quark mass to its physical
value $m_s$.
For the solid red line, the light sea mass is again shifted slightly,
but now $m_l=m_x+\hat m-m_u$, so that it takes its proper mass when $m_x=m_u$.
The bursts show the final results, and come from setting
$m_x=m_u$ for $\phi_{B^+}$ and $m_x=\hat m$ for $\phi_{B_s}$.
As before, the smaller, bright red, error bars, show purely statistical errors, and the
larger, dark red, error bars come from the fit with Bayesian priors and include
heavy-quark and generic light-quark discretization errors as well as statistical errors.

\begin{figure}[b]
\centering
    \includegraphics[width=11.0cm]{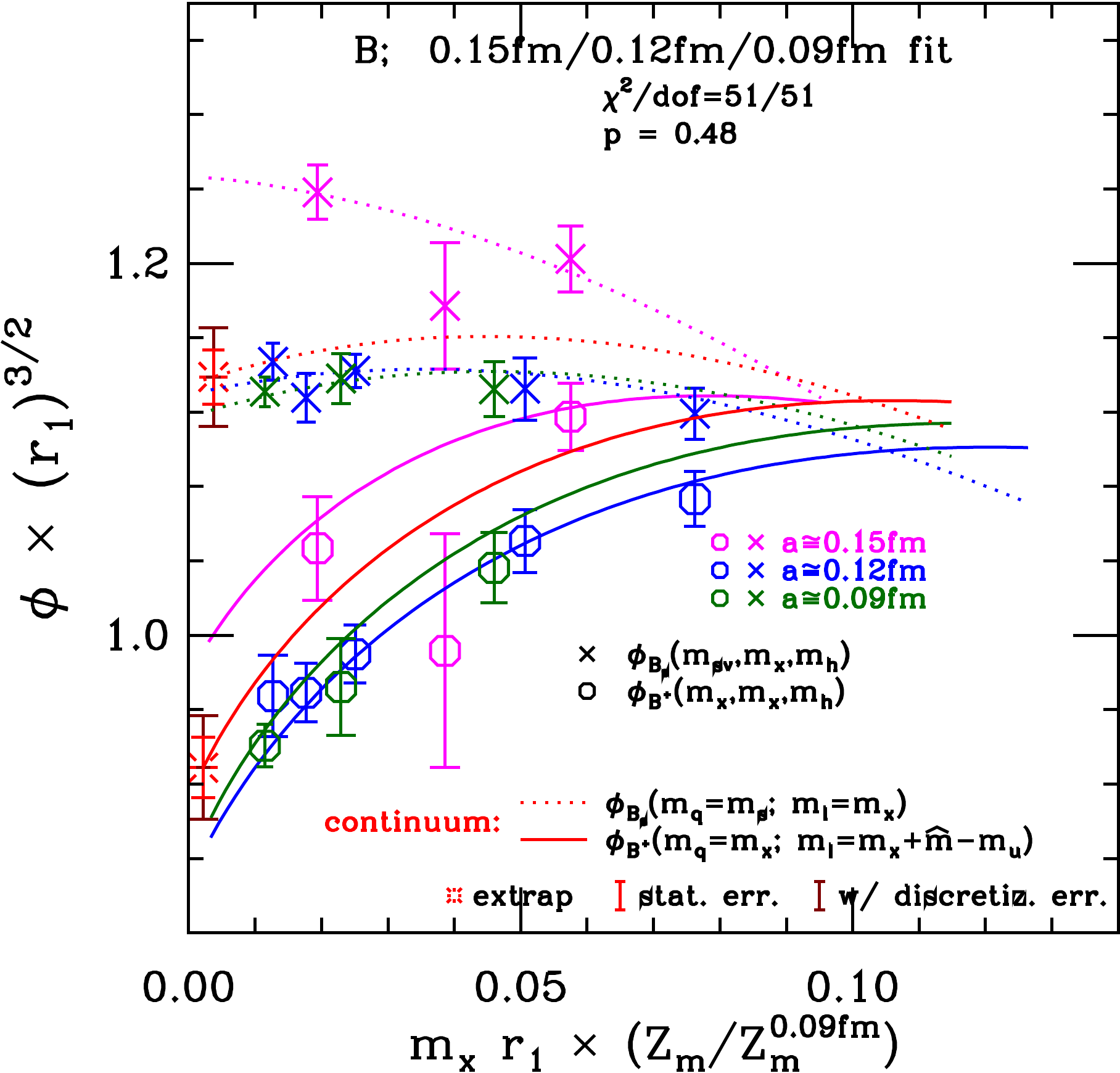}
    \caption{Same as \figref{fBcentral}, but including points at $a\approx\aMediumCoarse$~fm in the fit.}
    \label{fig:fBmcf}
\end{figure}

In \figref{fBcentral}, the $a\approx\aMediumCoarse$~fm data are both noisy and far from those of the finer lattice spacings. 
Therefore, these ensembles are again dropped from the central fit.
\Figref{fBmcf} shows the effect of including  the $a\approx\aMediumCoarse$~fm points.
Note that the resulting continuum-extrapolated line for $\phi_{B_s}$ (dotted red line)
now has what appears to be a rather unphysical shape, showing a significant initial increase
as the light sea-quark mass is decreased, starting at the right side of the graph.
Hence, the differences caused by including the $a \approx 0.15$~fm points is 10 to 20\% larger than the dark red error
bars in \figref{fBcentral}, and 40 to 60\% larger than discretization errors estimated by
removing the statistical errors from the dark red bars. Because the trend for  $a\approx\aMediumCoarse$~fm is so
different from the other spacings, and because of the unphysical behavior when these points
are included in the fit, we believe this difference overestimates the true discretization error,
and we do not enlarge the errors coming from the  fit.

\begin{figure}[b]
\centering
    \includegraphics[width=11.0cm]{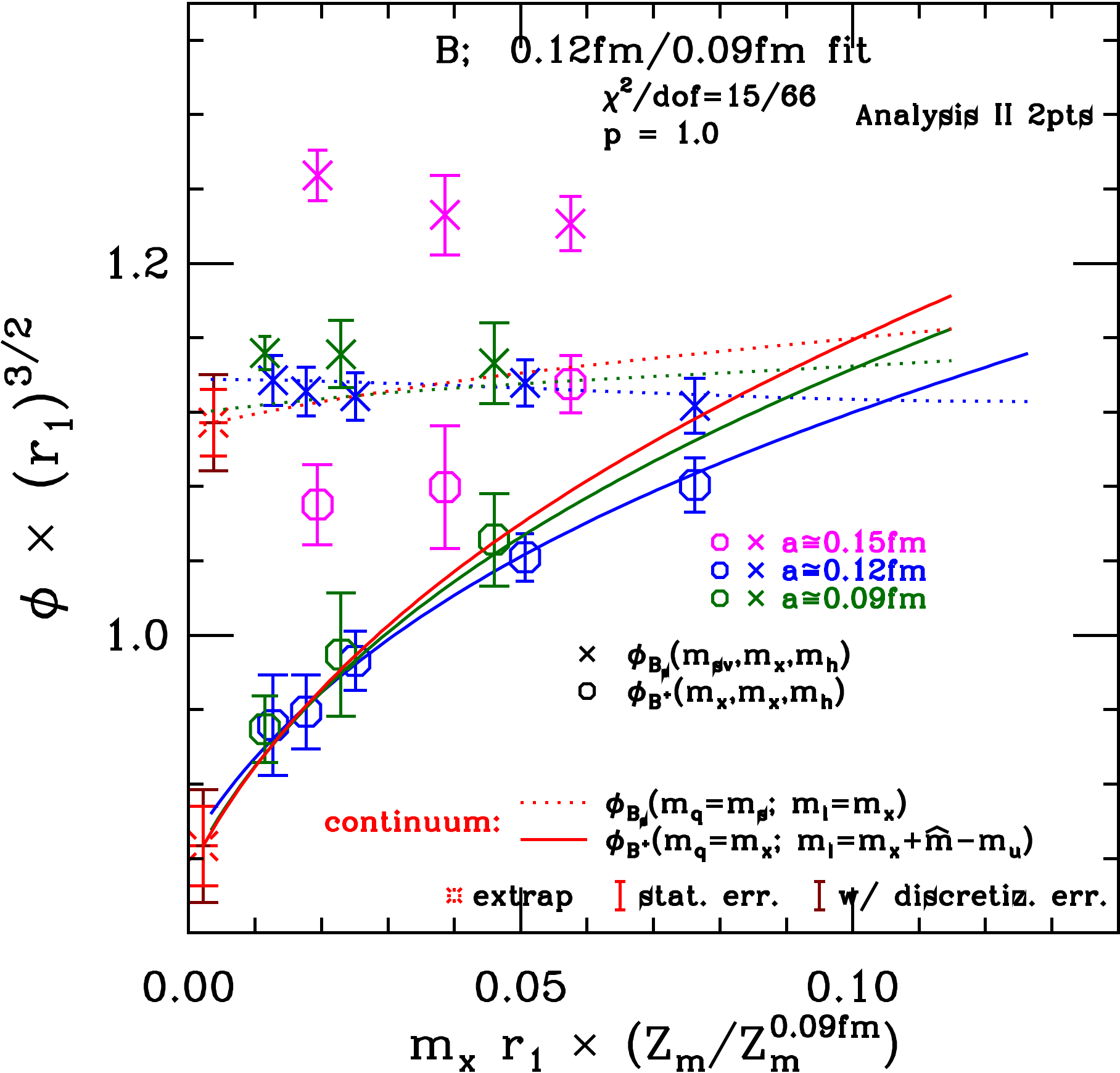}
    \caption{Same as \figref{fBcentral}, but using Analysis~II of the 2-point functions.}
    \label{fig:fBjim}
\end{figure}

\Figref{fBjim} shows the effect of using Analysis~II of the correlation functions.   In order to make these comparisons as direct as possible,
we first turn off all the Bayesian discretization terms in the fits.   
Compared to the results from \figref{fBcentral}, this fit gives a value of $f_{B_s}$ about 1 MeV higher 
and a value of $f_{B^+}$ about 2 MeV lower.  These differences are included in our estimate of
the fitting errors due to excited state contamination in \secref{Errors}.

\section{Estimation of systematic errors}
\label{sec:Errors}

In this section, we present a careful, quantitative accounting for the uncertainties in our calculation.
We consider in turn
discretization errors,
fitting errors,
errors from inputs $r_1$ and quark-mass tuning,
renormalization, 
and
finite-volume effects.
Table~\ref{tbl:errorBudget} details our error budget.
\begin{table}
    \caption{Total error budget for the heavy-light decay constants.
        Uncertainties are in MeV for decay constants.
        The total combines errors in quadrature.
        The first row includes statistics, heavy-quark discretization errors, and generic light-quark 
        discretization errors, as explained in the text.
        Errors in parentheses are approximate
        sub-parts of errors that are computed in combination.}
    \label{tbl:errorBudget}
    \begin{tabular*}{\textwidth}{l@{\extracolsep{\fill}}r@{.\!}l@{\extracolsep{\fill}}r@{.\!}l@{\extracolsep{\fill}}r@{.\!}l@{\extracolsep{\fill}}r@{.\!}l@{\extracolsep{\fill}}r@{.\!}l@{\extracolsep{\fill}}r@{.\!}l}
    \hline\hline
        Source & \multicolumn{2}{c}{$f_{D^+}$ (MeV)} & \multicolumn{2}{c}{$f_{D_s}$ (MeV)} & \multicolumn{2}{c}{$f_{D_s}/f_{D^+}$} 
               & \multicolumn{2}{c}{$f_{B^+}$ (MeV)} & \multicolumn{2}{c}{$f_{B_s}$ (MeV)} & \multicolumn{2}{c}{$f_{B_s}/f_{B^+}$} \\ 
        \hline
        Statistics $\oplus$ discretization & 9&2 & 8&9 &  0&014  &  5&5  &  5&1  &  0&013  \\
        \quad (statistics)             & (2&3) & (2&3) & (0&005) & (3&6) & (3&4) & (0&010) \\
        \quad (heavy-quark disc.)      & (8&2) & (8&3) & (0&007) & (3&7) & (3&8) & (0&004) \\
        \quad (light-quark disc.)      & (2&9) & (1&5) & (0&012) & (2&5) & (2&1) & (0&007) \\
        Chiral extrapolation           &  3&2  &  2&2  &  0&014  &  2&9  &  2&8  &  0&014  \\
        Two-point functions            &  3&3  &  1&6  &  0&013  &  3&0  &  4&1  &  0&015  \\
        Scale ($r_1$)                  &  1&0  &  1&0  &  0&001  &  1&0  &  1&4  &  0&001  \\
        Light quark masses             &  0&3  &  1&4  &  0&005  &  0&1  &  1&3  &  0&006  \\
        Heavy quark tuning             &  2&8  &  2&8  &  0&003  &  3&9  &  3&9  &  0&005  \\
        $u_0$ adjustment               &  1&8  &  2&0  &  0&001  &  2&5  &  2&8  &  0&001  \\
        Finite volume                  &  0&6  &  0&0  &  0&003  &  0&5  &  0&1  &  0&003  \\
        $\ZvQQ$ and $\Zvqq$            &  2&8  &  3&4  &  0&000  &  2&6  &  3&1  &  0&000  \\
        Higher-order $\rho_{A_4}^{Qq}$ &  1&5  &  1&8  &  0&001  &  1&4  &  1&7  &  0&001  \\
        \hline
        Total Error                    & 11&3  &  10&8 &  0&025  &  8&9  &  9&5  &  0&026  \\
        \hline\hline
        \end{tabular*}
\end{table}

\subsection{Heavy-quark and generic light-quark discretization effects}

As described in \secref{ChPT} and Appendix~\ref{app:HQcutoff}, we parameterize possible heavy-quark and
generic light-quark discretization effects and follow a Bayesian approach in including such
effects in our chiral fitting function.
Consequently, the raw ``statistical'' error that comes
from our fits is not a pure statistical error but includes an estimate of the errors coming
from the discretization effects.  
This inclusive error is shown with the
dark red error bars in the plots in \secref{ChPT}, and is listed in the first
line of \tabref{errorBudget}.  

For informational purposes, it is useful to break down this inclusive error into its component parts, at least approximately.
We can see what errors to expect and, hence, target for improvement in future simulations.
In particular, with our current actions, the light-quark and heavy-quark discretization errors should behave differently
as a function of lattice spacing, with heavy-quark errors decreasing more slowly as $a$ is reduced.
To extract the pure statistical errors, we rerun the fits with all the Bayesian discretization
terms set to zero.  We then find the pure heavy-quark (or pure light-quark) discretization
contributions, by turning back on the heavy-quark (light-quark) terms, and then subtracting
in quadrature the pure statistical errors from the resulting raw errors. These
individual errors are shown in \tabref{errorBudget} in parentheses.
Note that the total error at the bottom of the table includes the error on the first line, not the sum of the three errors in parentheses, when these differ.
Note also that the discretization errors are similar to what we would have obtained with less sophisticated power counting.

\subsection{Chiral extrapolation and taste-violating light-quark discretization effects}

\begin{figure}[b]
\centering
    \includegraphics[width=11.0cm]{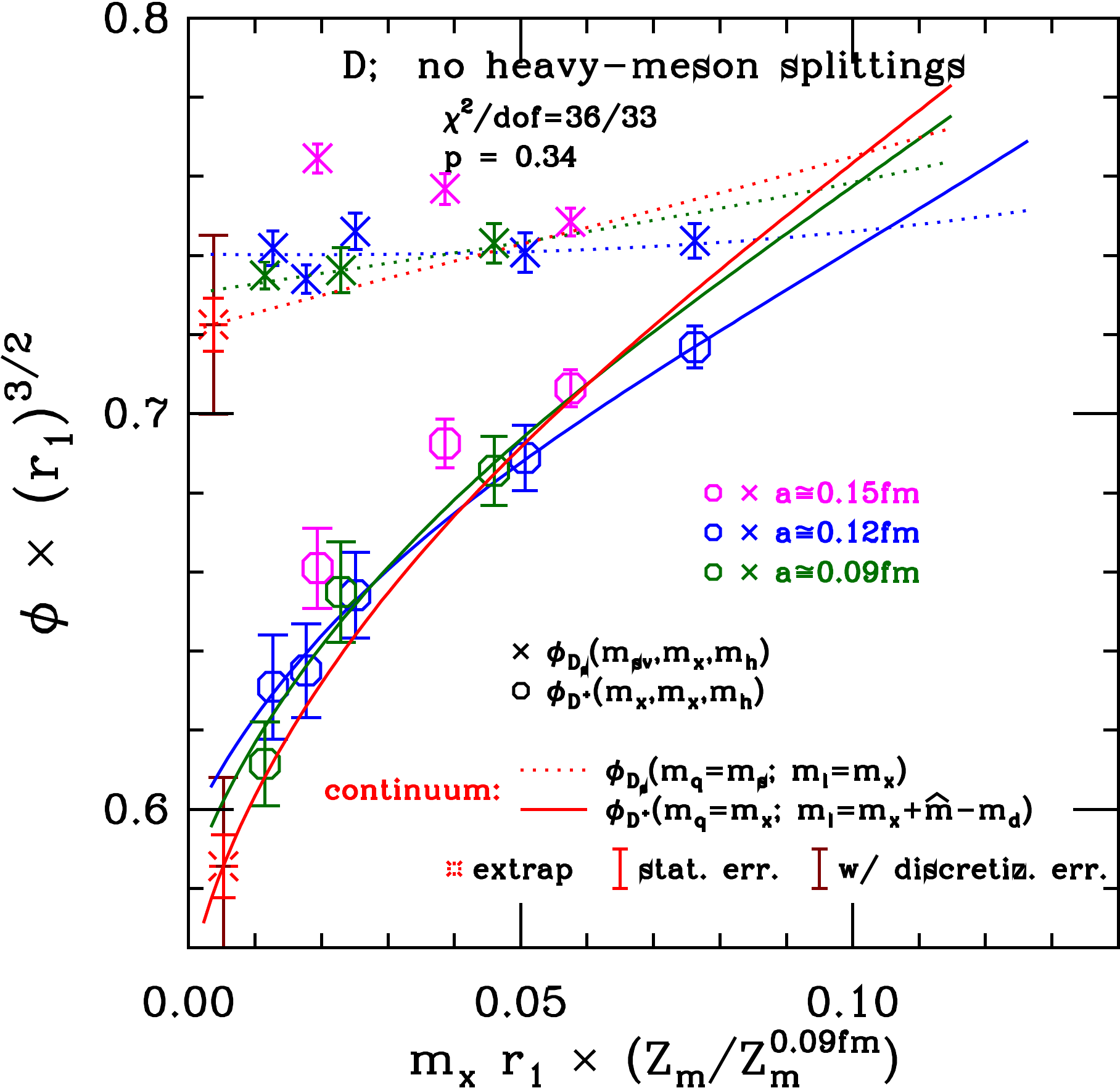}
    \caption{Same as \figref{fDcentral} but omitting heavy-light hyperfine and flavor splittings in the chiral fit function.}
    \label{fig:fDnosplit}
\end{figure}

As described in \secref{ChPT}, we modify the chiral fit function in a variety of ways to estimate
the error associated with the chiral extrapolation:
\begin{enumerate} 
    \renewcommand{\theenumi}{$\chi$\arabic{enumi}}
	\item Set the chiral coupling $f$ to $f_K$ instead of $f_\pi$. \label{item:f}

	\item Allow the chiral coupling $f$ to be a Bayesian fit parameter, 
        with prior value $f_\pi$ and prior width equal to $f_K-f_\pi$. \label{item:fBayes}

	\item Replace the $H$-$H^*$-$\pi$ coupling $g_\pi$ (which is 0.51 in the central fit) with 0.31 or 0.71, which are the extremes of the range  discussed in \secref{ChPT}. \label{item:gpi}

	\item Allow $g_\pi$ to be a constrained fit parameter, with prior value 0.51 and prior width 0.20.

	\item Fix to zero those NNLO analytic terms [$Q_2$ and/or $Q_4$ in \eq{Qs}] that may be eliminated without making the fit unacceptably poor.

	\item Use the chiral function without hyperfine and flavor splittings, \ie, use \eq{phiq}
        instead of \eq{phiq-Delta}. \label{item:nosplit}

	\item Use combinations of modifications~\ref{item:f} and~\ref{item:gpi} or modifications~\ref{item:fBayes} and~\ref{item:gpi}.
        These choices can produce significantly larger deviations since
        changes in $g_\pi$ have a similar effect on the fit function as changes in $f$.
\end{enumerate}
These modifications typically change the decay constant by 1--3 MeV, and the ratios by 1--1.5\%.  
We take the chiral extrapolation error of a given quantity to be the largest change (of either
sign) under the above modifications, and list it in \tabref{errorBudget}.  In several cases, ($f_{D^+}$, $f_{D_s}/f_{D^+}$, and
 $f_{B_s}/f_{B^+}$) the largest change comes from modification~\ref{item:nosplit}, eliminating 
the heavy-light splittings. 
The fit without the splittings is shown for the $D$ system in \figref{fDnosplit}.
It may be compared to \figref{fDcentral} to see the effects: the curvature at small mass for $\phi_{D^+}$ is slightly
greater without the splittings, which results in a decrease of $f_{D^+}$ of 
3.2 MeV. Note that the $p$ values of the two fits are almost identical, so the goodness-of-fit cannot
be used to choose one version of the chiral extrapolation over the other.

\begin{figure}[b]
\centering
    \includegraphics[width=11.0cm]{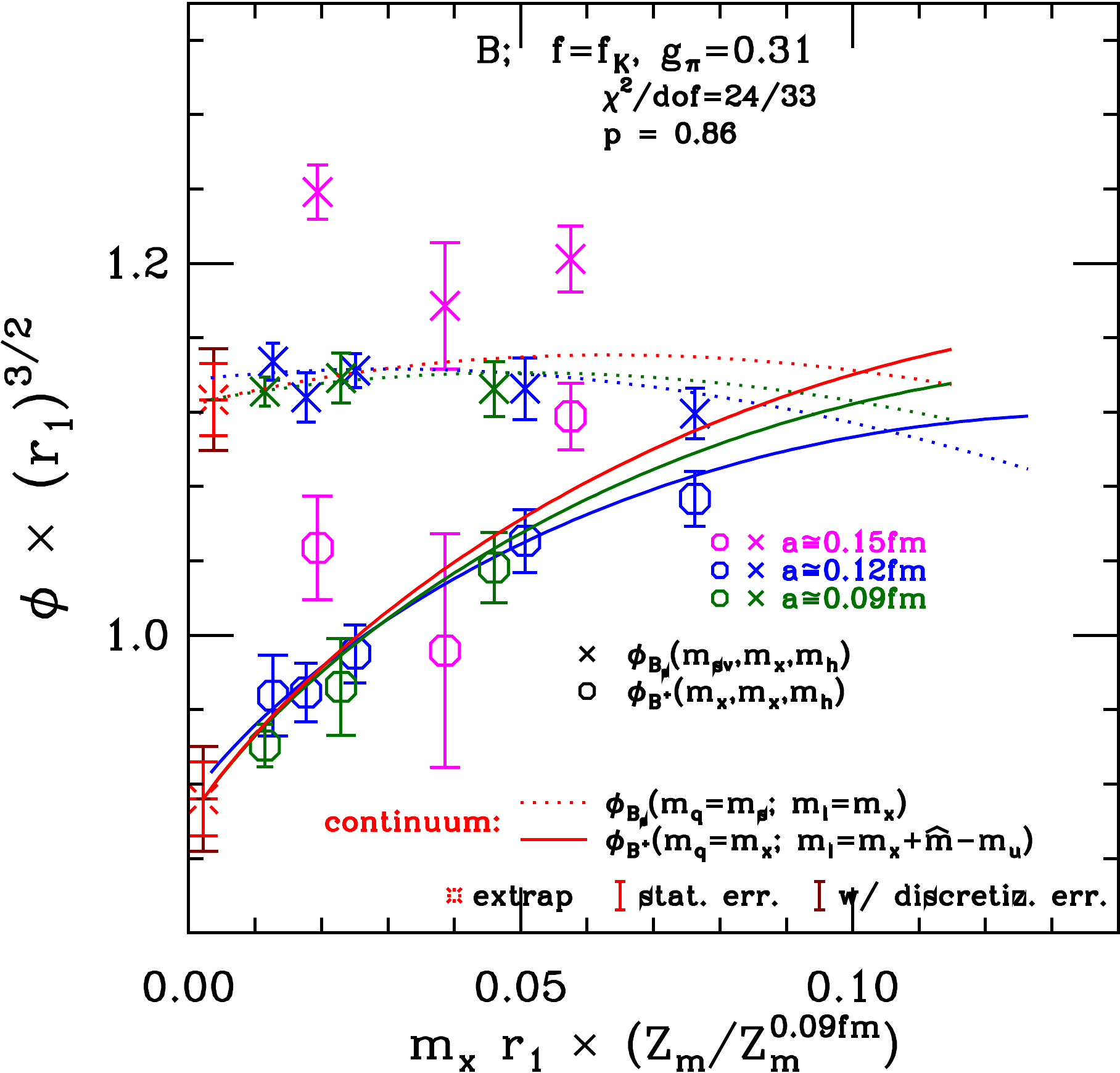}
    \caption{Same as \figref{fBcentral} but with $f=f_K$ and $g_\pi=0.31$ in the chiral fit function.}
    \label{fig:fBfKgsq0961}
\end{figure}

Modifications of $f$ and/or $g_\pi$ produce the largest changes in the other quantities, namely
$f_{D_s}$, $f_{B^+}$ and $f_{B_s}$.  In particular, putting $f=f_K$ and $g_\pi=0.31$ 
results in an increase of +2.9 for $f_{B^+}$ and +2.8 MeV for $f_{B_s}$.
The modified fit is shown in \figref{fBfKgsq0961},
and may be compared with \figref{fBcentral} to see the effects of the changes.
Increasing $f$ and decreasing $g_\pi$ both suppress the chiral logarithms [\cf\ \eq{phiq-Delta}]
and give fit functions with less curvature and smaller slope at low quark mass.

Since the  \rschpt\ fit functions in Eqs.~(\ref{eq:phiq}) and~(\ref{eq:phiq-Delta})
explicitly include one-loop discretization effects coming
from taste violations in the (rooted) staggered light quark action, the chiral error estimates we
describe here inherently include taste-violating discretization errors. However, it seems 
unlikely that the current data
can accurately distinguish between such taste-violating errors of order
$\alpha_s^2a^2$ and generic light-quark discretization effects of order $\alpha_sa^2$,
or even heavy-quark discretization effects.
Indeed, the taste-violating LEC $L_a$ [\cf\ \eq{pq}] is not well constrained by our fits and is
consistent with zero within large errors. The central fits give
\begin{eqnarray}
    L_a&=&+0.6\pm6.5\qquad (D {\rm \ system}),\\
    L_a&=&-1.9\pm8.8\qquad (B {\rm \ system}),
\eqn{La}
\end{eqnarray}
where the error is the raw statistical error. (Note that we do not constrain $L_a$ by any prior width.)
The errors in $L_a$ decrease by about 10\% if Bayesian parameters for
generic light-quark errors are removed, and an additional 10\% if the parameters for heavy-quark
errors are removed.  Thus, there is ``cross talk'' between various error sources, making it
difficult to completely distinguish the various types of discretization
errors.  Future work, with more and finer lattice spacings, should make a cleaner separation possible.

\subsection{Fitting errors}
\label{subsec:fitErrors}

The ``fitting errors'' are the errors introduced in the analysis of the two-point correlators.
They represent the effects of various choices of fit ranges and fitting functions, and are
an estimate of the systematic effect of the contamination by excited states.  
We compare results from the three choices of two-point
fitting (see \secref{2point}): Analysis~I, Analysis~II, and a modified Analysis~I using
1 simple + 1 oscillating state, but values of $t_\textit{min}$ larger 
than those described in \secref{jackknife}. 

\begin{figure}[b]
\centering
    \includegraphics[width=11.0cm]{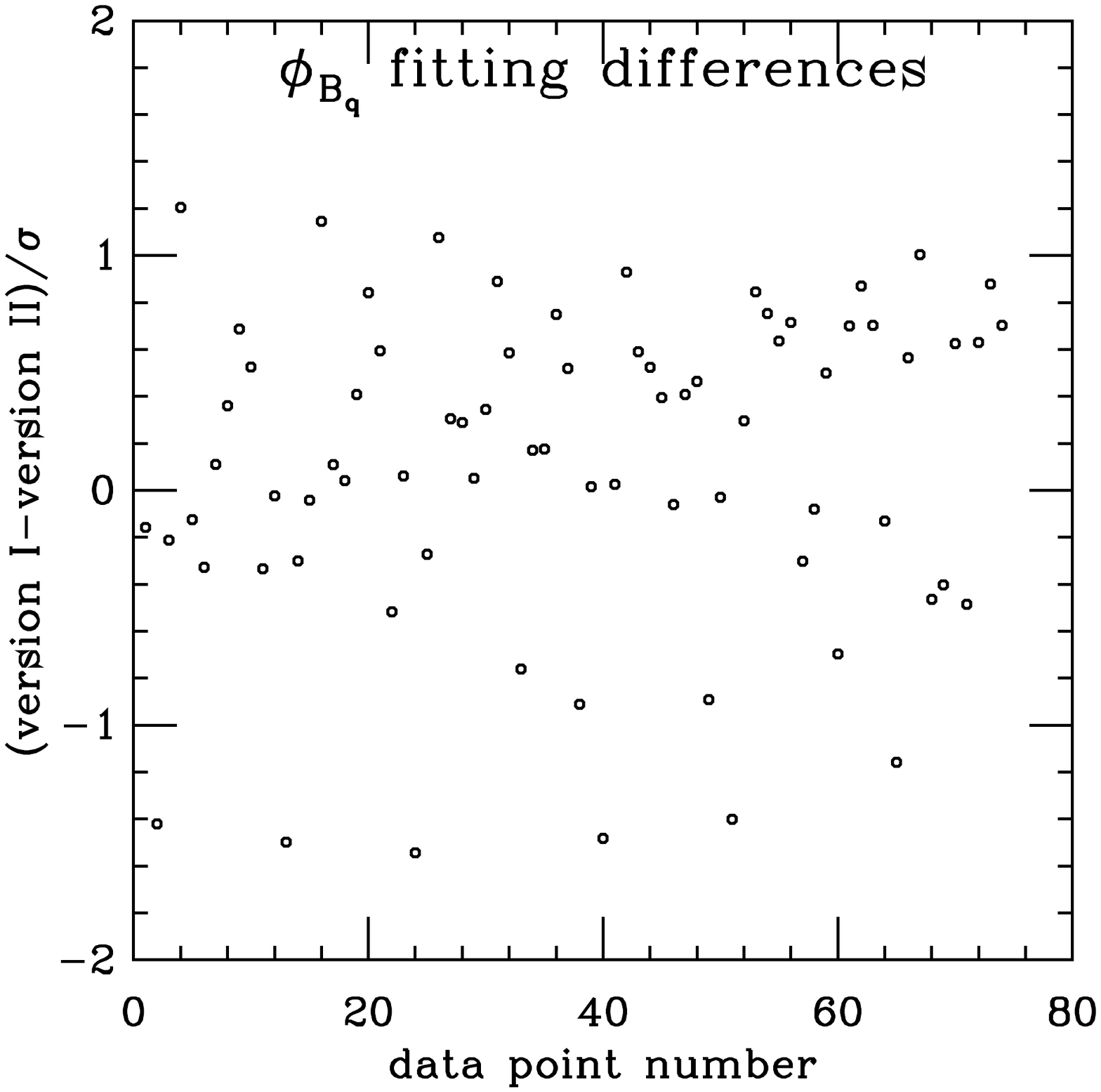}
    \caption{Difference of $\phi_{B_q}$ values from Analyses~I and~II, divided by the average statistical error
        at each of the common valence and sea mass points.
        The order along the abscissa is arbitrary.}
    \label{fig:Bscatter}
\end{figure}

Some of these differences may, in fact, be due simply to statistical effects, and hence already included in the statistical error.
\Figref{Bscatter} shows the differences between values of $\phi_{B_q}$ in Analyses~I and~II, divided by the average statistical error 
for each of the common partially quenched data points.
Only 10 of 74 differences
are greater than 1 statistical $\sigma$.  Nevertheless, there appears to be some significant
systematic trend in that 46 of 74 points are positive. 
To be conservative, 
we take the largest difference between the Analysis-I fits and the other two fits as the fitting error for each physical quantity, and list it in \tabref{errorBudget}.
For $f_{D_s}$ and $f_{B_s}$, the difference is largest for chiral fits based on 2-point Analysis~II, while, for the other four quantities, the difference is largest for
the modified Analysis~I.

\subsection{Scale uncertainty}

We use the scale $r_1=0.3117(22)$~fm to tune the values of the quark masses and convert the decay constants 
into physical units (see \secref{Params}).
To find the scale errors on the final results, we shift $r_1$ to 0.3139~fm or 0.3095~fm
and redo the analysis.  Although $\phi_H$ scales like $r_1^{-3/2}$, the change in the
results under a change in $r_1$ is smaller than pure dimensional analysis would imply, because our estimates
of the physical light masses and the heavy-quark \kapch\ and \kapbot\ also shift, producing partially 
compensating changes in~$\phi_H$.
At $r_1=0.3139$~fm, we shift the light masses in \tabref{params} upward by the scale error
shown in that table.  [The lattice light-quark masses scale like $r_1^2$, because they are
approximately linear in the squared meson masses $(r_1m_\pi)^2$ and $(r_1m_K)^2$.]
Similarly, we shift the tuned \kapch\ and \kapbot\ downward by the scale error in
\tabref{kappaFromMesonDispersion} 
because the bare heavy quark mass increases with $r_1$. We then adjust $\phi_{B_{(s)}}$ and
$\phi_{D_{(s)}}$ at each lattice spacing using \eq{phi-tune-adj} and the values of
 $d\phi/d\kappa$  given in \tabref{kappaFromMesonDispersion}.  Redoing the preferred chiral fits shown
in \figrefs{fDcentral}{fBcentral}, extrapolating to the continuum, and plugging in the
adjusted continuum light quark masses gives the scale error listed in \tabref{errorBudget}.

\subsection{Light-quark mass determinations}

To estimate the error from the light-quark mass determination, we follow a similar procedure to
that in the scale-error case.  We shift the continuum light quark masses in \tabref{params} by the
sum in quadrature of all errors except scale errors. This includes the statistical errors,
the chiral errors and, where relevant, the electromagnetic errors. We then plug the new masses
into the continuum-extrapolated 
chiral fits and take the difference from the central results to give the errors
listed in \tabref{errorBudget}.  The relative direction
of shifts on different masses makes little difference in the size of the errors on the decay constants
$f_{D_s}$, $f_{D^+}$, $f_{B_s}$, and $f_{B^+}$, since they
are sensitive primarily to the valence quark masses. However, it does affect the
error of the ratios $f_{D_s}/f_{D^+}$ and $f_{B_s}/f_{B^+}$. The largest effect clearly occurs when
the strange mass is shifted in the opposite direction from the lighter masses. To be conservative,
we take the size of change of the ratios in this case as the error, but this is almost certainly an 
overestimate because the statistical and chiral extrapolation errors 
on the light quark masses are positively correlated between the strange mass and the other masses.  

Note that the errors from the light-quark masses in \tabref{errorBudget} are much larger
for $f_{D_s}$ and $f_{B_s}$ than for $f_{D^+}$ and $f_{B^+}$.
That simply reflects the facts that
the decay constants have a nonzero limit when the quark masses vanish, and 
that the dependence on the quark masses is reasonably linear.
Thus a given percent error in the strange mass produces a much larger percent difference in 
$f_{D_s}$ and $f_{B_s}$, than the same percent error in the $d$ or $u$ mass does
in $f_{D^+}$ and $f_{B^+}$.

\subsection{Bottom and charm quark mass determinations}

The propagation of statistical errors in the tuned \kapch\ and \kapbot\ to the decay constants 
is complicated by the fact that the independent errors at  each lattice spacing affect the final results in a
nontrivial way through the continuum and chiral extrapolations. At each lattice spacing,
we choose 200 gaussian-distributed ensembles of trial $\kappa$
values with central value equal to the tuned values, and standard deviation equal to the
statistical error, taken from 
\tabref{kappaFromMesonDispersion}.
For a given choice of trial $\kappa$ values at each lattice spacing, we produce an adjusted
trial data sample by shifting the $\phi_H$ values according to \eq{phi-tune-adj}, but with the
trial values replacing the tuned values.
We then perform the complete chiral fit and extrapolation procedure on each of the
200 trial data sets. The
standard deviation over trials of a given decay constant or decay constant ratio
is taken to be the heavy quark tuning error, and is listed in \tabref{errorBudget}.

\subsection{Tadpole factor \boldmath ($u_0$) adjustment}

In order to improve the convergence of lattice perturbation theory, we use tadpole-improved actions for the 
gluons, light quarks, and heavy quarks~\cite{Lepage:1992xa}.  For the gluon and sea-quark actions we take 
the tadpole factor $u_0$ from the average plaquette.  On the $a \approx 0.15$~fm and $a \approx 0.09$~fm 
lattices we use the same choice for the light valence and heavy-quark actions.  On the $a \approx 0.12$~fm 
lattices, however, we use the tadpole factor $u_0$ taken from the Landau link in the valence-quark action 
and in the clover term in the heavy-quark action.  This results in a slight mismatch between the light 
valence and sea-quark actions on these ensembles, and also affects the values obtained for the tuned bottom- 
and charm-quark masses $\kappa_b$ and $\kappa_c$.  The difference between $u_0$ obtained from the average 
plaquette and the Landau link is approximately 3--4\% on the $a \approx 0.12$ fm ensembles.

We propagate this difference through the chiral/continuum extrapolation as follows.
First, we compute the heavy-strange meson decay amplitudes $\phi_{B_s}$ and $\phi_{D_s}$ with both choices 
for $u_0$ on 
the ensemble with $am_l/am_h=0.01/0.05$, $a\approx 0.12$~fm.
For each choice of $u_0$, we compute $\phi_{B_s}$ and $\phi_{D_s}$ directly at the tuned values 
of $\kappa_b$ and $\kappa_c$, thereby avoiding an interpolation in $\kappa$.
Next, we renormalize the lattice decay amplitudes using the nonperturbative, flavor-diagonal current 
renormalization factors \ZVqq\ and \ZVQQ\ obtained for each case.
(We neglect the slight difference in the perturbative correction \rhoAQq.)
Then, we calculate the ratio of the renormalized decay amplitudes, finding no difference within errors: 
\begin{eqnarray}
	 \phi_{c}^{\rm plaquette} / \phi_{c}^{\rm Landau} &=& 1.005(13), \label{eq:phic_u0} \\
	 \phi_{b}^{\rm plaquette} / \phi_{b}^{\rm Landau} &=& 1.014(20). \label{eq:phib_u0}
\end{eqnarray}
As expected, the $u_0$ dependence from the bare current and renormalization factors mostly cancels.  Finally, we repeat the chiral/continuum extrapolation shifting $\phi_c$ and $\phi_b$ on the $a\approx0.12$~fm ensembles by the statistical errors reported in Eqs.~(\ref{eq:phic_u0})--(\ref{eq:phib_u0}). We find that these percent-level errors in $\phi_c$ and $\phi_b$ lead to approximately 1\% errors in the extrapolated decay constants and approximately 0.1\% errors in the decay-constant ratios. These errors are listed as ``$u_0$ adjustment'' in the error budget in \tabref{errorBudget}.     

\subsection{Heavy-light current renormalization}

There are two sources of systematic error
in our heavy-light current renormalization.  The first is  due to the perturbative 
calculation of  \rhoAQq\  and the second is due to the nonperturbative calculation 
of $\ZvQQ$ and $\Zvqq$. 

The perturbative calculation of 
\rhoAQq\ has been carried out to one-loop
order.  Since \rhoAQq\ is defined from a ratio of
renormalization factors [see Eq.~(\ref{eq:rho})], 
its perturbative corrections are small by construction. Indeed, 
as can be seen from the results for
\rhoAQq\ given in Table~\ref{tab:rho}, we observe very small corrections. 
For bottom they range from $0.3$\% at $a \approx 0.09$ fm
to $0.8$\% at $a \approx 0.12$ fm and
$2.8$\% at $a \approx 0.15$ fm. 
For charm they range from less than $0.08$\%
at $a \approx 0.09$ fm to $0.4$\% at $a \approx 0.12$ fm and 
$0.6$\% at $a \approx 0.15$ fm.
As shown in Ref.~\cite{ElKhadra:2007qe} the perturbative corrections to the 
$\rho$-factors for the spatial currents, while still small, tend to be bigger than those for the temporal currents $A^4$ and $V^4$.  
We therefore estimate the error due to neglecting higher order terms as $\rhoVQq^{[1]} \, \alpha_s^2$. We take 
$\alpha_s$ at $a \approx 0.09$ fm and $\rhoVQq^{[1]} \approx 0.1$, which is the largest one-loop 
coefficient  for $\rhoVQq$ in the mass range $m_Qa \leq 3$.  This procedure yields a systematic error 
of $0.7$\%, which we take for both charm and bottom decay constants.

The decay constant ratios $f_{B_s}/f_{B^+}$ and $f_{D_s}/f_{D^+}$ depend on the corresponding ratios of $\ensuremath{\rho_{A^4_{Qs}}}/\ensuremath{\rho_{A^4_{Qq}}}$. 
These ratios differ from unity only because of the small variation of the $\rhoAQq$ with light valence mass, which is described in \secref{HLCurrents}. 
We take the variation of the $\rhoAQq$ with light valence mass at $a \approx 0.09$ fm as the error.
This yields an error of $0.1\%$ for both bottom and charm.

The dominant corrections in the heavy-light renormalization factor as defined in  Eq.~(\ref{eq:rho}) are due to $\ZvQQ$ and $\Zvqq$ which are calculated nonperturbatively.
The values (and errors) for $\Zvqq$ and $\ZvQQ$ are listed in \tabrefs{ZvqqResults}{ZvQQruns}, respectively.
To obtain the error in $\ZVQq=\sqrt{\ZVqq\ZVQQ}$ we add the statistical errors in $\ZVqq$ and $\ZVQQ$ in quadrature. 
The error on $\ZVQq$ is dominated by the error on $\ZVqq$.
The errors are largest, 1.3\%, on the $a\approx0.09~\fm$ ensemble and they are about the same for both charm and bottom on the two finest ensembles used to obtain our main decay constant results. 
Hence we use 1.3\% as our estimate for the uncertainty in $\ZVQq$.

\subsection{Finite volume effects}

To study finite volume effects, we use the chiral fit function with heavy-light hyperfine
and flavor splittings included (\eq{phiq-Delta}), since the effects are known to be
larger with the splittings than without \cite{Arndt:2004bg}. The central fit includes
the (one-loop) finite volume corrections, \eq{J-FV}, on the lattice data, and then takes
the infinite volume limit when extracting the final results for the decay constants.  
We then take the larger of the following two values as our
estimate of the finite volume error: 
\begin{enumerate} 
    \renewcommand{\theenumi}{V\arabic{enumi}}
	\item The difference between the central result and the result from a chiral fit
        in which the finite volume corrections are omitted.
    \label{item:noFV}

	\item The largest finite volume correction to the relevant data points, as 
        determined by the central fit.  For $\phi_{D^+}$ and $\phi_{B^+}$, the ``relevant data points''
        are the ones on each ensemble with the lightest valence mass, \ie, those closest to the
        chirally extrapolated point.  For $\phi_{D_s}$ and $\phi_{B_s}$, the relevant points are the
        ones on each ensemble with valence mass closest to $m_s$.
    \label{item:YFV}
\end{enumerate}

Method~\ref{item:noFV} gives a larger difference for $\phi_{D_s}$ and $\phi_{B_s}$; 
method~\ref{item:YFV}  for $\phi_{D^+}$ and $\phi_{B^+}$ and the ratios.  The resulting values 
are shown in \tabref{errorBudget}.  Note that our choices are conservative because 
we correct for the (one-loop) finite volume errors, but nevertheless take the
full size of these effects as our error.

\section{Results and Conclusions}
\label{sec:Results}

After adding the error estimates described in the previous section in quadrature, we obtain: 
\begin{eqnarray}
	f_{B^+} & = & 196.9(8.9)~\textrm{MeV},  \eqn{results:fB+} \\
	f_{B_s} & = & 242.0(9.5)~\textrm{MeV},  \eqn{results:fBs} \\
	f_{B_s}/f_{B^+} & = & 1.229(0.026),     \eqn{results:xiB} \\
	f_{D^+} & = & 218.9(11.3)~\textrm{MeV}, \eqn{results:fD+} \\
	f_{D_s} & = & 260.1(10.8)~\textrm{MeV}, \eqn{results:fDs} \\
	f_{D_s}/f_{D^+} & = & 1.188(0.025).     \eqn{results:xiD}
\end{eqnarray}
Since our most reliable method of determining discretization errors combines them with
statistical errors, we do not quote separate statistical and systematic
errors.

\begin{figure}[b]
\centering
    \includegraphics[width=0.80\textwidth]{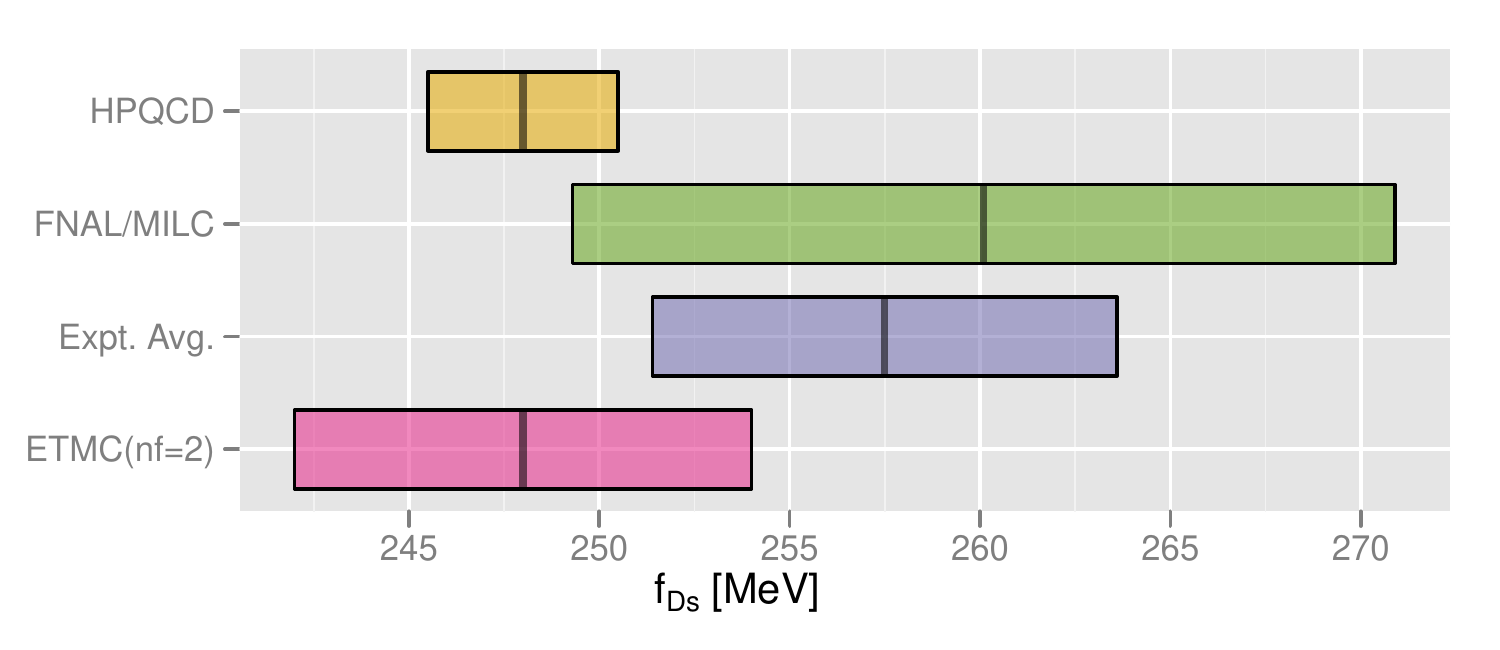}
    \includegraphics[width=0.80\textwidth]{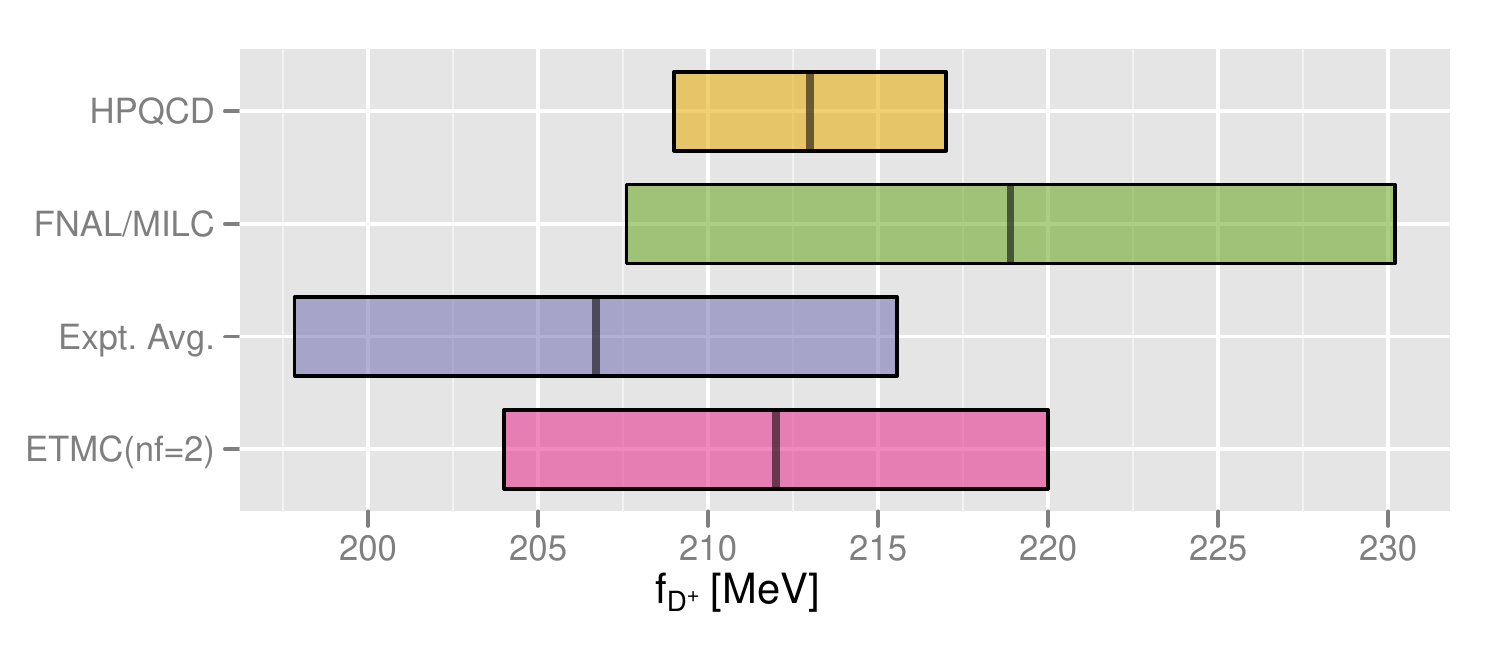}
    \includegraphics[width=0.80\textwidth]{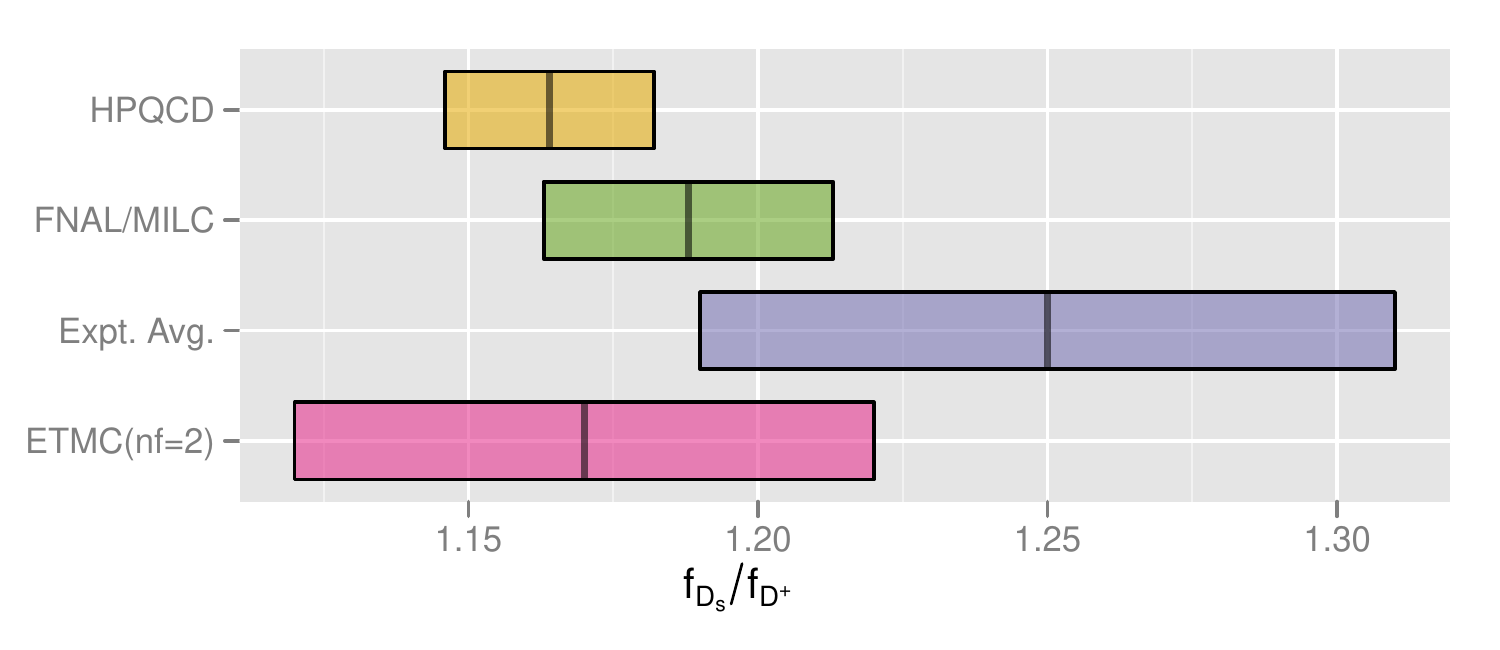}
    \caption{Comparison of $f_D$ and $f_{D_s}$ with other two- and three-flavor lattice QCD calculations and with experiment.
        Results shown come from Refs.~\cite{Davies:2010ip,:2011gx,:2008sq,Rosner:2010ak,Alexander:2009ux,:2007ws,Naik:2009tk,Onyisi:2009th,Lees:2010qj}. The HPQCD $f_D$ value is computed from their update
to $f_{D_s}$ and their earlier result for the ratio $f_{D_s}/f_D$.}
    \label{fig:fDdeterimations}
\end{figure}

Figure~\ref{fig:fDdeterimations} shows a comparison of our results for charmed decay constants with other 
lattice QCD calculations and with experiment.
Our results agree with the only other three-flavor lattice QCD determination from the HPQCD collaboration~\cite{Davies:2010ip}, which is obtained with HISQ staggered valence quarks and asqtad staggered sea quarks.
(The difference in $f_{D_s}$ is a bit greater than 1$\sigma$.)
They are also consistent with the two-flavor results of the ETM Collaboration using twisted-mass Wilson fermions~\cite{:2011gx}, although the ETM error budget does not include an estimate of the uncertainty due 
to quenching the strange quark.
One can also compare with ``experimental" determinations of $f_D$ and $f_{D_s}$ if one assumes CKM unitarity to obtain the matrix elements $|V_{cd}|$ and $|V_{cs}|$.
For the $D$ meson, Rosner and Stone combine CLEO's measurement of branching fraction ${\mathcal{B}}(D^+\to \mu^+ \nu)$~\cite{:2008sq} with the latest determination of $|V_{cd}|$ from the 
PDG~\cite{Nakamura:2010zzi} to obtain $f_D = 206.7(8.9)~\textrm{MeV}$~\cite{Rosner:2010ak}.
For the $D_s$ meson, they average CLEO and Belle results for ${\mathcal{B}}(D_s^+\to \mu^+ \nu)$~\cite{Alexander:2009ux,:2007ws} with CLEO and BABAR results for 
${\mathcal{B}}(D_s^+\to \tau^+ \nu)$~\cite{Alexander:2009ux,Naik:2009tk,Onyisi:2009th,Lees:2010qj} to obtain a combined average for the two decay channels of $f_{D_s} = 257.5(6.1)~\textrm{MeV}$~\cite{Rosner:2010ak}.
The Heavy Flavor Averaging Group obtains a similar average, $f_{D_s}=257.3(5.3)~\textrm{MeV}$~\cite{Asner:2010qj}.
Our results are consistent with these values, confirming Standard Model expectations at the $\sim 5\%$ level.

\begin{figure}[b]
\centering
    \includegraphics[width=0.80\textwidth]{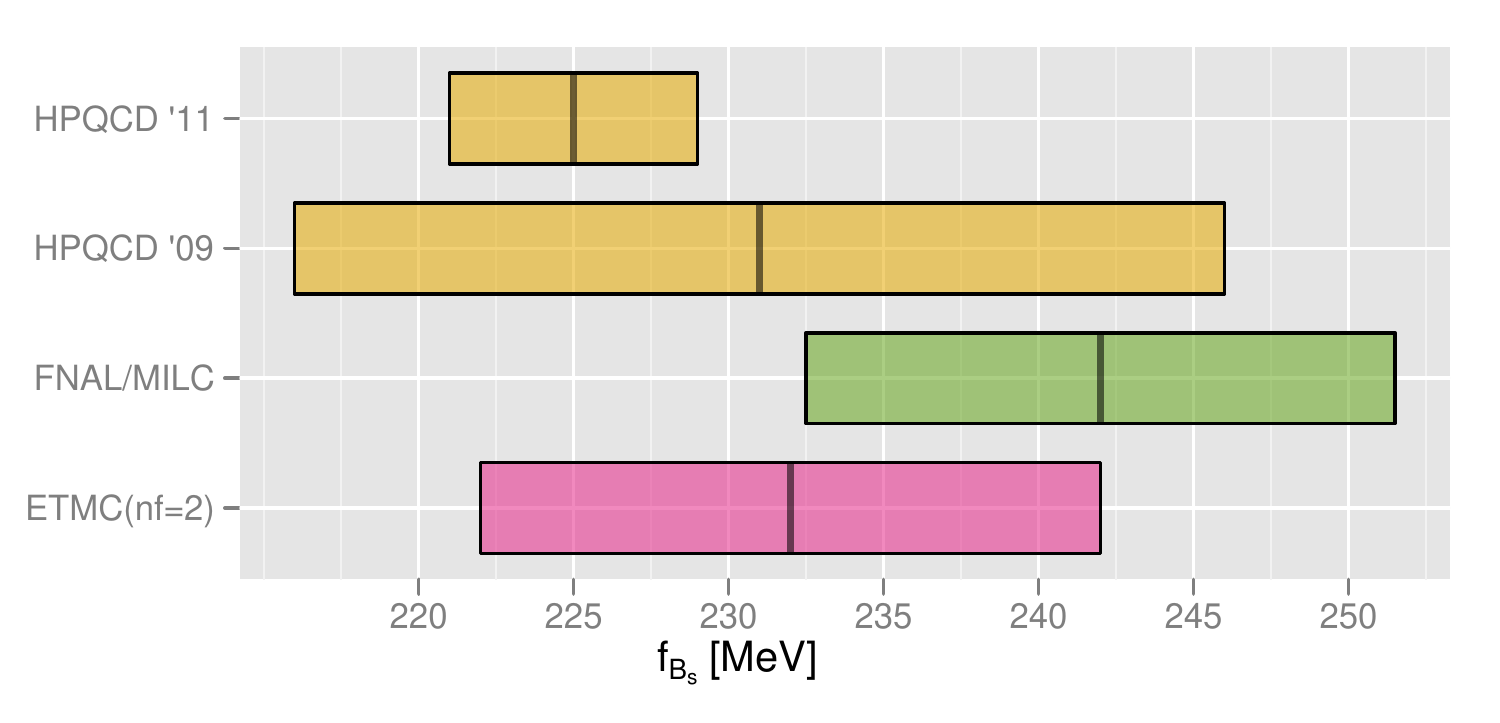}
    \includegraphics[width=0.80\textwidth]{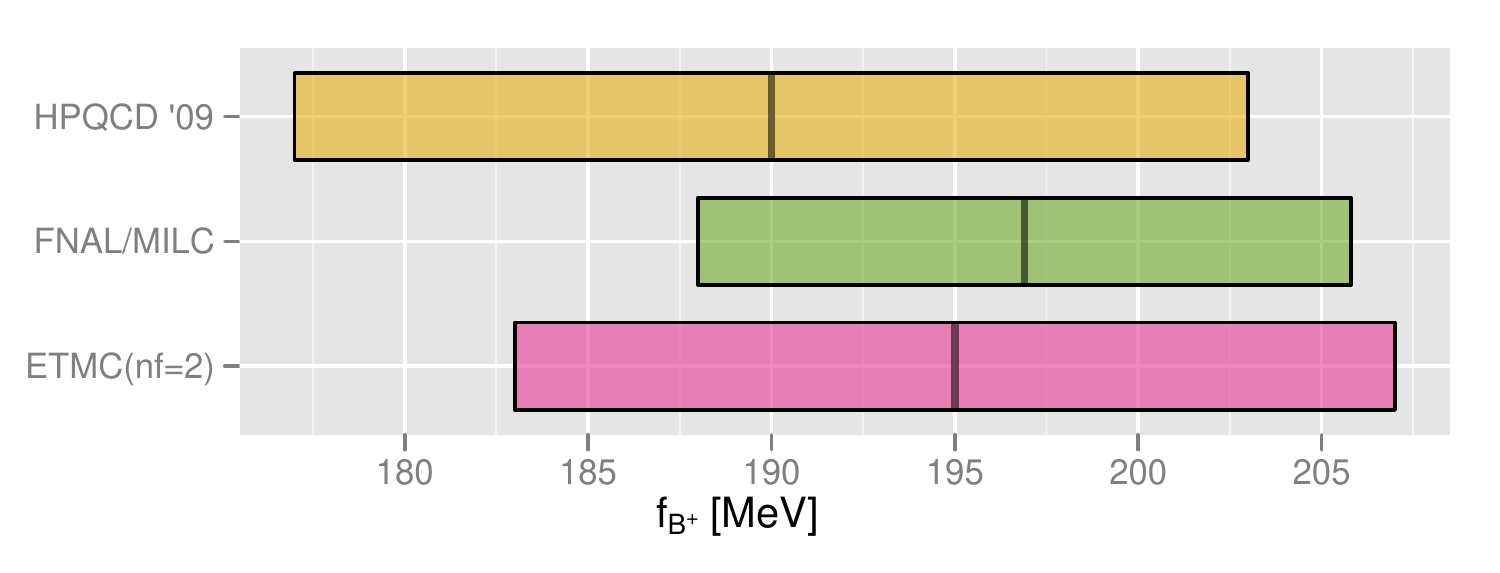}
    \includegraphics[width=0.80\textwidth]{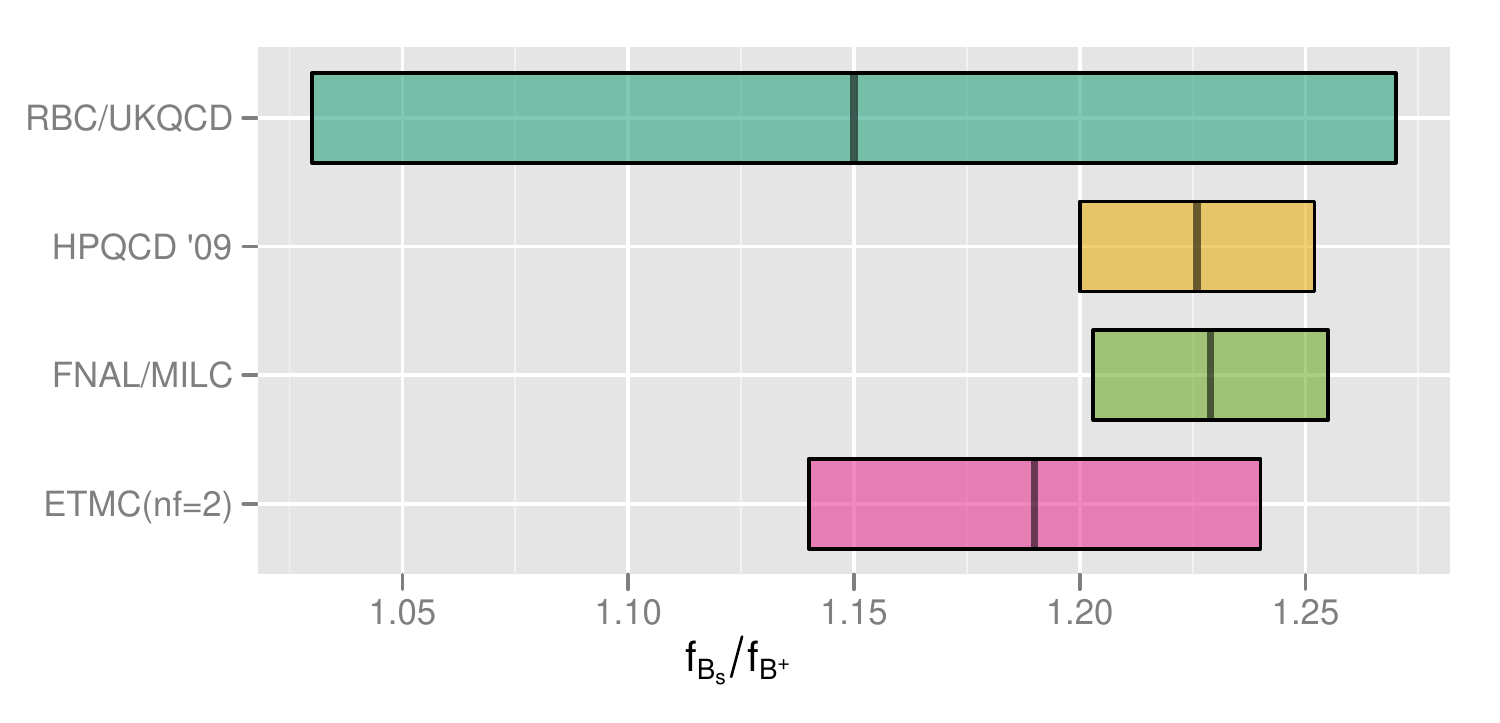}
    \caption{Comparison of $f_B$ and $f_{B_s}$ with other two- and three-flavor lattice QCD calculations.
        Results shown come from Refs.~\cite{McNeile:2011ng,Gamiz:2009ku,:2011gx,Albertus:2010nm}.  In the case of $f_{B_s}$ HPQCD has two separate calculations using NRQCD $b$ quarks and using HISQ $b$ quarks;  we show both the published NRQCD result (HPQCD '09) and the more recent HISQ result (HPQCD '11) in the plot above.}
    \label{fig:fBdeterimations}
\end{figure}

Figure~\ref{fig:fBdeterimations} shows a similar comparison of our results for bottom meson decay constants with other lattice QCD calculations.
Our results agree with the published three-flavor determination using NRQCD $b$-quarks and Asqtad staggered light quarks of the HPQCD collaboration~\cite{Gamiz:2009ku}, but are only marginally consistent with HPQCD's more recent calculation of $f_{B_s}$ using HISQ light valence quarks~\cite{McNeile:2011ng}.
Our results are also consistent with the two-flavor results of the ETM collaboration~\cite{:2011gx}, who use Wilson heavy quarks and interpolate between the charm-mass region and the static limit to 
obtain results for bottom.
Further, our result for the ratio $f_{B_s}/f_B$ also agrees with the significantly less precise three-flavor determination using static $b$-quarks and domain-wall light quarks by the RBC and UKQCD 
Collaborations~\cite{Albertus:2010nm}.

For the $D$ system the largest uncertainties in our current calculation stem from heavy-quark discretization, 
while the chiral extrapolation, the $Z_V$ factors, excited states, heavy-quark tuning, 
and the chiral-continuum extrapolation play important but subdominant roles.
For the $B$ system, heavy-quark tuning, statistics, and excited states are the sources of the largest errors, 
while the $Z_V$ factors and the chiral-continuum extrapolation (incorporating our estimate of heavy-quark 
discretization effects) are next in size.
Recall that a novel feature of our work is the treatment of heavy-quark discretization effects, via the 
functions $f_i$ in Eq.~(\ref{eq:methods:ZA=A}), and priors constraining the chiral-continuum fits to follow 
this form.
At tree level, we have explicit calculations of the mismatch, some of which appeared already in 
Ref.~\cite{ElKhadra:1996mp} and all of which are compiled in Ref.~\cite{Oktay:2008ex}.
Beyond the tree level, the continuum and static limits can be used to constrain the functional form.
That said, the theoretical guidance of the priors cannot be highly effective in an analysis, such as this, 
with only two lattice spacings. 
Indeed, the quoted heavy-quark discretization errors are similar to less sophisticated power-counting 
estimates. 

While completing this analysis, we have begun runs to generate data that will address the main sources of 
uncertainty reported here. 
The new data set will contain four times the configurations used here to reduce the statistical errors in 
the correlation functions and, thus, directly improve the decay amplitudes, the determinations of the  
hopping-parameters $\kapch$ and $\kapbot$, and the renormalization factors \ZVqq\ and \ZVQQ, all of which 
feed into the decay constant. 
Our new data will also encompass two finer lattice spacings of $a \approx 0.06$~fm and $a \approx 0.045$~fm, 
in order to explicitly reduce light- and heavy-quark discretization errors and better  
control the continuum extrapolation.
With four lattice spacings, our new method of heavy-quark discretization priors will be put to a more 
stringent test. 
The new runs will also include light valence- and sea-quark masses down to $\sim m_s/20$ in order to better 
control the chiral extrapolation to the physical $d$ and $u$ quark masses.    

In order to reduce errors further, we will have to eliminate the errors from the matching factors and from 
quenching the charmed quark.
The MILC Collaboration~\cite{Bazavov:2010pi} is generating ensembles with 2+1+1 flavors of sea quarks with 
the HISQ action, with plans to provide a range of lattice spacings and sea quark masses equal to or more 
extensive than the 2+1 asqtad ensembles.
Use of the HISQ action for the charm valence quark will allow us to further reduce many of the 
uncertainties, and provides the particularly nice advantage that one can use the local pseudoscalar  
density without multiplicative renormalization to obtain the continuum matrix element~\cite{Follana:2007uv}.
In several years, once the full suite of HISQ ensembles with several sea-quark masses and lattice spacings 
has been analyzed, we expect to obtain percent-level errors for both $B$- and $D$-meson decay constants. 
This will enable precise tests of the Standard Model and may help to reveal the presence of new physics in 
the quark-flavor sector. 

\begin{acknowledgments}
We thank David Lin for his finite-volume chiral-log {\it Mathematica} 
code, upon which our own code is based.
Computations for this work were carried out with resources provided by
the USQCD Collaboration, the Argonne Leadership Computing Facility,
the National Energy Research Scientific Computing Center, and the 
Los Alamos National Laboratory, which are funded by the Office of Science of the
U.S. Department of Energy; and with resources provided by the National Institute
for Computational Science, the Pittsburgh Supercomputer Center, the San Diego
Supercomputer Center, and the Texas Advanced Computing Center, which are funded
through the National Science Foundation's Teragrid/XSEDE Program.
This work was supported in part by the U.S. Department of Energy under 
Grants No.~DE-FC02-06ER41446 (C.D., L.L., M.B.O.),
No.~DE-FG02-91ER40661 (S.G.), 
No.~DE-FG02-91ER40677 (C.M.B, R.T.E., E.D.F., E.G., R.J., A.X.K.),
No.~DE-FG02-91ER40628 (C.B),
No.~DE-FG02-04ER-41298 (D.T.); 
by the National Science Foundation under Grants 
No.~PHY-0555243, No.~PHY-0757333, No.~PHY-0703296 (C.D., L.L., M.B.O.), 
No.~PHY-0757035 (R.S.), 
No.~PHY-0704171 (J.E.H.);
by the URA Visiting Scholars' program (C.M.B., R.T.E., E.G., M.B.O.);
by the Fermilab Fellowship in Theoretical Physics (C.M.B.);
by the M. Hildred Blewett Fellowship of the 
American Physical Society (E.D.F.);
and by the Science and Technology Facilities Council and the Scottish Universities Physics Alliance (J.L.).
This manuscript has been co-authored by employees of Brookhaven Science
Associates, LLC, under Contract No. DE-AC02-98CH10886 with the 
U.S. Department of Energy.
R.S.V. acknowledges support from BNL via the Goldhaber Distinguished Fellowship.
Fermilab is operated by Fermi Research Alliance, LLC, under Contract
No.~DE-AC02-07CH11359 with the United States Department of Energy.
\end{acknowledgments}

\appendix

\section{Heavy-quark Discretization Effects}
\label{app:HQcutoff}

We are using the heavy-quark Lagrangian as given
in~\cite{ElKhadra:1996mp}, with $\kappa_t=\kappa_s$ (or, equivalently,
$\zeta=1$), $r_s=1$, and $c_B=c_E=c_{\rm SW}$.
This amounts to the Sheikholeslami-Wohlert 
Lagrangian~\cite{Sheikholeslami:1985ij} 
for Wilson fermions~\cite{Wilson:1975id}.
The current has a heavy quark of this type, rotated as in Eq.~\eqref{eq:rotation}
(\cf\ Eqs.~(7.8)--(7.10) of Ref.~\cite{ElKhadra:1996mp}), and a staggered light quark.
At the tree level, the heavy-quark rotation is the same no matter what
the other quark is.
The discretization effects are estimated from a (continuum) effective
field theory~\cite{Kronfeld:2000ck,Harada:2001fi,Harada:2001fj},
as shown explicitly for decay constants in Eqs.~(8.7)--(8.12) of
Ref.~\cite{Kronfeld:2000ck}.

\subsection{Theory}

Both QCD and lattice gauge theory can be described via
\begin{eqnarray}
	{\cal L}_{\rm QCD} \doteq {\cal L}_{\text{HQET}} & = &
		\sum_i {\cal C}_i^{\rm cont}(m_Q) \mathcal{O}_i,
	\label{eq:QCDHQET} \\
	{\cal L}_{\rm LGT} \doteq {\cal L}_{\text{HQET}(m_0a)} & = &
		\sum_i {\cal C}_i^{\rm lat}(m_Q, m_0a) \mathcal{O}_i, 
	\label{eq:LGTHQET}
\end{eqnarray}
where the ${\cal C}_i$ are short-distance coefficients
and the $\mathcal{O}_i$ are operators describing the long-distance physics.
The coefficients have dimension $4-\dim\mathcal{O}_i$.
For lattice gauge theory. they depend on $m_0a$, which is a ratio of 
short distances $a$ and~$1/m_Q$.
The effective-theory operators~$\mathcal{O}_i$ in Eqs.~(\ref{eq:QCDHQET})
and~(\ref{eq:LGTHQET}) are the same.

The error from each term is simply the difference
\begin{equation}
	{\tt error}_i = \left|\left[
		{\cal C}_i^{\rm lat}(m_Q, m_0a) - {\cal C}_i^{\rm cont}(m_Q)
		\right] \mathcal{O}_i \right|.
\end{equation}
The relative error in our matrix elements can be estimated by setting
$\langle\mathcal{O}_i\rangle\sim\Lambda_{\text{QCD}}^{\dim\mathcal{O}_i-4}$;
choices for the QCD scale $\Lambda_{\text{QCD}}$ are discussed below.
The coefficient mismatch can be written
\begin{equation}
	{\cal C}_i^{\rm lat}(m_Q, m_0a) - {\cal C}_i^{\rm cont}(m_Q) =
		a^{\dim\mathcal{O}_i-4} f_i(m_0a).
\end{equation}
This recovers the usual counting of powers of $a$
(familiar from Symanzik~\cite{Symanzik:1983dc,Symanzik:1983gh}), 
but maintaining the full $m_0a$ dependence.
The final expression for the discretization errors is then
\begin{equation}
	{\tt error}_i \propto f_i(m_0a) (a\Lambda_{\text{QCD}})^{\dim\mathcal{O}_i-4}.
	\label{eq:error}
\end{equation}
For Wilson fermions, $\lim_{m_0a\to0}f_i=\text{constant}$
(whereas in lattice NRQCD without fine tuning this is not the case).
We have explicit calculations of the $f_i$ for the $O(a)$ and $O(a^2)$
errors at the tree level~\cite{ElKhadra:1996mp,Oktay:2008ex}.
The next subsection discusses how to use them to guide a 
continuum-limit extrapolation the $O(\alpha_s a)$ and $O(a^2)$ errors.

Equations~(\ref{eq:QCDHQET}) and~(\ref{eq:LGTHQET}) can be generalized
to currents.
For the axial-vector current,
\begin{eqnarray}
	\mathcal{A}^\mu & \doteq & C^{\rm cont}_{A_\perp}(m_Q)\bar{q}i\gamma_\perp^\mu\gamma_5 h_v - 
		C^{\rm cont}_{A_\parallel}(m_Q)v^\mu\bar{q}\gamma_5h_v - 
		\sum_i B^{\rm cont}_{Ai}(m_Q)\mathcal{Q}^\mu_{Ai}, \\
	A_{\rm lat}^\mu & \doteq & C^{\rm lat}_{A_\perp}(m_Q, m_0a)\bar{q}i\gamma_\perp^\mu\gamma_5 h_v -
		C^{\rm lat}_{A_\parallel}(m_Q, m_0a)v^\mu\bar{q}\gamma_5h_v -
		\sum_i B^{\rm lat}_{Ai}(m_Q, m_0a)\mathcal{Q}^\mu_{Ai} ,
\end{eqnarray}
and $\doteq$ again means in the sense of matrix elements.
Here $v^\mu$ selects the temporal component and $\perp$ the spatial, 
and the list of dimension-4 operators $\mathcal{Q}$ can be found in
Refs.~\cite{Harada:2001fi}.

The matrix element of the temporal component of the axial-vector 
current [\cf\ Eq.~\eqref{eq:methods:corr4phi}] is normalized by multiplying with 
$Z_{A^4}=C^{\rm cont}_{A_\parallel}/C^{\rm lat}_{A_\parallel}$.
The current mismatch then leads to errors
\begin{equation}
	a^{\dim\mathcal{Q}_i-3} f_i(m_0a) = Z_{A^4}B^{\rm lat}_{Ai}-B^{\rm cont}_{Ai},
\end{equation}
with the sum running over the two operators $\mathcal{Q}$ that
point in the temporal direction~\cite{Harada:2001fi}.

\subsection{Error Estimation}

The total error from heavy-quark discretization effects is then
\begin{equation}
	{\tt error} = \sum_i z_i\, (a\Lambda_{\rm QCD})^{s_i} f_i(m_0a)
\end{equation}
where the sum runs over
Lagrangian operators $\mathcal{O}_i$ of dimension 5 and~6 and current
operators $\mathcal{Q}_i$ of dimension 4 and~5,
$s_i=\dim\mathcal{O}_i-4$ or $\dim\mathcal{Q}_i-3$, and
the $z_i$ are unknown coefficients.
The functions $f_i$ (summarized below) have been computed for
$\mathrm{O}(a^2)$ and estimated for $\mathrm{O}(\alpha_sa)$.
We omit contributions of order $\alpha_s^la^2$, whether from extra
operators or from iterating to second order operators with
coefficients of order~$\alpha_sa$.

In the past, we have taken a very conservative 
$\Lambda_{\rm QCD}=700~{\rm MeV}$ and assumed a Gaussian distribution 
for the $z_i$ centered on 0 and of width~1.
This amounts to treating the discretization errors as independent and 
adding them in quadrature.
It also implicitly assumes that the data have nothing to say about the 
size or relative importance of the terms.

Here, however, we incorporate these errors into the chiral-continuum 
extrapolation, discussed in \secref{ChPT}.
This means that the $z_i$ are now constrained fit parameters,
with prior constraints discussed in  \secref{ChPT}.

The $f_i$ are collected next.

\subsubsection{$\mathrm{O}(a^2)$ errors}

We start with these, because explicit expressions for the functions 
$f_i(m_0a)$ are available.
The Lagrangian leads to two bilinears, $\bar{h}\bm{D}\cdot\bm{E}h$ and
$\bar{h}i\bm{\Sigma}\cdot[\bm{D}\times\bm{E}]h$, and many four-quark 
operators.
At the tree level the coefficients of all four-quark operators vanish.
At the tree level the coefficients of the two bilinears are the same, and
the mismatch function is
\begin{equation}
	f_E(m_0a) = \frac{1}{8m_E^2a^2} - \frac{1}{2(2m_2a)^2} .
\end{equation}
Using explicit expressions for $1/m_2$~\cite{ElKhadra:1996mp} and $1/m_E^2$~\cite{Oktay:2008ex}, one finds 
\begin{equation}
	f_E(m_0a) = \frac{1}{2}\left[
		\frac{c_E(1+m_0a)-1}{m_0a(2+m_0a)(1+m_0a)} -
		\frac{1}{4(1+m_0a)^2} \right].
\end{equation}
We are using $c_E=1$, so
\begin{equation}
	f_E(m_0a) = \frac{2+3m_0a}{8(2+m_0a)(1+m_0a)^2}.
\end{equation}
With no further assumptions, this term enters twice independently, 
so we take the width of this prior to be $\sqrt{2}$ rather than 1.

The current leads to three more terms with non-zero coefficients,
$\bar{q}\Gamma\bm{D}^2h$,
$\bar{q}\Gamma i\bm{\Sigma}\cdot\bm{B}h$, and
$\bar{q}\Gamma\bm{\alpha}\cdot\bm{E}h$, 
which can be deduced from Eq.~(A17) of Ref.~\cite{ElKhadra:1996mp}.
Their coefficients can be read off from Eq.~(A19).
When $c_B=r_s$ the first two share the same coefficient
\begin{eqnarray}
	f_X(m_0a) & = & \frac{1}{8m_X^2a^2} -
		\frac{\zeta d_1(1+m_0a)}{m_0a(2+m_0a)} - \frac{1}{2(2m_2a)^2},
		\nonumber \\
		& = & \frac{1}{2}\left[ \frac{1}{(2+m_0a)(1+m_0a)} +
			\frac{1}{2(1+m_0a)} - \frac{1}{4(1+m_0a)^2} - 
			\frac{1}{(2+m_0a)^2} \right], \nonumber \\
	& = & \frac{1}{2}\left[ \frac{1}{2(1+m_0a)} - \left(
		\frac{m_0a}{2(2+m_0a)(1+m_0a)} \right)^2\right],
\end{eqnarray}
where the last term on the second line comes from using the 
tree-level~$d_1$ (as we do in the simulations).
Because of the two-fold appearance, we again take the prior width to be~$\sqrt{2}$.

For $\bar{q}\Gamma\bm{\alpha}\cdot\bm{E}h$
\begin{eqnarray}
	f_Y(m_0a) & = & \frac{1}{2}\left[\frac{d_1}{m_2a} -
		\frac{\zeta(1-c_E)(1+m_0a)}{m_0a(2+m_0a)} \right], \nonumber \\
		& = & \frac{2 + 4 m_0a + (m_0a)^2}{4(1+m_0a)^2(2+m_0a)^2},
\end{eqnarray}
where the last line reflects the choices made for $c_E$ and~$d_1$.

\subsubsection{$\mathrm{O}(\alpha_s a)$ and $\mathrm{O}(a^3)$ errors}

Here the mismatch functions~$f_i(m_0a)$ start at order $\alpha_s$, 
and we do not have explicit expressions for them.
We take unimproved tree-level coefficients as a guide to the 
combinatoric factors and the asymptotic behavior as $m_0a\to0$ and 
$m_0a\to\infty$.

The Lagrangian leads to two bilinears, the kinetic energy 
$\mathcal{O}_2=\bar{h}\bm{D}^2h$ and the chromomagnetic moment 
$\mathcal{O}_B=\bar{h}i\bm{\Sigma}\cdot\bm{B}h$.
We match the former nonperturbatively, by identifying the meson's 
kinetic mass with the physical mass; the discretization error~$f_2$ 
stems, therefore, from discretization effects in $M_2$.

The computed kinetic meson mass is
\begin{equation}
    M_2 = m_2(\kappa) + \textrm{continuum binding energy} + \delta M_2,
\end{equation}
where \cite{Bernard:2010fr}
\begin{equation}
	\delta M_2 = \frac{\bar{\Lambda}^2}{6m_Q} 
		\left[5 \left(\frac{m_{2}^3}{m_{4}^3} - 1 \right) +
		4 w_{4}(m_{2}a)^3\right],
	\label{eq:error_nolightdisc}
\end{equation}
and $m_2$, $m_4$, and $w_4$ are functions of $m_0a$ and, hence, $\kappa$.
(See Refs.~\cite{ElKhadra:1996mp,Oktay:2008ex} for explicit expressions.)
Equating $M_2$ to a physical meson mass means that we choose $\kappa$ 
such that $m_2(\kappa)+\delta M_2=m_Q$, thereby making in $\phi$ a 
relative error
\begin{equation}
    \texttt{error}_2 =
        \bar{\Lambda}\left(\frac{1}{2m_2}-\frac{1}{2m_Q}\right) =
        \bar{\Lambda}\left(\frac{1}{2m_Q-2\delta M_2}-\frac{1}{2m_Q}\right) \approx
        \bar{\Lambda}\frac{\delta M_2}{2m_Q^2}.
\end{equation}
The right-most expression is  
$(a\bar{\Lambda})^3\,f_2(m_0a)$, $f_2=[~]/12(m_2a)^3$, where 
$[~]$ is the bracket in Eq.~(\ref{eq:error_nolightdisc}).
It is formally smaller than the other errors considered here---$f_2$ is 
of order 1 for all $m_0a$.
Numerically, however, it is not much smaller.

At the tree level the chromomagnetic mismatch is
\begin{equation}
	f^{[0]}_B(m_0a) = \frac{c_B-1}{2(1+m_0a)}.
\end{equation}
This has the right asymptotic behavior in both limits, so our Ansatz 
for the one-loop mismatch function is simply
\begin{equation}
	f_B(m_0a) = \frac{\alpha_s}{2(1+m_0a)},
	\label{eq:fB}
\end{equation}
and ${\tt error}_B$ is this function multiplied by $a\Lambda$.
We take $\alpha_s=0.288$ on the $a\approx0.12$ fm ensembles, which is the
value determined for $\alpha_V$ from the plaquette \cite{Lepage:1992xa} with one-loop
running to scale $q^*=2.5/a$. On other ensembles, $\alpha_s$ is found by assuming that
the measured average taste splitting goes like $\alpha_s^2 a^2$ (with $a$ determined
from $r_1/a$).  This gives $\alpha_s$ values that track $\alpha_V(q^*=2.5/a)$ quite well,
which is why we make  that $q^*$ choice.  The results are rather insensitive to the 
details here.  
For example, using $\alpha_s=0.325$ on the $a\approx0.12$ fm ensembles,
which corresponds to $q^*=2.0/a$, increases the error estimate by less than 0.6 MeV for
$f_{D^+}$, and less than 0.25 MeV for $f_{B^+}$.

The current leads to one more term, with tree-level mismatch function
\begin{equation}
	f^{[0]}_3(m_0a) = \frac{m_0a}{2(2+m_0a)(1+m_0a)} - d_1,
\end{equation}
and the tree-level $d_1$ is chosen so that $f_3^{[0]}=0$.
As with the mismatch function $f_B$, we would like to anticipate 
$f_3^{[1]}$ by setting $d_1^{[1]}=0$ and multiplying the rest with 
$\alpha_s$.
But it is not generic that this vanishes as $m_0a\to0$.
Therefore, we take
\begin{equation}
	f_3(m_0a) = \frac{\alpha_s}{2(2+m_0a)},
	\label{eq:f3}
\end{equation}
which has the right asymptotic behavior.
We take the prior width as  $\sqrt{2}$, because $A^4$ has two such 
corrections~\cite{Harada:2001fi}.

\subsection{Dispersion relation, Eq.~\eqref{eq:disprel}}
\label{app:disprel}

We take a similar approach to the dispersion relation,
Eq.~\eqref{eq:disprel}, with the difference that we now know the sign 
of the leading effect.

The tree-level functions are
\begin{eqnarray}
    a_4^{[0]} & = & \frac{1}{(m_2^{[0]}a)^2} - \frac{m_1^{[0]}a}{(m_4^{[0]}a)^3}, \\
    a_{4'}^{[0]} & = & m_1^{[0]}a\,w_4^{[0]}.
\end{eqnarray}
The binding energy enters $A_4$ and $A_{4'}$ via the meson's kinetic energy.
Hence, the binding contributions are
\begin{eqnarray}
    A'_4    & = & \frac{3m_1^{[0]}a}{m_2^{[0]}a\,(m_4^{[0]}a)^3} - \frac{2}{(m_2^{[0]}a)^3} - \frac{1}{(m_4^{[0]}a)^3}, \\
    A'_{4'} & = & w_4^{[0]}\left(1 - \frac{m_1^{[0]}a}{m_2^{[0]}a} \right),
\end{eqnarray}
and in Eq.~\eqref{eq:HQM:disprelprior} the binding energy floats within 
a Gaussian prior described by
$(\bar{\Lambda},\sigma_{\bar{\Lambda}})=(600,400)$~MeV.
This choice conservatively brackets the binding energy of a 
heavy-strange meson.
For the higher-order perturbative contribution to the coefficients, 
we take the Ans\"atze based on the asymptotic behavior:
\begin{eqnarray}
    a_4^{[1]} & = & \frac{y_4 + z_4\ln(1+m_0a)}{(1+m_0a)^2}, \\
    a_{4'}^{[1]} & = & \frac{y_{4'}m_0a + z_{4'}\ln(1+m_0a)}{1+m_0a},
\end{eqnarray}
where the $y$s and $z$s float within Gaussian priors described by
$(y_4,   \sigma_{y_4})    = (3,5)$, 
$(z_4,   \sigma_{z_4})    = (1,2)$,
$(y_{4'},\sigma_{y_{4'}}) = (0,0)$, and
$(z_{4'},\sigma_{z_{4'}}) = (0,2)$.
The terms proportional to $y_i$ stem from the $m_0a\to0$ limit, in 
which the renormalization of $m_4$ must coincide with that of $m_1$,
and $a_4=m_1a\,w_4$ must vanish like~$m_0a$.
The terms proportional to $z_i$ stem from the $m_0a\to\infty$ limit, 
where the static limit is obtained.
Except for $y_{4'}$, the numerical values have been chosen consistent 
with one-loop experience for $m_1$ and $m_2$~\cite{Mertens:1997wx}.
We have set $y_{4'}\equiv0$, because at small $m_0a$ it is 
indistinguishable from the other term in $a_{4'}^{[1]}$, and 
our range of $m_0a$ does not reach far into the region $m_0a\gg1$.

\clearpage

\section{Two point fit results from Analysis~I}
\label{apdx:twoPointFitsJackknife}

\newlength{\CRS}\setlength{\CRS}{-0.0ex}

\begin{table}[h]
\caption{Heavy-light pseudoscalar meson masses and renormalized decay amplitudes obtained from
Analysis~I fits of the charm correlators at lattice spacing $a\approx 0.09$~fm.}
\label{jackknife_charm_a09}
\begin{tabular}{llllll}
\hline\hline
$am_l/am_s$ & $am_q$ & $aM_H$ & $a^{3/2}\phi_H$ & $\chi^2/\mathrm{dof}$ & $p$ \\ \hline
0.0031/0.031	& 0.0031  	& 0.7523(0.0016)	& 0.0857(0.0015)	& 58/48	& 0.23	\\[\CRS]
0.0031/0.031	& 0.0044  	& 0.7553(0.0014)	& 0.0873(0.0013)	& 56/48	& 0.28	\\[\CRS]
0.0031/0.031	& 0.0062  	& 0.7589(0.0011)	& 0.0890(0.0011)	& 55/48	& 0.33	\\[\CRS]
0.0031/0.031	& 0.0087  	& 0.7634(0.0009)	& 0.0910(0.0009)	& 53/48	& 0.38	\\[\CRS]
0.0031/0.031	& 0.0124  	& 0.7699(0.0007)	& 0.0936(0.0007)	& 53/48	& 0.41	\\[\CRS]
0.0031/0.031	& 0.0186  	& 0.7807(0.0005)	& 0.0978(0.0006)	& 52/48	& 0.44	\\[\CRS]
0.0031/0.031	& 0.0272  	& 0.7954(0.0004)	& 0.1030(0.0005)	& 50/48	& 0.5	\\[\CRS]
0.0031/0.031	& 0.031  	& 0.8018(0.0004)	& 0.1052(0.0004)	& 50/48	& 0.5	\\[\CRS]
0.0062/0.031	& 0.0031  	& 0.7541(0.0030)	& 0.0875(0.0027)	& 56/48	& 0.37	\\[\CRS]
0.0062/0.031	& 0.0044  	& 0.7577(0.0023)	& 0.0899(0.0021)	& 52/48	& 0.49	\\[\CRS]
0.0062/0.031	& 0.0062  	& 0.7613(0.0019)	& 0.0917(0.0018)	& 50/48	& 0.58	\\[\CRS]
0.0062/0.031	& 0.0087  	& 0.7654(0.0015)	& 0.0933(0.0015)	& 58/51	& 0.43	\\[\CRS]
0.0062/0.031	& 0.0124  	& 0.7712(0.0012)	& 0.0952(0.0012)	& 52/48	& 0.48	\\[\CRS]
0.0062/0.031	& 0.0186  	& 0.7810(0.0009)	& 0.0985(0.0010)	& 56/48	& 0.37	\\[\CRS]
0.0062/0.031	& 0.0272  	& 0.7952(0.0006)	& 0.1032(0.0008)	& 59/48	& 0.28	\\[\CRS]
0.0062/0.031	& 0.031  	& 0.8015(0.0005)	& 0.1052(0.0007)	& 60/48	& 0.25	\\[\CRS]
0.0124/0.031	& 0.0031  	& 0.7551(0.0038)	& 0.0930(0.0036)	& 60/48	& 0.27	\\[\CRS]
0.0124/0.031	& 0.0042  	& 0.7554(0.0031)	& 0.0926(0.0028)	& 65/48	& 0.15	\\[\CRS]
0.0124/0.031	& 0.0062  	& 0.7574(0.0023)	& 0.0929(0.0021)	& 65/48	& 0.16	\\[\CRS]
0.0124/0.031	& 0.0087  	& 0.7608(0.0017)	& 0.0938(0.0015)	& 59/48	& 0.28	\\[\CRS]
0.0124/0.031	& 0.0124  	& 0.7666(0.0013)	& 0.0957(0.0012)	& 49/48	& 0.63	\\[\CRS]
0.0124/0.031	& 0.0186  	& 0.7766(0.0008)	& 0.0991(0.0009)	& 42/48	& 0.85	\\[\CRS]
0.0124/0.031	& 0.0272  	& 0.7907(0.0006)	& 0.1038(0.0007)	& 48/48	& 0.64	\\[\CRS]
0.0124/0.031	& 0.031  	& 0.7969(0.0005)	& 0.1058(0.0006)	& 53/48	& 0.47	\\[\CRS]
\hline\hline
\end{tabular}
\end{table}

\begin{table}
\caption{Heavy-light pseudoscalar meson masses and renormalized decay amplitudes obtained from Analysis~I
fits of the charm correlators at lattice spacing $a\approx 0.12$~fm.}
\label{jackknife_charm_a12}
\begin{tabular}{llllll}
\hline\hline
$am_l/am_s$ & $am_q$ & $aM_H$ & $a^{3/2}\phi_H$ & $\chi^2/\mathrm{dof}$ & $p$ \\ \hline
0.005/0.050	& 0.005  	& 0.9943(0.0032)	& 0.1436(0.0030)	& 30/30	& 0.52	\\[\CRS]
0.005/0.050	& 0.007  	& 0.9977(0.0024)	& 0.1453(0.0024)	& 29/30	& 0.6	\\[\CRS]
0.005/0.050	& 0.01  	& 1.0026(0.0018)	& 0.1477(0.0019)	& 28/30	& 0.64	\\[\CRS]
0.005/0.050	& 0.014  	& 1.0090(0.0016)	& 0.1508(0.0017)	& 28/30	& 0.64	\\[\CRS]
0.005/0.050	& 0.02  	& 1.0186(0.0013)	& 0.1551(0.0015)	& 29/30	& 0.58	\\[\CRS]
0.005/0.050	& 0.03  	& 1.0345(0.0010)	& 0.1620(0.0012)	& 33/30	& 0.42	\\[\CRS]
0.005/0.050	& 0.0415  	& 1.0526(0.0008)	& 0.1694(0.0010)	& 36/30	& 0.27	\\[\CRS]
0.007/0.050	& 0.005  	& 0.9948(0.0035)	& 0.1442(0.0035)	& 17/30	& 0.98	\\[\CRS]
0.007/0.050	& 0.007  	& 0.9975(0.0027)	& 0.1455(0.0028)	& 19/30	& 0.95	\\[\CRS]
0.007/0.050	& 0.01  	& 1.0019(0.0021)	& 0.1476(0.0021)	& 22/30	& 0.89	\\[\CRS]
0.007/0.050	& 0.014  	& 1.0081(0.0016)	& 0.1504(0.0017)	& 24/30	& 0.83	\\[\CRS]
0.007/0.050	& 0.02  	& 1.0178(0.0012)	& 0.1547(0.0014)	& 23/30	& 0.85	\\[\CRS]
0.007/0.050	& 0.03  	& 1.0338(0.0009)	& 0.1615(0.0010)	& 20/30	& 0.94	\\[\CRS]
0.007/0.050	& 0.0415  	& 1.0520(0.0007)	& 0.1687(0.0008)	& 19/30	& 0.95	\\[\CRS]
0.010/0.050	& 0.005  	& 0.9958(0.0039)	& 0.1461(0.0041)	& 15/30	& 0.99	\\[\CRS]
0.010/0.050	& 0.007  	& 1.0000(0.0031)	& 0.1486(0.0032)	& 20/30	& 0.94	\\[\CRS]
0.010/0.050	& 0.01  	& 1.0057(0.0024)	& 0.1516(0.0026)	& 26/30	& 0.75	\\[\CRS]
0.010/0.050	& 0.014  	& 1.0126(0.0019)	& 0.1549(0.0021)	& 29/27	& 0.41	\\[\CRS]
0.010/0.050	& 0.02  	& 1.0226(0.0015)	& 0.1594(0.0017)	& 33/30	& 0.39	\\[\CRS]
0.010/0.050	& 0.03  	& 1.0387(0.0011)	& 0.1662(0.0014)	& 31/30	& 0.5	\\[\CRS]
0.010/0.050	& 0.0415  	& 1.0567(0.0008)	& 0.1733(0.0011)	& 27/30	& 0.68	\\[\CRS]
0.020/0.050	& 0.005  	& 0.9942(0.0046)	& 0.1537(0.0050)	& 49/30	& 0.036	\\[\CRS]
0.020/0.050	& 0.007  	& 0.9959(0.0036)	& 0.1533(0.0039)	& 49/30	& 0.036	\\[\CRS]
0.020/0.050	& 0.01  	& 0.9987(0.0027)	& 0.1532(0.0031)	& 48/30	& 0.051	\\[\CRS]
0.020/0.050	& 0.014  	& 1.0037(0.0021)	& 0.1543(0.0024)	& 45/30	& 0.075	\\[\CRS]
0.020/0.050	& 0.02  	& 1.0124(0.0016)	& 0.1575(0.0019)	& 43/30	& 0.11	\\[\CRS]
0.020/0.050	& 0.03  	& 1.0274(0.0011)	& 0.1632(0.0014)	& 37/30	& 0.27	\\[\CRS]
0.020/0.050	& 0.0415  	& 1.0447(0.0009)	& 0.1695(0.0012)	& 32/30	& 0.48	\\[\CRS]
0.030/0.050	& 0.005  	& 0.9830(0.0042)	& 0.1475(0.0042)	& 33/30	& 0.39	\\[\CRS]
0.030/0.050	& 0.007  	& 0.9853(0.0033)	& 0.1485(0.0033)	& 33/30	& 0.4	\\[\CRS]
0.030/0.050	& 0.01  	& 0.9897(0.0025)	& 0.1505(0.0025)	& 32/30	& 0.47	\\[\CRS]
0.030/0.050	& 0.014  	& 0.9960(0.0020)	& 0.1534(0.0020)	& 31/30	& 0.53	\\[\CRS]
0.030/0.050	& 0.02  	& 1.0054(0.0015)	& 0.1574(0.0016)	& 32/30	& 0.46	\\[\CRS]
0.030/0.050	& 0.03  	& 1.0205(0.0011)	& 0.1633(0.0012)	& 37/30	& 0.27	\\[\CRS]
0.030/0.050	& 0.0415  	& 1.0376(0.0009)	& 0.1695(0.0010)	& 40/30	& 0.15	\\[\CRS]
\hline\hline
\end{tabular}
\end{table}

\begin{table}
\caption{Heavy-light pseudoscalar meson masses and renormalized decay amplitudes obtained from
Analysis~I fits of the charm correlators at lattice spacing $a\approx 0.15$~fm.}
\label{jackknife_charm-a15}
\begin{tabular}{llllll}
\hline\hline
$am_l/am_s$ & $am_q$ & $aM_H$ & $a^{3/2}\phi_H$ & $\chi^2/\mathrm{dof}$ & $p$ \\ \hline
0.0097/0.0484	& 0.0048  	& 1.1659(0.0044)	& 0.1979(0.0052)	& 20/20	& 0.5	\\[\CRS]
0.0097/0.0484	& 0.007  	& 1.1710(0.0034)	& 0.2017(0.0040)	& 22/20	& 0.37	\\[\CRS]
0.0097/0.0484	& 0.0097  	& 1.1768(0.0027)	& 0.2054(0.0032)	& 25/20	& 0.26	\\[\CRS]
0.0097/0.0484	& 0.0194  	& 1.1951(0.0016)	& 0.2159(0.0020)	& 25/20	& 0.26	\\[\CRS]
0.0097/0.0484	& 0.029  	& 1.2117(0.0012)	& 0.2242(0.0015)	& 20/20	& 0.51	\\[\CRS]
0.0097/0.0484	& 0.0484  	& 1.2432(0.0009)	& 0.2385(0.0012)	& 15/20	& 0.79	\\[\CRS]
0.0194/0.0484	& 0.0048  	& 1.1726(0.0046)	& 0.2106(0.0052)	& 23/20	& 0.35	\\[\CRS]
0.0194/0.0484	& 0.007  	& 1.1749(0.0036)	& 0.2105(0.0041)	& 23/20	& 0.35	\\[\CRS]
0.0194/0.0484	& 0.0097  	& 1.1785(0.0028)	& 0.2113(0.0031)	& 23/20	& 0.32	\\[\CRS]
0.0194/0.0484	& 0.0194  	& 1.1935(0.0016)	& 0.2174(0.0020)	& 30/20	& 0.092	\\[\CRS]
0.0194/0.0484	& 0.029  	& 1.2091(0.0013)	& 0.2244(0.0016)	& 32/20	& 0.055	\\[\CRS]
0.0194/0.0484	& 0.0484  	& 1.2400(0.0010)	& 0.2381(0.0013)	& 27/20	& 0.17	\\[\CRS]
0.0290/0.0484	& 0.0048  	& 1.1613(0.0044)	& 0.1975(0.0049)	& 17/20	& 0.72	\\[\CRS]
0.0290/0.0484	& 0.007  	& 1.1660(0.0034)	& 0.2010(0.0040)	& 18/20	& 0.64	\\[\CRS]
0.0290/0.0484	& 0.0097  	& 1.1717(0.0026)	& 0.2049(0.0031)	& 21/20	& 0.47	\\[\CRS]
0.0290/0.0484	& 0.0194  	& 1.1896(0.0015)	& 0.2151(0.0019)	& 24/20	& 0.3	\\[\CRS]
0.0290/0.0484	& 0.029  	& 1.2058(0.0011)	& 0.2229(0.0015)	& 23/20	& 0.32	\\[\CRS]
0.0290/0.0484	& 0.0484  	& 1.2368(0.0008)	& 0.2364(0.0011)	& 20/20	& 0.49	\\[\CRS]
\hline\hline
\end{tabular}
\end{table}

\begin{table}
\caption{Heavy-light pseudoscalar meson masses and renormalized decay amplitudes obtained from Analysis~I
fits of the bottom correlators at lattice spacing $a\approx 0.09$~fm.}
\label{jackknife_bottom_a09}
\begin{tabular}{llllll}
\hline\hline
$am_l/am_s$ & $am_q$ & $aM_H$ & $a^{3/2}\phi_H$ & $\chi^2/\mathrm{dof}$ & $p$ \\ \hline
0.0031/0.031	& 0.0031  	& 1.6509(0.0018)	& 0.1359(0.0016)	& 41/39	& 0.48	\\
0.0031/0.031	& 0.0044  	& 1.6532(0.0016)	& 0.1378(0.0015)	& 40/39	& 0.51	\\
0.0031/0.031	& 0.0062  	& 1.6562(0.0015)	& 0.1402(0.0014)	& 40/39	& 0.49	\\
0.0031/0.031	& 0.0087  	& 1.6601(0.0013)	& 0.1433(0.0014)	& 42/39	& 0.42	\\
0.0031/0.031	& 0.0124  	& 1.6659(0.0012)	& 0.1475(0.0013)	& 45/39	& 0.31	\\
0.0031/0.031	& 0.0186  	& 1.6752(0.0011)	& 0.1542(0.0012)	& 47/39	& 0.23	\\
0.0031/0.031	& 0.0272  	& 1.6879(0.0009)	& 0.1628(0.0011)	& 49/39	& 0.19	\\
0.0031/0.031	& 0.031  	& 1.6934(0.0009)	& 0.1664(0.0011)	& 49/39	& 0.18	\\
0.0062/0.031	& 0.0031  	& 1.6539(0.0046)	& 0.1358(0.0051)	& 40/39	& 0.56	\\
0.0062/0.031	& 0.0044  	& 1.6557(0.0039)	& 0.1377(0.0044)	& 37/39	& 0.68	\\
0.0062/0.031	& 0.0062  	& 1.6584(0.0032)	& 0.1402(0.0037)	& 34/39	& 0.77	\\
0.0062/0.031	& 0.0087  	& 1.6620(0.0027)	& 0.1434(0.0031)	& 34/39	& 0.8	\\
0.0062/0.031	& 0.0124  	& 1.6675(0.0022)	& 0.1480(0.0026)	& 36/39	& 0.72	\\
0.0062/0.031	& 0.0186  	& 1.6767(0.0018)	& 0.1550(0.0022)	& 41/39	& 0.53	\\
0.0062/0.031	& 0.0272  	& 1.6892(0.0014)	& 0.1637(0.0019)	& 45/39	& 0.37	\\
0.0062/0.031	& 0.031  	& 1.6946(0.0014)	& 0.1672(0.0018)	& 45/39	& 0.35	\\
0.0124/0.031	& 0.0031  	& 1.6532(0.0036)	& 0.1387(0.0038)	& 52/39	& 0.16	\\
0.0124/0.031	& 0.0042  	& 1.6550(0.0033)	& 0.1407(0.0034)	& 48/39	& 0.27	\\
0.0124/0.031	& 0.0062  	& 1.6576(0.0030)	& 0.1432(0.0031)	& 40/39	& 0.55	\\
0.0124/0.031	& 0.0087  	& 1.6606(0.0027)	& 0.1456(0.0029)	& 35/39	& 0.77	\\
0.0124/0.031	& 0.0124  	& 1.6650(0.0024)	& 0.1488(0.0027)	& 33/39	& 0.84	\\
0.0124/0.031	& 0.0186  	& 1.6730(0.0019)	& 0.1544(0.0023)	& 36/39	& 0.73	\\
0.0124/0.031	& 0.0272  	& 1.6847(0.0016)	& 0.1623(0.0021)	& 42/39	& 0.48	\\
0.0124/0.031	& 0.031  	& 1.6900(0.0015)	& 0.1657(0.0020)	& 45/39	& 0.38	\\
\hline\hline
\end{tabular}
\end{table}

\begin{table}
\caption{Heavy-light pseudoscalar meson masses and renormalized decay amplitudes obtained from Analysis~I
fits of the bottom correlators at lattice spacing $a\approx 0.12$~fm.}
\label{jackknife_bottom_a12}
\begin{tabular}{llllll}
\hline\hline
$am_l/am_s$ & $am_q$ & $aM_H$ & $a^{3/2}\phi_H$ & $\chi^2/\mathrm{dof}$ & $p$ \\ \hline
0.005/0.050	& 0.005  	& 1.9170(0.0044)	& 0.2236(0.0050)	& 45/27	& 0.03	\\
0.005/0.050	& 0.007  	& 1.9197(0.0039)	& 0.2263(0.0046)	& 46/27	& 0.022	\\
0.005/0.050	& 0.01  	& 1.9235(0.0033)	& 0.2300(0.0040)	& 46/27	& 0.021	\\
0.005/0.050	& 0.014  	& 1.9287(0.0029)	& 0.2347(0.0036)	& 45/27	& 0.027	\\
0.005/0.050	& 0.02  	& 1.9367(0.0024)	& 0.2418(0.0031)	& 43/27	& 0.046	\\
0.005/0.050	& 0.03  	& 1.9503(0.0020)	& 0.2532(0.0026)	& 39/27	& 0.096	\\
0.005/0.050	& 0.0415  	& 1.9657(0.0017)	& 0.2654(0.0023)	& 36/27	& 0.17	\\
0.007/0.050	& 0.005  	& 1.9147(0.0036)	& 0.2224(0.0039)	& 37/27	& 0.12	\\
0.007/0.050	& 0.007  	& 1.9177(0.0033)	& 0.2254(0.0037)	& 35/27	& 0.17	\\
0.007/0.050	& 0.01  	& 1.9219(0.0030)	& 0.2292(0.0036)	& 34/27	& 0.2	\\
0.007/0.050	& 0.014  	& 1.9272(0.0028)	& 0.2337(0.0037)	& 35/27	& 0.19	\\
0.007/0.050	& 0.02  	& 1.9351(0.0026)	& 0.2401(0.0037)	& 36/27	& 0.15	\\
0.007/0.050	& 0.03  	& 1.9485(0.0022)	& 0.2508(0.0035)	& 38/27	& 0.096	\\
0.007/0.050	& 0.0415  	& 1.9638(0.0019)	& 0.2628(0.0031)	& 40/27	& 0.07	\\
0.010/0.050	& 0.005  	& 1.9182(0.0047)	& 0.2254(0.0047)	& 30/27	& 0.4	\\
0.010/0.050	& 0.007  	& 1.9207(0.0041)	& 0.2284(0.0042)	& 32/27	& 0.29	\\
0.010/0.050	& 0.01  	& 1.9250(0.0035)	& 0.2328(0.0037)	& 36/27	& 0.18	\\
0.010/0.050	& 0.014  	& 1.9307(0.0030)	& 0.2383(0.0033)	& 39/27	& 0.097	\\
0.010/0.050	& 0.02  	& 1.9391(0.0025)	& 0.2457(0.0028)	& 43/27	& 0.048	\\
0.010/0.050	& 0.03  	& 1.9527(0.0020)	& 0.2569(0.0024)	& 47/27	& 0.02	\\
0.010/0.050	& 0.0415  	& 1.9682(0.0017)	& 0.2689(0.0021)	& 51/27	& 0.0092	\\
0.020/0.050	& 0.005  	& 1.9136(0.0060)	& 0.2278(0.0069)	& 33/27	& 0.27	\\
0.020/0.050	& 0.007  	& 1.9163(0.0050)	& 0.2305(0.0059)	& 33/27	& 0.28	\\
0.020/0.050	& 0.01  	& 1.9200(0.0042)	& 0.2340(0.0050)	& 31/27	& 0.36	\\
0.020/0.050	& 0.014  	& 1.9249(0.0036)	& 0.2381(0.0043)	& 29/27	& 0.47	\\
0.020/0.050	& 0.02  	& 1.9322(0.0031)	& 0.2437(0.0039)	& 28/27	& 0.52	\\
0.020/0.050	& 0.03  	& 1.9445(0.0027)	& 0.2526(0.0038)	& 30/27	& 0.42	\\
0.020/0.050	& 0.0415  	& 1.9590(0.0025)	& 0.2627(0.0039)	& 33/27	& 0.3	\\
0.030/0.050	& 0.005  	& 1.9030(0.0058)	& 0.2196(0.0073)	& 38/27	& 0.12	\\
0.030/0.050	& 0.007  	& 1.9058(0.0049)	& 0.2223(0.0064)	& 32/27	& 0.29	\\
0.030/0.050	& 0.01  	& 1.9099(0.0041)	& 0.2258(0.0056)	& 27/27	& 0.56	\\
0.030/0.050	& 0.014  	& 1.9155(0.0034)	& 0.2306(0.0048)	& 23/27	& 0.74	\\
0.030/0.050	& 0.02  	& 1.9239(0.0028)	& 0.2376(0.0040)	& 22/27	& 0.77	\\
0.030/0.050	& 0.03  	& 1.9372(0.0022)	& 0.2479(0.0034)	& 25/27	& 0.64	\\
0.030/0.050	& 0.0415  	& 1.9518(0.0019)	& 0.2585(0.0032)	& 28/27	& 0.49	\\
\hline\hline
\end{tabular}
\end{table}

\begin{table}
\caption{Heavy-light pseudoscalar meson masses and renormalized decay amplitudes obtained from Analysis~I
fits of the bottom correlators at lattice spacing $a\approx 0.15$~fm.}
\label{jackknife_bottom_a15}
\begin{tabular}{lllllll}
\hline\hline
$am_l/am_s$ & $am_q$ & $aM_H$ & $a^{3/2}\phi_H$ & $\chi^2/\mathrm{dof}$ & $p$ \\ \hline
0.0097/0.0484	& 0.0048  	& 2.2553(0.0071)	& 0.3311(0.0115)	& 36/25	& 0.097	\\
0.0097/0.0484	& 0.007  	& 2.2576(0.0061)	& 0.3341(0.0102)	& 37/25	& 0.09	\\
0.0097/0.0484	& 0.0097  	& 2.2611(0.0052)	& 0.3389(0.0089)	& 36/25	& 0.1	\\
0.0097/0.0484	& 0.0194  	& 2.2757(0.0036)	& 0.3568(0.0063)	& 34/25	& 0.16	\\
0.0097/0.0484	& 0.029  	& 2.2901(0.0030)	& 0.3727(0.0053)	& 33/25	& 0.16	\\
0.0097/0.0484	& 0.0484  	& 2.3175(0.0023)	& 0.4002(0.0046)	& 35/25	& 0.12	\\
0.0194/0.0484	& 0.0048  	& 2.2296(0.0175)	& 0.2743(0.0416)	& 32/25	& 0.2	\\
0.0194/0.0484	& 0.007  	& 2.2349(0.0142)	& 0.2823(0.0357)	& 34/25	& 0.15	\\
0.0194/0.0484	& 0.0097  	& 2.2416(0.0118)	& 0.2917(0.0309)	& 36/25	& 0.1	\\
0.0194/0.0484	& 0.0194  	& 2.2639(0.0072)	& 0.3243(0.0202)	& 36/25	& 0.1	\\
0.0194/0.0484	& 0.029  	& 2.2819(0.0054)	& 0.3482(0.0152)	& 30/25	& 0.27	\\
0.0194/0.0484	& 0.0484  	& 2.3124(0.0038)	& 0.3839(0.0109)	& 24/25	& 0.59	\\
0.0290/0.0484	& 0.0048  	& 2.2402(0.0073)	& 0.3101(0.0123)	& 29/25	& 0.32	\\
0.0290/0.0484	& 0.007  	& 2.2464(0.0061)	& 0.3199(0.0104)	& 30/25	& 0.28	\\
0.0290/0.0484	& 0.0097  	& 2.2524(0.0052)	& 0.3289(0.0089)	& 31/25	& 0.25	\\
0.0290/0.0484	& 0.0194  	& 2.2695(0.0036)	& 0.3502(0.0066)	& 27/25	& 0.42	\\
0.0290/0.0484	& 0.029  	& 2.2847(0.0030)	& 0.3665(0.0058)	& 21/25	& 0.72	\\
0.0290/0.0484	& 0.0484  	& 2.3125(0.0025)	& 0.3939(0.0057)	& 18/25	& 0.87	\\
\hline\hline
\end{tabular}
\end{table}

\clearpage


\bibliography{leptonic_B_D}

18 uid=1388364422
20 ctime=1323810800
20 atime=1323811061
24 SCHILY.dev=234881026
23 SCHILY.ino=12652108
18 SCHILY.nlink=1


\begin{thebibliography}{100}%
\makeatletter
\providecommand \@ifxundefined [1]{%
 \ifx #1\undefined \expandafter \@firstoftwo
 \else \expandafter \@secondoftwo
\fi
}%
\providecommand \@ifnum [1]{%
 \ifnum #1\expandafter \@firstoftwo
 \else \expandafter \@secondoftwo
\fi
}%
\providecommand \enquote [1]{``#1''}%
\providecommand \bibnamefont  [1]{#1}%
\providecommand \bibfnamefont [1]{#1}%
\providecommand \citenamefont [1]{#1}%
\providecommand\href[0]{\@sanitize\@href}%
\providecommand\@href[1]{\endgroup\@@startlink{#1}\endgroup\@@href}%
\providecommand\@@href[1]{#1\@@endlink}%
\providecommand \@sanitize [0]{\begingroup\catcode`\&12\catcode`\#12\relax}%
\@ifxundefined \pdfoutput {\@firstoftwo}{%
 \@ifnum{\z@=\pdfoutput}{\@firstoftwo}{\@secondoftwo}%
}{%
 \providecommand\@@startlink[1]{\leavevmode\special{html:<a href="#1">}}%
 \providecommand\@@endlink[0]{\special{html:</a>}}%
}{%
 \providecommand\@@startlink[1]{%
  \leavevmode
  \pdfstartlink
   attr{/Border[0 0 1 ]/H/I/C[0 1 1]}%
   user{/Subtype/Link/A<</Type/Action/S/URI/URI(#1)>>}%
  \relax
 }%
 \providecommand\@@endlink[0]{\pdfendlink}%
}%
\providecommand \url  [0]{\begingroup\@sanitize \@url }%
\providecommand \@url [1]{\endgroup\@href {#1}{\urlprefix}}%
\providecommand \urlprefix [0]{URL }%
\providecommand \Eprint[0]{\href }%
\@ifxundefined \urlstyle {%
  \providecommand \doi [1]{doi:\discretionary{}{}{}#1}%
}{%
  \providecommand \doi [0]{doi:\discretionary{}{}{}\begingroup
  \urlstyle{rm}\Url }%
}%
\providecommand \doibase [0]{http://dx.doi.org/}%
\providecommand \Doi[1]{\href{\doibase#1}}%
\providecommand \bibAnnote [3]{%
  \BibitemShut{#1}%
  \begin{quotation}\noindent
    \textsc{Key:}\ #2\\\textsc{Annotation:}\ #3%
  \end{quotation}%
}%
\providecommand \bibAnnoteFile [2]{%
  \IfFileExists{#2}{\bibAnnote {#1} {#2} {\input{#2}}}{}%
}%
\providecommand \typeout [0]{\immediate \write \m@ne }%
\providecommand \selectlanguage [0]{\@gobble}%
\providecommand \bibinfo [0]{\@secondoftwo}%
\providecommand \bibfield [0]{\@secondoftwo}%
\providecommand \translation [1]{[#1]}%
\providecommand \BibitemOpen[0]{}%
\providecommand \bibitemStop [0]{}%
\providecommand \bibitemNoStop [0]{.\EOS\space}%
\providecommand \EOS [0]{\spacefactor3000\relax}%
\providecommand \BibitemShut [1]{\csname bibitem#1\endcsname}%
\bibitem{Lenz:2010gu}%
  \BibitemOpen
  \bibfield{author}{%
  \bibinfo {author} {\bibfnamefont{A.}~\bibnamefont{Lenz}} \emph{et~al.}
  (\bibinfo {collaboration} {CKMfitter}),\ }%
  \bibfield{journal}{%
  \Doi{10.1103/PhysRevD.83.036004}{\bibinfo {journal} {Phys. Rev.}}\ }%
  \textbf{\bibinfo {volume} {D83}},\ \bibinfo {pages} {036004} (\bibinfo {year}
  {2011}),\ \Eprint{http://arxiv.org/abs/1008.1593}{arXiv:1008.1593 [hep-ph]}%
  \bibAnnoteFile{NoStop}{Lenz:2010gu}%
\bibitem{Lunghi:2010gv}%
  \BibitemOpen
  \bibfield{author}{%
  \bibinfo {author} {\bibfnamefont{E.}~\bibnamefont{Lunghi}}\ and\ \bibinfo
  {author} {\bibfnamefont{A.}~\bibnamefont{Soni}},\ }%
  \bibfield{journal}{%
  \Doi{10.1016/j.physletb.2011.02.016}{\bibinfo {journal} {Phys. Lett.}}\ }%
  \textbf{\bibinfo {volume} {B697}},\ \bibinfo {pages} {323} (\bibinfo {year}
  {2011}),\ \Eprint{http://arxiv.org/abs/1010.6069}{arXiv:1010.6069 [hep-ph]}%
  \bibAnnoteFile{NoStop}{Lunghi:2010gv}%
\bibitem{Laiho:2011nz}%
  \BibitemOpen
  \bibfield{author}{%
  \bibinfo {author} {\bibfnamefont{J.}~\bibnamefont{Laiho}}, \bibinfo {author}
  {\bibfnamefont{E.}~\bibnamefont{Lunghi}},\ and\ \bibinfo {author}
  {\bibfnamefont{R.}~\bibnamefont{Van~de Water}},\ }%
  \bibfield{journal}{%
  \bibinfo {journal} {PoS}\ }%
  \textbf{\bibinfo {volume} {FPCP2010}},\ \bibinfo {pages} {040} (\bibinfo
  {year} {2010}),\ \Eprint{http://arxiv.org/abs/1102.3917}{arXiv:1102.3917
  [hep-ph]}%
  \bibAnnoteFile{NoStop}{Laiho:2011nz}%
\bibitem{Hara:2010dk}%
  \BibitemOpen
  \bibfield{author}{%
  \bibinfo {author} {\bibfnamefont{K.}~\bibnamefont{Hara}} \emph{et~al.}
  (\bibinfo {collaboration} {Belle}),\ }%
  \bibfield{journal}{%
  \Doi{10.1103/PhysRevD.82.071101}{\bibinfo {journal} {Phys. Rev.}}\ }%
  \textbf{\bibinfo {volume} {D82}},\ \bibinfo {pages} {071101} (\bibinfo {year}
  {2010}),\ \Eprint{http://arxiv.org/abs/1006.4201}{arXiv:1006.4201 [hep-ex]}%
  \bibAnnoteFile{NoStop}{Hara:2010dk}%
\bibitem{:2010rt}%
  \BibitemOpen
  \bibfield{author}{%
  \bibinfo {author} {\bibfnamefont{P.}~\bibnamefont{del Amo~Sanchez}}
  \emph{et~al.} (\bibinfo {collaboration} {BaBar})}%
   (\bibinfo {year} {2010}),\
  \Eprint{http://arxiv.org/abs/1008.0104}{arXiv:1008.0104 [hep-ex]}%
  \bibAnnoteFile{NoStop}{:2010rt}%
\bibitem{Masuzawa:2010zz}%
  \BibitemOpen
  \bibfield{author}{%
  \bibinfo {author} {\bibfnamefont{M.}~\bibnamefont{Masuzawa}},\ }%
  in\ \emph{\bibinfo {booktitle} {Proceedings of IPAC 2010}},\ \bibinfo
  {editor} {edited by\ \bibinfo {editor}
  {\bibfnamefont{K.}~\bibnamefont{Oide}}}\ (\bibinfo {year} {2010})\
  \url{http://accelconf.web.cern.ch/AccelConf/IPAC10/}%
  \bibAnnoteFile{NoStop}{Masuzawa:2010zz}%
\bibitem{Iijima:HINTS09}%
  \BibitemOpen
  \bibfield{author}{%
  \bibinfo {author} {\bibfnamefont{T.}~\bibnamefont{Iijima}},\ }%
  \enquote{\bibinfo {title} {{Search for Charged Higgs in $B\to\tau\nu$ and
  $D\to\tau\nu$ Decays}},}\  (\bibinfo {year} {2009}),\
  \url{http://kds.kek.jp/materialDisplay.py?contribId=5&sessionId=2&materialId%
=slides&confId=2865}%
  \bibAnnoteFile{NoStop}{Iijima:HINTS09}%
\bibitem{Lunghi:2009ke}%
  \BibitemOpen
  \bibfield{author}{%
  \bibinfo {author} {\bibfnamefont{E.}~\bibnamefont{Lunghi}}\ and\ \bibinfo
  {author} {\bibfnamefont{A.}~\bibnamefont{Soni}},\ }%
  \bibfield{journal}{%
  \Doi{10.1103/PhysRevLett.104.251802}{\bibinfo {journal} {Phys. Rev. Lett.}}\
  }%
  \textbf{\bibinfo {volume} {104}},\ \bibinfo {pages} {251802} (\bibinfo {year}
  {2010}),\ \Eprint{http://arxiv.org/abs/0912.0002}{arXiv:0912.0002 [hep-ph]}%
  \bibAnnoteFile{NoStop}{Lunghi:2009ke}%
\bibitem{Davies:2003ik}%
  \BibitemOpen
  \bibfield{author}{%
  \bibinfo {author} {\bibfnamefont{C.~T.~H.}\ \bibnamefont{Davies}}
  \emph{et~al.} (\bibinfo {collaboration} {HPQCD, MILC, and Fermilab
  Lattice}),\ }%
  \bibfield{journal}{%
  \bibinfo {journal} {Phys. Rev. Lett.}\ }%
  \textbf{\bibinfo {volume} {92}},\ \bibinfo {pages} {022001} (\bibinfo {year}
  {2004}),\
  \Eprint{http://arxiv.org/abs/hep-lat/0304004}{arXiv:hep-lat/0304004}%
  \bibAnnoteFile{NoStop}{Davies:2003ik}%
\bibitem{Susskind:1976jm}%
  \BibitemOpen
  \bibfield{author}{%
  \bibinfo {author} {\bibfnamefont{L.}~\bibnamefont{Susskind}},\ }%
  \bibfield{journal}{%
  \bibinfo {journal} {Phys. Rev.}\ }%
  \textbf{\bibinfo {volume} {D16}},\ \bibinfo {pages} {3031} (\bibinfo {year}
  {1977})%
  \bibAnnoteFile{NoStop}{Susskind:1976jm}%
\bibitem{Sharatchandra:1981si}%
  \BibitemOpen
  \bibfield{author}{%
  \bibinfo {author} {\bibfnamefont{H.~S.}\ \bibnamefont{Sharatchandra}},
  \bibinfo {author} {\bibfnamefont{H.~J.}\ \bibnamefont{Thun}},\ and\ \bibinfo
  {author} {\bibfnamefont{P.}~\bibnamefont{Weisz}},\ }%
  \bibfield{journal}{%
  \Doi{10.1016/0550-3213(81)90200-5}{\bibinfo {journal} {Nucl. Phys.}}\ }%
  \textbf{\bibinfo {volume} {B192}},\ \bibinfo {pages} {205} (\bibinfo {year}
  {1981})%
  \bibAnnoteFile{NoStop}{Sharatchandra:1981si}%
\bibitem{Aubin:2004fs}%
  \BibitemOpen
  \bibfield{author}{%
  \bibinfo {author} {\bibfnamefont{C.}~\bibnamefont{Aubin}} \emph{et~al.}
  (\bibinfo {collaboration} {MILC}),\ }%
  \bibfield{journal}{%
  \Doi{10.1103/PhysRevD.70.114501}{\bibinfo {journal} {Phys. Rev.}}\ }%
  \textbf{\bibinfo {volume} {D70}},\ \bibinfo {pages} {114501} (\bibinfo {year}
  {2004}),\ \Eprint{http://arxiv.org/abs/hep-lat/0407028}{arXiv:hep-lat/0407028
  [hep-lat]}%
  \bibAnnoteFile{NoStop}{Aubin:2004fs}%
\bibitem{Allison:2004be}%
  \BibitemOpen
  \bibfield{author}{%
  \bibinfo {author} {\bibfnamefont{I.~F.}\ \bibnamefont{Allison}} \emph{et~al.}
  (\bibinfo {collaboration} {HPQCD}),\ }%
  \bibfield{journal}{%
  \bibinfo {journal} {Phys. Rev. Lett.}\ }%
  \textbf{\bibinfo {volume} {94}},\ \bibinfo {pages} {172001} (\bibinfo {year}
  {2005}),\
  \Eprint{http://arxiv.org/abs/hep-lat/0411027}{arXiv:hep-lat/0411027}%
  \bibAnnoteFile{NoStop}{Allison:2004be}%
\bibitem{Prelovsek:2005rf}%
  \BibitemOpen
  \bibfield{author}{%
  \bibinfo {author} {\bibfnamefont{S.}~\bibnamefont{{Prelov\v{s}ek}}},\ }%
  \bibfield{journal}{%
  \Doi{10.1103/PhysRevD.73.014506}{\bibinfo {journal} {Phys. Rev.}}\ }%
  \textbf{\bibinfo {volume} {D73}},\ \bibinfo {pages} {014506} (\bibinfo {year}
  {2006}),\
  \Eprint{http://arxiv.org/abs/hep-lat/0510080}{arXiv:hep-lat/0510080}%
  \bibAnnoteFile{NoStop}{Prelovsek:2005rf}%
\bibitem{Bernard:2006zw}%
  \BibitemOpen
  \bibfield{author}{%
  \bibinfo {author} {\bibfnamefont{C.}~\bibnamefont{Bernard}},\ }%
  \bibfield{journal}{%
  \Doi{10.1103/PhysRevD.73.114503}{\bibinfo {journal} {Phys. Rev.}}\ }%
  \textbf{\bibinfo {volume} {D73}},\ \bibinfo {pages} {114503} (\bibinfo {year}
  {2006}),\
  \Eprint{http://arxiv.org/abs/hep-lat/0603011}{arXiv:hep-lat/0603011}%
  \bibAnnoteFile{NoStop}{Bernard:2006zw}%
\bibitem{Bernard:2007qf}%
  \BibitemOpen
  \bibfield{author}{%
  \bibinfo {author} {\bibfnamefont{C.}~\bibnamefont{Bernard}}, \bibinfo
  {author} {\bibfnamefont{C.~E.}\ \bibnamefont{DeTar}}, \bibinfo {author}
  {\bibfnamefont{Z.}~\bibnamefont{Fu}},\ and\ \bibinfo {author}
  {\bibfnamefont{S.}~\bibnamefont{{Prelov\v{s}ek}}},\ }%
  \bibfield{journal}{%
  \Doi{10.1103/PhysRevD.76.094504}{\bibinfo {journal} {Phys. Rev.}}\ }%
  \textbf{\bibinfo {volume} {D76}},\ \bibinfo {pages} {094504} (\bibinfo {year}
  {2007}),\ \Eprint{http://arxiv.org/abs/0707.2402}{arXiv:0707.2402 [hep-lat]}%
  \bibAnnoteFile{NoStop}{Bernard:2007qf}%
\bibitem{Aubin:2008wk}%
  \BibitemOpen
  \bibfield{author}{%
  \bibinfo {author} {\bibfnamefont{C.}~\bibnamefont{Aubin}}, \bibinfo {author}
  {\bibfnamefont{J.}~\bibnamefont{Laiho}},\ and\ \bibinfo {author}
  {\bibfnamefont{R.~S.}\ \bibnamefont{Van~de Water}},\ }%
  \bibfield{journal}{%
  \Doi{10.1103/PhysRevD.77.114501}{\bibinfo {journal} {Phys. Rev.}}\ }%
  \textbf{\bibinfo {volume} {D77}},\ \bibinfo {pages} {114501} (\bibinfo {year}
  {2008}),\ \Eprint{http://arxiv.org/abs/0803.0129}{arXiv:0803.0129 [hep-lat]}%
  \bibAnnoteFile{NoStop}{Aubin:2008wk}%
\bibitem{Bernard:2006ee}%
  \BibitemOpen
  \bibfield{author}{%
  \bibinfo {author} {\bibfnamefont{C.}~\bibnamefont{Bernard}}, \bibinfo
  {author} {\bibfnamefont{M.}~\bibnamefont{Golterman}},\ and\ \bibinfo {author}
  {\bibfnamefont{Y.}~\bibnamefont{Shamir}},\ }%
  \bibfield{journal}{%
  \Doi{10.1103/PhysRevD.73.114511}{\bibinfo {journal} {Phys. Rev.}}\ }%
  \textbf{\bibinfo {volume} {D73}},\ \bibinfo {pages} {114511} (\bibinfo {year}
  {2006}),\ \Eprint{http://arxiv.org/abs/hep-lat/0604017}{arXiv:hep-lat/0604017
  [hep-lat]}%
  \bibAnnoteFile{NoStop}{Bernard:2006ee}%
\bibitem{Shamir:2004zc}%
  \BibitemOpen
  \bibfield{author}{%
  \bibinfo {author} {\bibfnamefont{Y.}~\bibnamefont{Shamir}},\ }%
  \bibfield{journal}{%
  \Doi{10.1103/PhysRevD.71.034509}{\bibinfo {journal} {Phys. Rev.}}\ }%
  \textbf{\bibinfo {volume} {D71}},\ \bibinfo {pages} {034509} (\bibinfo {year}
  {2005}),\
  \Eprint{http://arxiv.org/abs/hep-lat/0412014}{arXiv:hep-lat/0412014}%
  \bibAnnoteFile{NoStop}{Shamir:2004zc}%
\bibitem{Shamir:2006nj}%
  \BibitemOpen
  \bibfield{author}{%
  \bibinfo {author} {\bibfnamefont{Y.}~\bibnamefont{Shamir}},\ }%
  \bibfield{journal}{%
  \Doi{10.1103/PhysRevD.75.054503}{\bibinfo {journal} {Phys. Rev.}}\ }%
  \textbf{\bibinfo {volume} {D75}},\ \bibinfo {pages} {054503} (\bibinfo {year}
  {2007}),\
  \Eprint{http://arxiv.org/abs/hep-lat/0607007}{arXiv:hep-lat/0607007}%
  \bibAnnoteFile{NoStop}{Shamir:2006nj}%
\bibitem{Bernard:2007ma}%
  \BibitemOpen
  \bibfield{author}{%
  \bibinfo {author} {\bibfnamefont{C.}~\bibnamefont{Bernard}}, \bibinfo
  {author} {\bibfnamefont{M.}~\bibnamefont{Golterman}},\ and\ \bibinfo {author}
  {\bibfnamefont{Y.}~\bibnamefont{Shamir}},\ }%
  \bibfield{journal}{%
  \Doi{10.1103/PhysRevD.77.074505}{\bibinfo {journal} {Phys. Rev.}}\ }%
  \textbf{\bibinfo {volume} {D77}},\ \bibinfo {pages} {074505} (\bibinfo {year}
  {2008}),\ \Eprint{http://arxiv.org/abs/0712.2560}{arXiv:0712.2560 [hep-lat]}%
  \bibAnnoteFile{NoStop}{Bernard:2007ma}%
\bibitem{Lee:1999zxa}%
  \BibitemOpen
  \bibfield{author}{%
  \bibinfo {author} {\bibfnamefont{W.-J.}\ \bibnamefont{Lee}}\ and\ \bibinfo
  {author} {\bibfnamefont{S.~R.}\ \bibnamefont{Sharpe}},\ }%
  \bibfield{journal}{%
  \Doi{10.1103/PhysRevD.60.114503}{\bibinfo {journal} {Phys. Rev.}}\ }%
  \textbf{\bibinfo {volume} {D60}},\ \bibinfo {pages} {114503} (\bibinfo {year}
  {1999}),\
  \Eprint{http://arxiv.org/abs/hep-lat/9905023}{arXiv:hep-lat/9905023}%
  \bibAnnoteFile{NoStop}{Lee:1999zxa}%
\bibitem{Aubin:2003mg}%
  \BibitemOpen
  \bibfield{author}{%
  \bibinfo {author} {\bibfnamefont{C.}~\bibnamefont{Aubin}}\ and\ \bibinfo
  {author} {\bibfnamefont{C.}~\bibnamefont{Bernard}},\ }%
  \bibfield{journal}{%
  \bibinfo {journal} {Phys. Rev.}\ }%
  \textbf{\bibinfo {volume} {D68}},\ \bibinfo {pages} {034014} (\bibinfo {year}
  {2003}),\ \Eprint{http://arxiv.org/abs/hep-lat/0304014}{hep-lat/0304014}%
  \bibAnnoteFile{NoStop}{Aubin:2003mg}%
\bibitem{Sharpe:2004is}%
  \BibitemOpen
  \bibfield{author}{%
  \bibinfo {author} {\bibfnamefont{S.~R.}\ \bibnamefont{Sharpe}}\ and\ \bibinfo
  {author} {\bibfnamefont{R.~S.}\ \bibnamefont{Van~de Water}},\ }%
  \bibfield{journal}{%
  \bibinfo {journal} {Phys. Rev.}\ }%
  \textbf{\bibinfo {volume} {D71}},\ \bibinfo {pages} {114505} (\bibinfo {year}
  {2005}),\
  \Eprint{http://arxiv.org/abs/hep-lat/0409018}{arXiv:hep-lat/0409018}%
  \bibAnnoteFile{NoStop}{Sharpe:2004is}%
\bibitem{Sharpe:2006re}%
  \BibitemOpen
  \bibfield{author}{%
  \bibinfo {author} {\bibfnamefont{S.~R.}\ \bibnamefont{Sharpe}},\ }%
  \bibfield{journal}{%
  \bibinfo {journal} {PoS}\ }%
  \textbf{\bibinfo {volume} {LAT2006}},\ \bibinfo {pages} {022} (\bibinfo
  {year} {2006}),\
  \Eprint{http://arxiv.org/abs/hep-lat/0610094}{arXiv:hep-lat/0610094}%
  \bibAnnoteFile{NoStop}{Sharpe:2006re}%
\bibitem{Kronfeld:2007ek}%
  \BibitemOpen
  \bibfield{author}{%
  \bibinfo {author} {\bibfnamefont{A.~S.}\ \bibnamefont{Kronfeld}},\ }%
  \bibfield{journal}{%
  \bibinfo {journal} {PoS}\ }%
  \textbf{\bibinfo {volume} {LAT2007}},\ \bibinfo {pages} {016} (\bibinfo
  {year} {2007}),\ \Eprint{http://arxiv.org/abs/0711.0699}{arXiv:0711.0699
  [hep-lat]}%
  \bibAnnoteFile{NoStop}{Kronfeld:2007ek}%
\bibitem{Golterman:2008gt}%
  \BibitemOpen
  \bibfield{author}{%
  \bibinfo {author} {\bibfnamefont{M.}~\bibnamefont{Golterman}},\ }%
  \bibfield{journal}{%
  \bibinfo {journal} {PoS}\ }%
  \textbf{\bibinfo {volume} {CONFINEMENT8}},\ \bibinfo {pages} {014} (\bibinfo
  {year} {2008}),\ \Eprint{http://arxiv.org/abs/0812.3110}{arXiv:0812.3110
  [hep-ph]}%
  \bibAnnoteFile{NoStop}{Golterman:2008gt}%
\bibitem{Bazavov:2009bb}%
  \BibitemOpen
  \bibfield{author}{%
  \bibinfo {author} {\bibfnamefont{A.}~\bibnamefont{Bazavov}} \emph{et~al.},\
  }%
  \bibfield{journal}{%
  \Doi{10.1103/RevModPhys.82.1349}{\bibinfo {journal} {Rev. Mod. Phys.}}\ }%
  \textbf{\bibinfo {volume} {82}},\ \bibinfo {pages} {1349} (\bibinfo {year}
  {2010}),\ \Eprint{http://arxiv.org/abs/0903.3598}{arXiv:0903.3598 [hep-lat]}%
  \bibAnnoteFile{NoStop}{Bazavov:2009bb}%
\bibitem{Donald:2011if}%
  \BibitemOpen
  \bibfield{author}{%
  \bibinfo {author} {\bibfnamefont{G.~C.}\ \bibnamefont{Donald}}, \bibinfo
  {author} {\bibfnamefont{C.~T.~H.}\ \bibnamefont{Davies}}, \bibinfo {author}
  {\bibfnamefont{E.}~\bibnamefont{Follana}},\ and\ \bibinfo {author}
  {\bibfnamefont{A.~S.}\ \bibnamefont{Kronfeld}},\ }%
  \bibfield{journal}{%
  \bibinfo {journal} {Phys. Rev.}\ }%
  \textbf{\bibinfo {volume} {D84}},\ \bibinfo {pages} {054501} (\bibinfo {year}
  {2011}),\ \Eprint{http://arxiv.org/abs/1106.2412}{arXiv:1106.2412 [hep-lat]}%
  \bibAnnoteFile{NoStop}{Donald:2011if}%
\bibitem{Laiho:2009eu}%
  \BibitemOpen
  \bibfield{author}{%
  \bibinfo {author} {\bibfnamefont{J.}~\bibnamefont{Laiho}}, \bibinfo {author}
  {\bibfnamefont{E.}~\bibnamefont{Lunghi}},\ and\ \bibinfo {author}
  {\bibfnamefont{R.~S.}\ \bibnamefont{Van~de Water}},\ }%
  \bibfield{journal}{%
  \Doi{10.1103/PhysRevD.81.034503}{\bibinfo {journal} {Phys. Rev.}}\ }%
  \textbf{\bibinfo {volume} {D81}},\ \bibinfo {pages} {034503} (\bibinfo {year}
  {2010}),\ \Eprint{http://arxiv.org/abs/0910.2928}{arXiv:0910.2928 [hep-ph]}%
  \bibAnnoteFile{NoStop}{Laiho:2009eu}%
\bibitem{Bernard:2001av}%
  \BibitemOpen
  \bibfield{author}{%
  \bibinfo {author} {\bibfnamefont{C.~W.}\ \bibnamefont{Bernard}}
  \emph{et~al.},\ }%
  \bibfield{journal}{%
  \bibinfo {journal} {Phys. Rev.}\ }%
  \textbf{\bibinfo {volume} {D64}},\ \bibinfo {pages} {054506} (\bibinfo {year}
  {2001}),\
  \Eprint{http://arxiv.org/abs/hep-lat/0104002}{arXiv:hep-lat/0104002}%
  \bibAnnoteFile{NoStop}{Bernard:2001av}%
\bibitem{Sheikholeslami:1985ij}%
  \BibitemOpen
  \bibfield{author}{%
  \bibinfo {author} {\bibfnamefont{B.}~\bibnamefont{Sheikholeslami}}\ and\
  \bibinfo {author} {\bibfnamefont{R.}~\bibnamefont{Wohlert}},\ }%
  \bibfield{journal}{%
  \Doi{10.1016/0550-3213(85)90002-1}{\bibinfo {journal} {Nucl. Phys.}}\ }%
  \textbf{\bibinfo {volume} {B259}},\ \bibinfo {pages} {572} (\bibinfo {year}
  {1985})%
  \bibAnnoteFile{NoStop}{Sheikholeslami:1985ij}%
\bibitem{ElKhadra:1996mp}%
  \BibitemOpen
  \bibfield{author}{%
  \bibinfo {author} {\bibfnamefont{A.~X.}\ \bibnamefont{El-Khadra}}, \bibinfo
  {author} {\bibfnamefont{A.~S.}\ \bibnamefont{Kronfeld}},\ and\ \bibinfo
  {author} {\bibfnamefont{P.~B.}\ \bibnamefont{Mackenzie}},\ }%
  \bibfield{journal}{%
  \Doi{10.1103/PhysRevD.55.3933}{\bibinfo {journal} {Phys. Rev.}}\ }%
  \textbf{\bibinfo {volume} {D55}},\ \bibinfo {pages} {3933} (\bibinfo {year}
  {1997}),\
  \Eprint{http://arxiv.org/abs/hep-lat/9604004}{arXiv:hep-lat/9604004}%
  \bibAnnoteFile{NoStop}{ElKhadra:1996mp}%
\bibitem{Symanzik:1983dc}%
  \BibitemOpen
  \bibfield{author}{%
  \bibinfo {author} {\bibfnamefont{K.}~\bibnamefont{Symanzik}},\ }%
  \bibfield{journal}{%
  \Doi{10.1016/0550-3213(83)90468-6}{\bibinfo {journal} {Nucl. Phys.}}\ }%
  \textbf{\bibinfo {volume} {B226}},\ \bibinfo {pages} {187} (\bibinfo {year}
  {1983})%
  \bibAnnoteFile{NoStop}{Symanzik:1983dc}%
\bibitem{Symanzik:1983gh}%
  \BibitemOpen
  \bibfield{author}{%
  \bibinfo {author} {\bibfnamefont{K.}~\bibnamefont{Symanzik}},\ }%
  \bibfield{journal}{%
  \Doi{10.1016/0550-3213(83)90469-8}{\bibinfo {journal} {Nucl. Phys.}}\ }%
  \textbf{\bibinfo {volume} {B226}},\ \bibinfo {pages} {205} (\bibinfo {year}
  {1983})%
  \bibAnnoteFile{NoStop}{Symanzik:1983gh}%
\bibitem{Kronfeld:2000ck}%
  \BibitemOpen
  \bibfield{author}{%
  \bibinfo {author} {\bibfnamefont{A.~S.}\ \bibnamefont{Kronfeld}},\ }%
  \bibfield{journal}{%
  \Doi{10.1103/PhysRevD.62.014505}{\bibinfo {journal} {Phys. Rev.}}\ }%
  \textbf{\bibinfo {volume} {D62}},\ \bibinfo {pages} {014505} (\bibinfo {year}
  {2000}),\
  \Eprint{http://arxiv.org/abs/hep-lat/0002008}{arXiv:hep-lat/0002008}%
  \bibAnnoteFile{NoStop}{Kronfeld:2000ck}%
\bibitem{Harada:2001fi}%
  \BibitemOpen
  \bibfield{author}{%
  \bibinfo {author} {\bibfnamefont{J.}~\bibnamefont{Harada}} \emph{et~al.},\ }%
  \bibfield{journal}{%
  \Doi{10.1103/PhysRevD.65.094513}{\bibinfo {journal} {Phys. Rev.}}\ }%
  \textbf{\bibinfo {volume} {D65}},\ \bibinfo {pages} {094513} (\bibinfo {year}
  {2002}),\ \bibinfo {note} {[Erratum-ibid.\ {\bf D71}, 019903 (2005)]},\
  \Eprint{http://arxiv.org/abs/hep-lat/0112044}{arXiv:hep-lat/0112044}%
  \bibAnnoteFile{NoStop}{Harada:2001fi}%
\bibitem{Harada:2001fj}%
  \BibitemOpen
  \bibfield{author}{%
  \bibinfo {author} {\bibfnamefont{J.}~\bibnamefont{Harada}} \emph{et~al.},\ }%
  \bibfield{journal}{%
  \Doi{10.1103/PhysRevD.65.094514}{\bibinfo {journal} {Phys. Rev.}}\ }%
  \textbf{\bibinfo {volume} {D65}},\ \bibinfo {pages} {094514} (\bibinfo {year}
  {2002}),\
  \Eprint{http://arxiv.org/abs/hep-lat/0112045}{arXiv:hep-lat/0112045}%
  \bibAnnoteFile{NoStop}{Harada:2001fj}%
\bibitem{Aubin:2005aq}%
  \BibitemOpen
  \bibfield{author}{%
  \bibinfo {author} {\bibfnamefont{C.}~\bibnamefont{Aubin}}\ and\ \bibinfo
  {author} {\bibfnamefont{C.}~\bibnamefont{Bernard}},\ }%
  \bibfield{journal}{%
  \Doi{10.1103/PhysRevD.73.014515}{\bibinfo {journal} {Phys. Rev.}}\ }%
  \textbf{\bibinfo {volume} {D73}},\ \bibinfo {pages} {014515} (\bibinfo {year}
  {2006}),\
  \Eprint{http://arxiv.org/abs/hep-lat/0510088}{arXiv:hep-lat/0510088}%
  \bibAnnoteFile{NoStop}{Aubin:2005aq}%
\bibitem{Laiho:2005ue}%
  \BibitemOpen
  \bibfield{author}{%
  \bibinfo {author} {\bibfnamefont{J.}~\bibnamefont{Laiho}}\ and\ \bibinfo
  {author} {\bibfnamefont{R.~S.}\ \bibnamefont{Van~de Water}},\ }%
  \bibfield{journal}{%
  \Doi{10.1103/PhysRevD.73.054501}{\bibinfo {journal} {Phys. Rev.}}\ }%
  \textbf{\bibinfo {volume} {D73}},\ \bibinfo {pages} {054501} (\bibinfo {year}
  {2006}),\
  \Eprint{http://arxiv.org/abs/hep-lat/0512007}{arXiv:hep-lat/0512007}%
  \bibAnnoteFile{NoStop}{Laiho:2005ue}%
\bibitem{Aubin:2007mc}%
  \BibitemOpen
  \bibfield{author}{%
  \bibinfo {author} {\bibfnamefont{C.}~\bibnamefont{Aubin}}\ and\ \bibinfo
  {author} {\bibfnamefont{C.}~\bibnamefont{Bernard}},\ }%
  \bibfield{journal}{%
  \Doi{10.1103/PhysRevD.76.014002}{\bibinfo {journal} {Phys. Rev.}}\ }%
  \textbf{\bibinfo {volume} {D76}},\ \bibinfo {pages} {014002} (\bibinfo {year}
  {2007}),\ \Eprint{http://arxiv.org/abs/0704.0795}{arXiv:0704.0795 [hep-lat]}%
  \bibAnnoteFile{NoStop}{Aubin:2007mc}%
\bibitem{ElKhadra:2001rv}%
  \BibitemOpen
  \bibfield{author}{%
  \bibinfo {author} {\bibfnamefont{A.~X.}\ \bibnamefont{El-Khadra}}, \bibinfo
  {author} {\bibfnamefont{A.~S.}\ \bibnamefont{Kronfeld}}, \bibinfo {author}
  {\bibfnamefont{P.~B.}\ \bibnamefont{Mackenzie}}, \bibinfo {author}
  {\bibfnamefont{S.~M.}\ \bibnamefont{Ryan}},\ and\ \bibinfo {author}
  {\bibfnamefont{J.~N.}\ \bibnamefont{Simone}},\ }%
  \bibfield{journal}{%
  \Doi{10.1103/PhysRevD.64.014502}{\bibinfo {journal} {Phys. Rev.}}\ }%
  \textbf{\bibinfo {volume} {D64}},\ \bibinfo {pages} {014502} (\bibinfo {year}
  {2001}),\ \Eprint{http://arxiv.org/abs/hep-ph/0101023}{arXiv:hep-ph/0101023}%
  \bibAnnoteFile{NoStop}{ElKhadra:2001rv}%
\bibitem{ElKhadra:2007qe}%
  \BibitemOpen
  \bibfield{author}{%
  \bibinfo {author} {\bibfnamefont{A.~X.}\ \bibnamefont{El-Khadra}}, \bibinfo
  {author} {\bibfnamefont{E.}~\bibnamefont{{G\'amiz}}}, \bibinfo {author}
  {\bibfnamefont{A.~S.}\ \bibnamefont{Kronfeld}},\ and\ \bibinfo {author}
  {\bibfnamefont{M.~A.}\ \bibnamefont{Nobes}},\ }%
  \bibfield{journal}{%
  \bibinfo {journal} {PoS}\ }%
  \textbf{\bibinfo {volume} {LAT2007}},\ \bibinfo {pages} {242} (\bibinfo
  {year} {2007}),\ \Eprint{http://arxiv.org/abs/0710.1437}{arXiv:0710.1437
  [hep-lat]}%
  \bibAnnoteFile{NoStop}{ElKhadra:2007qe}%
\bibitem{Aubin:2005ar}%
  \BibitemOpen
  \bibfield{author}{%
  \bibinfo {author} {\bibfnamefont{C.}~\bibnamefont{Aubin}} \emph{et~al.}
  (\bibinfo {collaboration} {Fermilab Lattice and MILC}),\ }%
  \bibfield{journal}{%
  \Doi{10.1103/PhysRevLett.95.122002}{\bibinfo {journal} {Phys. Rev. Lett.}}\
  }%
  \textbf{\bibinfo {volume} {95}},\ \bibinfo {pages} {122002} (\bibinfo {year}
  {2005}),\
  \Eprint{http://arxiv.org/abs/hep-lat/0506030}{arXiv:hep-lat/0506030}%
  \bibAnnoteFile{NoStop}{Aubin:2005ar}%
\bibitem{Bernard:2006zz}%
  \BibitemOpen
  \bibfield{author}{%
  \bibinfo {author} {\bibfnamefont{C.}~\bibnamefont{Bernard}} \emph{et~al.}
  (\bibinfo {collaboration} {Fermilab Lattice and MILC}),\ }%
  \bibfield{journal}{%
  \bibinfo {journal} {PoS}\ }%
  \textbf{\bibinfo {volume} {LAT2006}},\ \bibinfo {pages} {094} (\bibinfo
  {year} {2006})%
  \bibAnnoteFile{NoStop}{Bernard:2006zz}%
\bibitem{Bernard:2007zz}%
  \BibitemOpen
  \bibfield{author}{%
  \bibinfo {author} {\bibfnamefont{C.}~\bibnamefont{Bernard}} \emph{et~al.}
  (\bibinfo {collaboration} {Fermilab Lattice and MILC}),\ }%
  \bibfield{journal}{%
  \bibinfo {journal} {PoS}\ }%
  \textbf{\bibinfo {volume} {LAT2007}},\ \bibinfo {pages} {370} (\bibinfo
  {year} {2007})%
  \bibAnnoteFile{NoStop}{Bernard:2007zz}%
\bibitem{Bernard:2009wr}%
  \BibitemOpen
  \bibfield{author}{%
  \bibinfo {author} {\bibfnamefont{C.}~\bibnamefont{Bernard}} \emph{et~al.}
  (\bibinfo {collaboration} {Fermilab Lattice and MILC}),\ }%
  \bibfield{journal}{%
  \bibinfo {journal} {PoS}\ }%
  \textbf{\bibinfo {volume} {LATTICE2008}},\ \bibinfo {pages} {278} (\bibinfo
  {year} {2008}),\ \Eprint{http://arxiv.org/abs/0904.1895}{arXiv:0904.1895
  [hep-lat]}%
  \bibAnnoteFile{NoStop}{Bernard:2009wr}%
\bibitem{Bazavov:2009ii}%
  \BibitemOpen
  \bibfield{author}{%
  \bibinfo {author} {\bibfnamefont{A.}~\bibnamefont{Bazavov}} \emph{et~al.}
  (\bibinfo {collaboration} {Fermilab Lattice and MILC}),\ }%
  \bibfield{journal}{%
  \bibinfo {journal} {PoS}\ }%
  \textbf{\bibinfo {volume} {LAT2009}},\ \bibinfo {pages} {249} (\bibinfo
  {year} {2009}),\ \Eprint{http://arxiv.org/abs/0912.5221}{arXiv:0912.5221
  [hep-lat]}%
  \bibAnnoteFile{NoStop}{Bazavov:2009ii}%
\bibitem{Simone:2010zz}%
  \BibitemOpen
  \bibfield{author}{%
  \bibinfo {author} {\bibfnamefont{J.}~\bibnamefont{Simone}} \emph{et~al.}
  (\bibinfo {collaboration} {Fermilab Lattice and MILC}),\ }%
  \bibfield{journal}{%
  \bibinfo {journal} {PoS}\ }%
  \textbf{\bibinfo {volume} {LATTICE2010}},\ \bibinfo {pages} {317} (\bibinfo
  {year} {2010})%
  \bibAnnoteFile{NoStop}{Simone:2010zz}%
\bibitem{Rosner:2010ak}%
  \BibitemOpen
  \bibfield{author}{%
  \bibinfo {author} {\bibfnamefont{J.~L.}\ \bibnamefont{Rosner}}\ and\ \bibinfo
  {author} {\bibfnamefont{S.}~\bibnamefont{Stone}}}%
   (\bibinfo {year} {2010}),\ \bibinfo {note} {in
  Ref.~\cite{Nakamura:2010zzi}},\
  \Eprint{http://arxiv.org/abs/1002.1655}{arXiv:1002.1655 [hep-ex]}%
  \bibAnnoteFile{NoStop}{Rosner:2010ak}%
\bibitem{Nakamura:2010zzi}%
  \BibitemOpen
  \bibfield{author}{%
  \bibinfo {author} {\bibfnamefont{K.}~\bibnamefont{Nakamura}} \emph{et~al.}
  (\bibinfo {collaboration} {Particle Data Group}),\ }%
  \bibfield{journal}{%
  \Doi{10.1088/0954-3899/37/7A/075021}{\bibinfo {journal} {J. Phys.}}\ }%
  \textbf{\bibinfo {volume} {G37}},\ \bibinfo {pages} {075021} (\bibinfo {year}
  {2010}),\ \bibinfo {note} {and 2011 partial update for the 2012 edition}%
  \bibAnnoteFile{NoStop}{Nakamura:2010zzi}%
\bibitem{Aubin:2004ej}%
  \BibitemOpen
  \bibfield{author}{%
  \bibinfo {author} {\bibfnamefont{C.}~\bibnamefont{Aubin}} \emph{et~al.}
  (\bibinfo {collaboration} {Fermilab Lattice and MILC}),\ }%
  \bibfield{journal}{%
  \Doi{10.1103/PhysRevLett.94.011601}{\bibinfo {journal} {Phys. Rev. Lett.}}\
  }%
  \textbf{\bibinfo {volume} {94}},\ \bibinfo {pages} {011601} (\bibinfo {year}
  {2005}),\ \Eprint{http://arxiv.org/abs/hep-ph/0408306}{arXiv:hep-ph/0408306}%
  \bibAnnoteFile{NoStop}{Aubin:2004ej}%
\bibitem{Bernard:2008dn}%
  \BibitemOpen
  \bibfield{author}{%
  \bibinfo {author} {\bibfnamefont{C.}~\bibnamefont{Bernard}} \emph{et~al.}
  (\bibinfo {collaboration} {Fermilab Lattice and MILC}),\ }%
  \bibfield{journal}{%
  \Doi{10.1103/PhysRevD.79.014506}{\bibinfo {journal} {Phys. Rev.}}\ }%
  \textbf{\bibinfo {volume} {D79}},\ \bibinfo {pages} {014506} (\bibinfo {year}
  {2009}),\ \Eprint{http://arxiv.org/abs/0808.2519}{arXiv:0808.2519 [hep-lat]}%
  \bibAnnoteFile{NoStop}{Bernard:2008dn}%
\bibitem{Bailey:2008wp}%
  \BibitemOpen
  \bibfield{author}{%
  \bibinfo {author} {\bibfnamefont{J.~A.}\ \bibnamefont{Bailey}} \emph{et~al.}
  (\bibinfo {collaboration} {Fermilab Lattice and MILC}),\ }%
  \bibfield{journal}{%
  \Doi{10.1103/PhysRevD.79.054507}{\bibinfo {journal} {Phys. Rev.}}\ }%
  \textbf{\bibinfo {volume} {D79}},\ \bibinfo {pages} {054507} (\bibinfo {year}
  {2009}),\ \Eprint{http://arxiv.org/abs/0811.3640}{arXiv:0811.3640 [hep-lat]}%
  \bibAnnoteFile{NoStop}{Bailey:2008wp}%
\bibitem{Bernard:2010fr}%
  \BibitemOpen
  \bibfield{author}{%
  \bibinfo {author} {\bibfnamefont{C.}~\bibnamefont{Bernard}} \emph{et~al.}
  (\bibinfo {collaboration} {Fermilab Lattice and MILC}),\ }%
  \bibfield{journal}{%
  \Doi{10.1103/PhysRevD.83.034503}{\bibinfo {journal} {Phys. Rev.}}\ }%
  \textbf{\bibinfo {volume} {D83}},\ \bibinfo {pages} {034503} (\bibinfo {year}
  {2011}),\ \Eprint{http://arxiv.org/abs/1003.1937}{arXiv:1003.1937 [hep-lat]}%
  \bibAnnoteFile{NoStop}{Bernard:2010fr}%
\bibitem{Lepage:1998vj}%
  \BibitemOpen
  \bibfield{author}{%
  \bibinfo {author} {\bibfnamefont{G.~P.}\ \bibnamefont{Lepage}},\ }%
  \bibfield{journal}{%
  \Doi{10.1103/PhysRevD.59.074502}{\bibinfo {journal} {Phys. Rev.}}\ }%
  \textbf{\bibinfo {volume} {D59}},\ \bibinfo {pages} {074502} (\bibinfo {year}
  {1999}),\
  \Eprint{http://arxiv.org/abs/hep-lat/9809157}{arXiv:hep-lat/9809157}%
  \bibAnnoteFile{NoStop}{Lepage:1998vj}%
\bibitem{Luscher:1985zq}%
  \BibitemOpen
  \bibfield{author}{%
  \bibinfo {author} {\bibfnamefont{M.}~\bibnamefont{{L\"uscher}}}\ and\
  \bibinfo {author} {\bibfnamefont{P.}~\bibnamefont{Weisz}},\ }%
  \bibfield{journal}{%
  \Doi{10.1016/0370-2693(85)90966-9}{\bibinfo {journal} {Phys. Lett.}}\ }%
  \textbf{\bibinfo {volume} {B158}},\ \bibinfo {pages} {250} (\bibinfo {year}
  {1985})%
  \bibAnnoteFile{NoStop}{Luscher:1985zq}%
\bibitem{Hao:2007iz}%
  \BibitemOpen
  \bibfield{author}{%
  \bibinfo {author} {\bibfnamefont{Z.}~\bibnamefont{Hao}}, \bibinfo {author}
  {\bibfnamefont{G.~M.}\ \bibnamefont{von Hippel}}, \bibinfo {author}
  {\bibfnamefont{R.~R.}\ \bibnamefont{Horgan}}, \bibinfo {author}
  {\bibfnamefont{Q.~J.}\ \bibnamefont{Mason}},\ and\ \bibinfo {author}
  {\bibfnamefont{H.~D.}\ \bibnamefont{Trottier}},\ }%
  \bibfield{journal}{%
  \Doi{10.1103/PhysRevD.76.034507}{\bibinfo {journal} {Phys. Rev.}}\ }%
  \textbf{\bibinfo {volume} {D76}},\ \bibinfo {pages} {034507} (\bibinfo {year}
  {2007}),\ \Eprint{http://arxiv.org/abs/0705.4660}{arXiv:0705.4660 [hep-lat]}%
  \bibAnnoteFile{NoStop}{Hao:2007iz}%
\bibitem{Wilson:1975id}%
  \BibitemOpen
  \bibfield{author}{%
  \bibinfo {author} {\bibfnamefont{K.~G.}\ \bibnamefont{Wilson}},\ }%
  \enquote{\bibinfo {title} {{Quarks and Strings on a Lattice}},}\ in\
  \emph{\bibinfo {booktitle} {New Phenomena In Subnuclear Physics}},\ \bibinfo
  {editor} {edited by\ \bibinfo {editor}
  {\bibfnamefont{A.}~\bibnamefont{Zichichi}}}\ (\bibinfo {publisher} {Plenum},\
  \bibinfo {address} {New York},\ \bibinfo {year} {1975})\ p.~\bibinfo {pages}
  {69},\ \bibinfo {note} {cLNS-321}%
  \bibAnnoteFile{NoStop}{Wilson:1975id}%
\bibitem{Oktay:2008ex}%
  \BibitemOpen
  \bibfield{author}{%
  \bibinfo {author} {\bibfnamefont{M.~B.}\ \bibnamefont{Oktay}}\ and\ \bibinfo
  {author} {\bibfnamefont{A.~S.}\ \bibnamefont{Kronfeld}},\ }%
  \bibfield{journal}{%
  \Doi{10.1103/PhysRevD.78.014504}{\bibinfo {journal} {Phys. Rev.}}\ }%
  \textbf{\bibinfo {volume} {D78}},\ \bibinfo {pages} {014504} (\bibinfo {year}
  {2008}),\ \Eprint{http://arxiv.org/abs/0803.0523}{arXiv:0803.0523 [hep-lat]}%
  \bibAnnoteFile{NoStop}{Oktay:2008ex}%
\bibitem{Lin:2006ur}%
  \BibitemOpen
  \bibfield{author}{%
  \bibinfo {author} {\bibfnamefont{H.-W.}\ \bibnamefont{Lin}}\ and\ \bibinfo
  {author} {\bibfnamefont{N.}~\bibnamefont{Christ}},\ }%
  \bibfield{journal}{%
  \Doi{10.1103/PhysRevD.76.074506}{\bibinfo {journal} {Phys. Rev.}}\ }%
  \textbf{\bibinfo {volume} {D76}},\ \bibinfo {pages} {074506} (\bibinfo {year}
  {2007}),\
  \Eprint{http://arxiv.org/abs/hep-lat/0608005}{arXiv:hep-lat/0608005}%
  \bibAnnoteFile{NoStop}{Lin:2006ur}%
\bibitem{Lepage:1992tx}%
  \BibitemOpen
  \bibfield{author}{%
  \bibinfo {author} {\bibfnamefont{G.~P.}\ \bibnamefont{Lepage}}, \bibinfo
  {author} {\bibfnamefont{L.}~\bibnamefont{Magnea}}, \bibinfo {author}
  {\bibfnamefont{C.}~\bibnamefont{Nakhleh}}, \bibinfo {author}
  {\bibfnamefont{U.}~\bibnamefont{Magnea}},\ and\ \bibinfo {author}
  {\bibfnamefont{K.}~\bibnamefont{Hornbostel}},\ }%
  \bibfield{journal}{%
  \Doi{10.1103/PhysRevD.46.4052}{\bibinfo {journal} {Phys. Rev.}}\ }%
  \textbf{\bibinfo {volume} {D46}},\ \bibinfo {pages} {4052} (\bibinfo {year}
  {1992}),\
  \Eprint{http://arxiv.org/abs/hep-lat/9205007}{arXiv:hep-lat/9205007}%
  \bibAnnoteFile{NoStop}{Lepage:1992tx}%
\bibitem{Sommer:1993ce}%
  \BibitemOpen
  \bibfield{author}{%
  \bibinfo {author} {\bibfnamefont{R.}~\bibnamefont{Sommer}},\ }%
  \bibfield{journal}{%
  \Doi{10.1016/0550-3213(94)90473-1}{\bibinfo {journal} {Nucl. Phys.}}\ }%
  \textbf{\bibinfo {volume} {B411}},\ \bibinfo {pages} {839} (\bibinfo {year}
  {1994}),\
  \Eprint{http://arxiv.org/abs/hep-lat/9310022}{arXiv:hep-lat/9310022}%
  \bibAnnoteFile{NoStop}{Sommer:1993ce}%
\bibitem{Bernard:2000gd}%
  \BibitemOpen
  \bibfield{author}{%
  \bibinfo {author} {\bibfnamefont{C.~W.}\ \bibnamefont{Bernard}}
  \emph{et~al.},\ }%
  \bibfield{journal}{%
  \Doi{10.1103/PhysRevD.62.034503}{\bibinfo {journal} {Phys. Rev.}}\ }%
  \textbf{\bibinfo {volume} {D62}},\ \bibinfo {pages} {034503} (\bibinfo {year}
  {2000}),\
  \Eprint{http://arxiv.org/abs/hep-lat/0002028}{arXiv:hep-lat/0002028}%
  \bibAnnoteFile{NoStop}{Bernard:2000gd}%
\bibitem{Allton:1996kr}%
  \BibitemOpen
  \bibfield{author}{%
  \bibinfo {author} {\bibfnamefont{C.~R.}\ \bibnamefont{Allton}}}%
   (\bibinfo {year} {1996}),\
  \Eprint{http://arxiv.org/abs/hep-lat/9610016}{hep-lat/9610016}%
  \bibAnnoteFile{NoStop}{Allton:1996kr}%
\bibitem{Bazavov:2009fk}%
  \BibitemOpen
  \bibfield{author}{%
  \bibinfo {author} {\bibfnamefont{A.}~\bibnamefont{Bazavov}} \emph{et~al.}
  (\bibinfo {collaboration} {MILC}),\ }%
  \bibfield{journal}{%
  \bibinfo {journal} {PoS}\ }%
  \textbf{\bibinfo {volume} {CD09}},\ \bibinfo {pages} {007} (\bibinfo {year}
  {2009}),\ \Eprint{http://arxiv.org/abs/0910.2966}{arXiv:0910.2966 [hep-ph]}%
  \bibAnnoteFile{NoStop}{Bazavov:2009fk}%
\bibitem{Davies:2009tsa}%
  \BibitemOpen
  \bibfield{author}{%
  \bibinfo {author} {\bibfnamefont{C.~T.~H.}\ \bibnamefont{Davies}}, \bibinfo
  {author} {\bibfnamefont{E.}~\bibnamefont{Follana}}, \bibinfo {author}
  {\bibfnamefont{I.~D.}\ \bibnamefont{Kendall}}, \bibinfo {author}
  {\bibfnamefont{G.~P.}\ \bibnamefont{Lepage}},\ and\ \bibinfo {author}
  {\bibfnamefont{C.}~\bibnamefont{McNeile}} (\bibinfo {collaboration}
  {HPQCD}),\ }%
  \bibfield{journal}{%
  \Doi{10.1103/PhysRevD.81.034506}{\bibinfo {journal} {Phys. Rev.}}\ }%
  \textbf{\bibinfo {volume} {D81}},\ \bibinfo {pages} {034506} (\bibinfo {year}
  {2010}),\ \Eprint{http://arxiv.org/abs/0910.1229}{arXiv:0910.1229 [hep-lat]}%
  \bibAnnoteFile{NoStop}{Davies:2009tsa}%
\bibitem{Lepage:2001ym}%
  \BibitemOpen
  \bibfield{author}{%
  \bibinfo {author} {\bibfnamefont{G.~P.}\ \bibnamefont{Lepage}}
  \emph{et~al.},\ }%
  \bibfield{journal}{%
  \Doi{10.1016/S0920-5632(01)01638-3}{\bibinfo {journal} {Nucl. Phys. Proc.
  Suppl.}}\ }%
  \textbf{\bibinfo {volume} {106}},\ \bibinfo {pages} {12} (\bibinfo {year}
  {2002}),\
  \Eprint{http://arxiv.org/abs/hep-lat/0110175}{arXiv:hep-lat/0110175}%
  \bibAnnoteFile{NoStop}{Lepage:2001ym}%
\bibitem{Morningstar:2001je}%
  \BibitemOpen
  \bibfield{author}{%
  \bibinfo {author} {\bibfnamefont{C.}~\bibnamefont{Morningstar}},\ }%
  \bibfield{journal}{%
  \bibinfo {journal} {Nucl. Phys. Proc. Suppl.}\ }%
  \textbf{\bibinfo {volume} {109A}},\ \bibinfo {pages} {185} (\bibinfo {year}
  {2002}),\
  \Eprint{http://arxiv.org/abs/hep-lat/0112023}{arXiv:hep-lat/0112023}%
  \bibAnnoteFile{NoStop}{Morningstar:2001je}%
\bibitem{Kronfeld:1996uy}%
  \BibitemOpen
  \bibfield{author}{%
  \bibinfo {author} {\bibfnamefont{A.~S.}\ \bibnamefont{Kronfeld}},\ }%
  \bibfield{journal}{%
  \Doi{10.1016/S0920-5632(96)00671-8}{\bibinfo {journal} {Nucl. Phys. Proc.
  Suppl.}}\ }%
  \textbf{\bibinfo {volume} {53}},\ \bibinfo {pages} {401} (\bibinfo {year}
  {1997}),\
  \Eprint{http://arxiv.org/abs/hep-lat/9608139}{arXiv:hep-lat/9608139}%
  \bibAnnoteFile{NoStop}{Kronfeld:1996uy}%
\bibitem{Kawamoto:1981hw}%
  \BibitemOpen
  \bibfield{author}{%
  \bibinfo {author} {\bibfnamefont{N.}~\bibnamefont{Kawamoto}}\ and\ \bibinfo
  {author} {\bibfnamefont{J.}~\bibnamefont{Smit}},\ }%
  \bibfield{journal}{%
  \Doi{10.1016/0550-3213(81)90196-6}{\bibinfo {journal} {Nucl. Phys.}}\ }%
  \textbf{\bibinfo {volume} {B192}},\ \bibinfo {pages} {100} (\bibinfo {year}
  {1981})%
  \bibAnnoteFile{NoStop}{Kawamoto:1981hw}%
\bibitem{Richardson:1978bt}%
  \BibitemOpen
  \bibfield{author}{%
  \bibinfo {author} {\bibfnamefont{J.~L.}\ \bibnamefont{Richardson}},\ }%
  \bibfield{journal}{%
  \Doi{10.1016/0370-2693(79)90753-6}{\bibinfo {journal} {Phys. Lett.}}\ }%
  \textbf{\bibinfo {volume} {B82}},\ \bibinfo {pages} {272} (\bibinfo {year}
  {1979})%
  \bibAnnoteFile{NoStop}{Richardson:1978bt}%
\bibitem{Menscher:2005kj}%
  \BibitemOpen
  \bibfield{author}{%
  \bibinfo {author} {\bibfnamefont{D.~P.}\ \bibnamefont{Menscher}},\ }%
  \enquote{\bibinfo {title} {{Charmonium and Charmed Mesons with Improved
  Lattice QCD}},}\  (\bibinfo {year} {2005}),\ \bibinfo {note} {{PhD} thesis}%
  \bibAnnoteFile{NoStop}{Menscher:2005kj}%
\bibitem{Wingate:2002fh}%
  \BibitemOpen
  \bibfield{author}{%
  \bibinfo {author} {\bibfnamefont{M.}~\bibnamefont{Wingate}}, \bibinfo
  {author} {\bibfnamefont{J.}~\bibnamefont{Shigemitsu}}, \bibinfo {author}
  {\bibfnamefont{C.~T.~H.}\ \bibnamefont{Davies}}, \bibinfo {author}
  {\bibfnamefont{G.~P.}\ \bibnamefont{Lepage}},\ and\ \bibinfo {author}
  {\bibfnamefont{H.~D.}\ \bibnamefont{Trottier}},\ }%
  \bibfield{journal}{%
  \Doi{10.1103/PhysRevD.67.054505}{\bibinfo {journal} {Phys. Rev.}}\ }%
  \textbf{\bibinfo {volume} {D67}},\ \bibinfo {pages} {054505} (\bibinfo {year}
  {2003}),\
  \Eprint{http://arxiv.org/abs/hep-lat/0211014}{arXiv:hep-lat/0211014}%
  \bibAnnoteFile{NoStop}{Wingate:2002fh}%
\bibitem{Pierro:2001uq}%
  \BibitemOpen
  \bibfield{author}{%
  \bibinfo {author} {\bibfnamefont{M.~D.}\ \bibnamefont{Pierro}}\ and\ \bibinfo
  {author} {\bibfnamefont{E.}~\bibnamefont{Eichten}},\ }%
  \bibfield{journal}{%
  \bibinfo {journal} {Phys. Rev.}\ }%
  \textbf{\bibinfo {volume} {D64}},\ \bibinfo {pages} {114004} (\bibinfo {year}
  {2001}),\ \Eprint{http://arxiv.org/abs/hep-ph/0104208}{hep-ph/0104208},\
  \url{http://arxiv.org/abs/hep-ph/0104208}%
  \bibAnnoteFile{NoStop}{Pierro:2001uq}%
\bibitem{Lepage:1992xa}%
  \BibitemOpen
  \bibfield{author}{%
  \bibinfo {author} {\bibfnamefont{G.~P.}\ \bibnamefont{Lepage}}\ and\ \bibinfo
  {author} {\bibfnamefont{P.~B.}\ \bibnamefont{Mackenzie}},\ }%
  \bibfield{journal}{%
  \Doi{10.1103/PhysRevD.48.2250}{\bibinfo {journal} {Phys. Rev.}}\ }%
  \textbf{\bibinfo {volume} {D48}},\ \bibinfo {pages} {2250} (\bibinfo {year}
  {1993}),\
  \Eprint{http://arxiv.org/abs/hep-lat/9209022}{arXiv:hep-lat/9209022}%
  \bibAnnoteFile{NoStop}{Lepage:1992xa}%
\bibitem{Mason:2005zx}%
  \BibitemOpen
  \bibfield{author}{%
  \bibinfo {author} {\bibfnamefont{Q.}~\bibnamefont{Mason}} \emph{et~al.}
  (\bibinfo {collaboration} {HPQCD}),\ }%
  \bibfield{journal}{%
  \Doi{10.1103/PhysRevLett.95.052002}{\bibinfo {journal} {Phys. Rev. Lett.}}\
  }%
  \textbf{\bibinfo {volume} {95}},\ \bibinfo {pages} {052002} (\bibinfo {year}
  {2005}),\
  \Eprint{http://arxiv.org/abs/hep-lat/0503005}{arXiv:hep-lat/0503005}%
  \bibAnnoteFile{NoStop}{Mason:2005zx}%
\bibitem{Hornbostel:2002af}%
  \BibitemOpen
  \bibfield{author}{%
  \bibinfo {author} {\bibfnamefont{K.}~\bibnamefont{Hornbostel}}, \bibinfo
  {author} {\bibfnamefont{G.~P.}\ \bibnamefont{Lepage}},\ and\ \bibinfo
  {author} {\bibfnamefont{C.}~\bibnamefont{Morningstar}},\ }%
  \bibfield{journal}{%
  \Doi{10.1103/PhysRevD.67.034023}{\bibinfo {journal} {Phys. Rev.}}\ }%
  \textbf{\bibinfo {volume} {D67}},\ \bibinfo {pages} {034023} (\bibinfo {year}
  {2003}),\ \Eprint{http://arxiv.org/abs/hep-ph/0208224}{arXiv:hep-ph/0208224}%
  \bibAnnoteFile{NoStop}{Hornbostel:2002af}%
\bibitem{Arndt:2004bg}%
  \BibitemOpen
  \bibfield{author}{%
  \bibinfo {author} {\bibfnamefont{D.}~\bibnamefont{Arndt}}\ and\ \bibinfo
  {author} {\bibfnamefont{C.~J.~D.}\ \bibnamefont{Lin}},\ }%
  \bibfield{journal}{%
  \Doi{10.1103/PhysRevD.70.014503}{\bibinfo {journal} {Phys. Rev.}}\ }%
  \textbf{\bibinfo {volume} {D70}},\ \bibinfo {pages} {014503} (\bibinfo {year}
  {2004}),\
  \Eprint{http://arxiv.org/abs/hep-lat/0403012}{arXiv:hep-lat/0403012}%
  \bibAnnoteFile{NoStop}{Arndt:2004bg}%
\bibitem{Bernard:2001yj}%
  \BibitemOpen
  \bibfield{author}{%
  \bibinfo {author} {\bibfnamefont{C.}~\bibnamefont{Bernard}} (\bibinfo
  {collaboration} {MILC}),\ }%
  \bibfield{journal}{%
  \Doi{10.1103/PhysRevD.65.054031}{\bibinfo {journal} {Phys. Rev.}}\ }%
  \textbf{\bibinfo {volume} {D65}},\ \bibinfo {pages} {054031} (\bibinfo {year}
  {2002}),\
  \Eprint{http://arxiv.org/abs/hep-lat/0111051}{arXiv:hep-lat/0111051}%
  \bibAnnoteFile{NoStop}{Bernard:2001yj}%
\bibitem{Kronfeld:2002ab}%
  \BibitemOpen
  \bibfield{author}{%
  \bibinfo {author} {\bibfnamefont{A.~S.}\ \bibnamefont{Kronfeld}}\ and\
  \bibinfo {author} {\bibfnamefont{S.~M.}\ \bibnamefont{Ryan}},\ }%
  \bibfield{journal}{%
  \Doi{10.1016/S0370-2693(02)02407-3}{\bibinfo {journal} {Phys. Lett.}}\ }%
  \textbf{\bibinfo {volume} {B543}},\ \bibinfo {pages} {59} (\bibinfo {year}
  {2002}),\ \Eprint{http://arxiv.org/abs/hep-ph/0206058}{arXiv:hep-ph/0206058}%
  \bibAnnoteFile{NoStop}{Kronfeld:2002ab}%
\bibitem{Boyd:1994pa}%
  \BibitemOpen
  \bibfield{author}{%
  \bibinfo {author} {\bibfnamefont{C.~G.}\ \bibnamefont{Boyd}}\ and\ \bibinfo
  {author} {\bibfnamefont{B.}~\bibnamefont{Grinstein}},\ }%
  \bibfield{journal}{%
  \Doi{10.1016/S0550-3213(95)00005-4}{\bibinfo {journal} {Nucl. Phys.}}\ }%
  \textbf{\bibinfo {volume} {B442}},\ \bibinfo {pages} {205} (\bibinfo {year}
  {1995}),\ \Eprint{http://arxiv.org/abs/hep-ph/9402340}{arXiv:hep-ph/9402340}%
  \bibAnnoteFile{NoStop}{Boyd:1994pa}%
\bibitem{Stewart:1998ke}%
  \BibitemOpen
  \bibfield{author}{%
  \bibinfo {author} {\bibfnamefont{I.~W.}\ \bibnamefont{Stewart}},\ }%
  \bibfield{journal}{%
  \Doi{10.1016/S0550-3213(98)00374-5}{\bibinfo {journal} {Nucl. Phys.}}\ }%
  \textbf{\bibinfo {volume} {B529}},\ \bibinfo {pages} {62} (\bibinfo {year}
  {1998}),\ \Eprint{http://arxiv.org/abs/hep-ph/9803227}{arXiv:hep-ph/9803227}%
  \bibAnnoteFile{NoStop}{Stewart:1998ke}%
\bibitem{Becirevic:2003ad}%
  \BibitemOpen
  \bibfield{author}{%
  \bibinfo {author} {\bibfnamefont{D.}~\bibnamefont{{Be\'cirevi\'c}}}, \bibinfo
  {author} {\bibfnamefont{S.}~\bibnamefont{{Prelov\v{s}ek}}},\ and\ \bibinfo
  {author} {\bibfnamefont{J.}~\bibnamefont{Zupan}},\ }%
  \bibfield{journal}{%
  \Doi{10.1103/PhysRevD.68.074003}{\bibinfo {journal} {Phys. Rev.}}\ }%
  \textbf{\bibinfo {volume} {D68}},\ \bibinfo {pages} {074003} (\bibinfo {year}
  {2003}),\
  \Eprint{http://arxiv.org/abs/hep-lat/0305001}{arXiv:hep-lat/0305001}%
  \bibAnnoteFile{NoStop}{Becirevic:2003ad}%
\bibitem{Bernard:2007ps}%
  \BibitemOpen
  \bibfield{author}{%
  \bibinfo {author} {\bibfnamefont{C.}~\bibnamefont{Bernard}} \emph{et~al.}
  (\bibinfo {collaboration} {MILC}),\ }%
  \bibfield{journal}{%
  \bibinfo {journal} {PoS}\ }%
  \textbf{\bibinfo {volume} {LATTICE2007}},\ \bibinfo {pages} {090} (\bibinfo
  {year} {2007}),\ \Eprint{http://arxiv.org/abs/0710.1118}{arXiv:0710.1118
  [hep-lat]}%
  \bibAnnoteFile{NoStop}{Bernard:2007ps}%
\bibitem{Casalbuoni:1996pg}%
  \BibitemOpen
  \bibfield{author}{%
  \bibinfo {author} {\bibfnamefont{R.}~\bibnamefont{Casalbuoni}}
  \emph{et~al.},\ }%
  \bibfield{journal}{%
  \Doi{10.1016/S0370-1573(96)00027-0}{\bibinfo {journal} {Phys. Rept.}}\ }%
  \textbf{\bibinfo {volume} {281}},\ \bibinfo {pages} {145} (\bibinfo {year}
  {1997}),\ \Eprint{http://arxiv.org/abs/hep-ph/9605342}{arXiv:hep-ph/9605342
  [hep-ph]}%
  \bibAnnoteFile{NoStop}{Casalbuoni:1996pg}%
\bibitem{Anastassov:2001cw}%
  \BibitemOpen
  \bibfield{author}{%
  \bibinfo {author} {\bibfnamefont{A.}~\bibnamefont{Anastassov}} \emph{et~al.}
  (\bibinfo {collaboration} {CLEO}),\ }%
  \bibfield{journal}{%
  \Doi{10.1103/PhysRevD.65.032003}{\bibinfo {journal} {Phys. Rev.}}\ }%
  \textbf{\bibinfo {volume} {D65}},\ \bibinfo {pages} {032003} (\bibinfo {year}
  {2002}),\ \Eprint{http://arxiv.org/abs/hep-ex/0108043}{arXiv:hep-ex/0108043}%
  \bibAnnoteFile{NoStop}{Anastassov:2001cw}%
\bibitem{Abada:2002vj}%
  \BibitemOpen
  \bibfield{author}{%
  \bibinfo {author} {\bibfnamefont{A.}~\bibnamefont{Abada}} \emph{et~al.},\ }%
  \bibfield{journal}{%
  \Doi{10.1016/S0920-5632(03)01611-6}{\bibinfo {journal} {Nucl. Phys. Proc.
  Suppl.}}\ }%
  \textbf{\bibinfo {volume} {119}},\ \bibinfo {pages} {641} (\bibinfo {year}
  {2003}),\
  \Eprint{http://arxiv.org/abs/hep-lat/0209092}{arXiv:hep-lat/0209092}%
  \bibAnnoteFile{NoStop}{Abada:2002vj}%
\bibitem{Arnesen:2005ez}%
  \BibitemOpen
  \bibfield{author}{%
  \bibinfo {author} {\bibfnamefont{M.~C.}\ \bibnamefont{Arnesen}}, \bibinfo
  {author} {\bibfnamefont{B.}~\bibnamefont{Grinstein}}, \bibinfo {author}
  {\bibfnamefont{I.~Z.}\ \bibnamefont{Rothstein}},\ and\ \bibinfo {author}
  {\bibfnamefont{I.~W.}\ \bibnamefont{Stewart}},\ }%
  \bibfield{journal}{%
  \Doi{10.1103/PhysRevLett.95.071802}{\bibinfo {journal} {Phys. Rev. Lett.}}\
  }%
  \textbf{\bibinfo {volume} {95}},\ \bibinfo {pages} {071802} (\bibinfo {year}
  {2005}),\ \Eprint{http://arxiv.org/abs/hep-ph/0504209}{arXiv:hep-ph/0504209}%
  \bibAnnoteFile{NoStop}{Arnesen:2005ez}%
\bibitem{Ohki:2008py}%
  \BibitemOpen
  \bibfield{author}{%
  \bibinfo {author} {\bibfnamefont{H.}~\bibnamefont{Ohki}}, \bibinfo {author}
  {\bibfnamefont{H.}~\bibnamefont{Matsufuru}},\ and\ \bibinfo {author}
  {\bibfnamefont{T.}~\bibnamefont{Onogi}},\ }%
  \bibfield{journal}{%
  \Doi{10.1103/PhysRevD.77.094509}{\bibinfo {journal} {Phys. Rev.}}\ }%
  \textbf{\bibinfo {volume} {D77}},\ \bibinfo {pages} {094509} (\bibinfo {year}
  {2008}),\ \Eprint{http://arxiv.org/abs/0802.1563}{arXiv:0802.1563 [hep-lat]}%
  \bibAnnoteFile{NoStop}{Ohki:2008py}%
\bibitem{Bulava:2010ej}%
  \BibitemOpen
  \bibfield{author}{%
  \bibinfo {author} {\bibfnamefont{J.}~\bibnamefont{Bulava}}, \bibinfo {author}
  {\bibfnamefont{M.~A.}\ \bibnamefont{Donnellan}},\ and\ \bibinfo {author}
  {\bibfnamefont{R.}~\bibnamefont{Sommer}} (\bibinfo {collaboration} {ALPHA}),\
  }%
  \bibfield{journal}{%
  \bibinfo {journal} {PoS}\ }%
  \textbf{\bibinfo {volume} {LATTICE2010}},\ \bibinfo {pages} {303} (\bibinfo
  {year} {2010}),\ \Eprint{http://arxiv.org/abs/1011.4393}{arXiv:1011.4393
  [hep-lat]}%
  \bibAnnoteFile{NoStop}{Bulava:2010ej}%
\bibitem{Aubin:2004ck}%
  \BibitemOpen
  \bibfield{author}{%
  \bibinfo {author} {\bibfnamefont{C.}~\bibnamefont{Aubin}} \emph{et~al.}
  (\bibinfo {collaboration} {HPQCD}),\ }%
  \bibfield{journal}{%
  \Doi{10.1103/PhysRevD.70.031504}{\bibinfo {journal} {Phys. Rev.}}\ }%
  \textbf{\bibinfo {volume} {D70}},\ \bibinfo {pages} {031504} (\bibinfo {year}
  {2004}),\
  \Eprint{http://arxiv.org/abs/hep-lat/0405022}{arXiv:hep-lat/0405022}%
  \bibAnnoteFile{NoStop}{Aubin:2004ck}%
\bibitem{Bernard:2002pc}%
  \BibitemOpen
  \bibfield{author}{%
  \bibinfo {author} {\bibfnamefont{C.}~\bibnamefont{Bernard}} \emph{et~al.}
  (\bibinfo {collaboration} {MILC}),\ }%
  \bibfield{journal}{%
  \Doi{10.1103/PhysRevD.66.094501}{\bibinfo {journal} {Phys. Rev.}}\ }%
  \textbf{\bibinfo {volume} {D66}},\ \bibinfo {pages} {094501} (\bibinfo {year}
  {2002}),\
  \Eprint{http://arxiv.org/abs/hep-lat/0206016}{arXiv:hep-lat/0206016}%
  \bibAnnoteFile{NoStop}{Bernard:2002pc}%
\bibitem{Lepage:TASI}%
  \BibitemOpen
  \bibfield{author}{%
  \bibinfo {author} {\bibfnamefont{G.~P.}\ \bibnamefont{Lepage}},\ }%
  in\ \emph{\bibinfo {booktitle} {{From actions to answers. Proceedings,
  Theoretical Advanced Study Institute in Elementary Particle Physics, Boulder,
  USA, June 5-30, 1989}}},\ \bibinfo {editor} {edited by\ \bibinfo {editor}
  {\bibfnamefont{T.~A.}\ \bibnamefont{DeGrand}}\ and\ \bibinfo {editor}
  {\bibfnamefont{D.}~\bibnamefont{Toussaint}}}\ (\bibinfo {publisher} {World
  Scientific},\ \bibinfo {year} {1989})%
  \bibAnnoteFile{NoStop}{Lepage:TASI}%
\bibitem{Davies:2010ip}%
  \BibitemOpen
  \bibfield{author}{%
  \bibinfo {author} {\bibfnamefont{C.~T.~H.}\ \bibnamefont{Davies}}, \bibinfo
  {author} {\bibfnamefont{C.}~\bibnamefont{McNeile}}, \bibinfo {author}
  {\bibfnamefont{E.}~\bibnamefont{Follana}}, \bibinfo {author}
  {\bibfnamefont{G.~P.}\ \bibnamefont{Lepage}}, \bibinfo {author}
  {\bibfnamefont{H.}~\bibnamefont{Na}}, \emph{et~al.} (\bibinfo {collaboration}
  {HPQCD}),\ }%
  \bibfield{journal}{%
  \Doi{10.1103/PhysRevD.82.114504}{\bibinfo {journal} {Phys. Rev.}}\ }%
  \textbf{\bibinfo {volume} {D82}},\ \bibinfo {pages} {114504} (\bibinfo {year}
  {2010}),\ \Eprint{http://arxiv.org/abs/1008.4018}{arXiv:1008.4018 [hep-lat]}%
  \bibAnnoteFile{NoStop}{Davies:2010ip}%
\bibitem{:2011gx}%
  \BibitemOpen
  \bibfield{author}{%
  \bibinfo {author} {\bibfnamefont{P.}~\bibnamefont{Dimopoulos}} \emph{et~al.}
  (\bibinfo {collaboration} {ETM})}%
   (\bibinfo {year} {2011}),\
  \Eprint{http://arxiv.org/abs/1107.1441}{arXiv:1107.1441 [hep-lat]}%
  \bibAnnoteFile{NoStop}{:2011gx}%
\bibitem{:2008sq}%
  \BibitemOpen
  \bibfield{author}{%
  \bibinfo {author} {\bibfnamefont{B.~I.}\ \bibnamefont{Eisenstein}}
  \emph{et~al.} (\bibinfo {collaboration} {CLEO}),\ }%
  \bibfield{journal}{%
  \Doi{10.1103/PhysRevD.78.052003}{\bibinfo {journal} {Phys. Rev.}}\ }%
  \textbf{\bibinfo {volume} {D78}},\ \bibinfo {pages} {052003} (\bibinfo {year}
  {2008}),\ \Eprint{http://arxiv.org/abs/0806.2112}{arXiv:0806.2112 [hep-ex]}%
  \bibAnnoteFile{NoStop}{:2008sq}%
\bibitem{Alexander:2009ux}%
  \BibitemOpen
  \bibfield{author}{%
  \bibinfo {author} {\bibfnamefont{J.~P.}\ \bibnamefont{Alexander}}
  \emph{et~al.} (\bibinfo {collaboration} {CLEO}),\ }%
  \bibfield{journal}{%
  \Doi{10.1103/PhysRevD.79.052001}{\bibinfo {journal} {Phys. Rev.}}\ }%
  \textbf{\bibinfo {volume} {D79}},\ \bibinfo {pages} {052001} (\bibinfo {year}
  {2009}),\ \Eprint{http://arxiv.org/abs/0901.1216}{arXiv:0901.1216 [hep-ex]}%
  \bibAnnoteFile{NoStop}{Alexander:2009ux}%
\bibitem{:2007ws}%
  \BibitemOpen
  \bibfield{author}{%
  \bibinfo {author} {\bibfnamefont{L.}~\bibnamefont{Widhalm}} \emph{et~al.}
  (\bibinfo {collaboration} {Belle}),\ }%
  \bibfield{journal}{%
  \Doi{10.1103/PhysRevLett.100.241801}{\bibinfo {journal} {Phys. Rev. Lett.}}\
  }%
  \textbf{\bibinfo {volume} {100}},\ \bibinfo {pages} {241801} (\bibinfo {year}
  {2008}),\ \Eprint{http://arxiv.org/abs/0709.1340}{arXiv:0709.1340 [hep-ex]}%
  \bibAnnoteFile{NoStop}{:2007ws}%
\bibitem{Naik:2009tk}%
  \BibitemOpen
  \bibfield{author}{%
  \bibinfo {author} {\bibfnamefont{P.}~\bibnamefont{Naik}} \emph{et~al.}
  (\bibinfo {collaboration} {CLEO}),\ }%
  \bibfield{journal}{%
  \Doi{10.1103/PhysRevD.80.112004}{\bibinfo {journal} {Phys. Rev.}}\ }%
  \textbf{\bibinfo {volume} {D80}},\ \bibinfo {pages} {112004} (\bibinfo {year}
  {2009}),\ \Eprint{http://arxiv.org/abs/0910.3602}{arXiv:0910.3602 [hep-ex]}%
  \bibAnnoteFile{NoStop}{Naik:2009tk}%
\bibitem{Onyisi:2009th}%
  \BibitemOpen
  \bibfield{author}{%
  \bibinfo {author} {\bibfnamefont{P.~U.~E.}\ \bibnamefont{Onyisi}}
  \emph{et~al.} (\bibinfo {collaboration} {CLEO}),\ }%
  \bibfield{journal}{%
  \Doi{10.1103/PhysRevD.79.052002}{\bibinfo {journal} {Phys. Rev.}}\ }%
  \textbf{\bibinfo {volume} {D79}},\ \bibinfo {pages} {052002} (\bibinfo {year}
  {2009}),\ \Eprint{http://arxiv.org/abs/0901.1147}{arXiv:0901.1147 [hep-ex]}%
  \bibAnnoteFile{NoStop}{Onyisi:2009th}%
\bibitem{Lees:2010qj}%
  \BibitemOpen
  \bibfield{author}{%
  \bibinfo {author} {\bibfnamefont{J.~P.}\ \bibnamefont{Lees}} \emph{et~al.}
  (\bibinfo {collaboration} {BaBar}),\ }%
  \bibfield{journal}{%
  \bibinfo {journal} {Phys. Rev.}\ }%
  \textbf{\bibinfo {volume} {D82}},\ \bibinfo {pages} {091103} (\bibinfo {year}
  {2010}),\ \Eprint{http://arxiv.org/abs/1003.3063}{arXiv:1003.3063 [hep-ex]}%
  \bibAnnoteFile{NoStop}{Lees:2010qj}%
\bibitem{Asner:2010qj}%
  \BibitemOpen
  \bibfield{author}{%
  \bibinfo {author} {\bibfnamefont{D.}~\bibnamefont{Asner}} \emph{et~al.}
  (\bibinfo {collaboration} {Heavy Flavor Averaging Group})}%
   (\bibinfo {year} {2010}),\ \bibinfo {note}
  {\url{http://www.slac.stanford.edu/xorg/hfag/charm/CHARM10/f_ds/results_20ja%
n11.html} with future updates at
  \url{http://www.slac.stanford.edu/xorg/hfag/charm/}},\
  \Eprint{http://arxiv.org/abs/1010.1589}{arXiv:1010.1589 [hep-ex]}%
  \bibAnnoteFile{NoStop}{Asner:2010qj}%
\bibitem{McNeile:2011ng}%
  \BibitemOpen
  \bibfield{author}{%
  \bibinfo {author} {\bibfnamefont{C.}~\bibnamefont{McNeile}}, \bibinfo
  {author} {\bibfnamefont{C.~T.~H.}\ \bibnamefont{Davies}}, \bibinfo {author}
  {\bibfnamefont{E.}~\bibnamefont{Follana}}, \bibinfo {author}
  {\bibfnamefont{K.}~\bibnamefont{Hornbostel}},\ and\ \bibinfo {author}
  {\bibfnamefont{G.~P.}\ \bibnamefont{Lepage}} (\bibinfo {collaboration}
  {HPQCD})}%
   (\bibinfo {year} {2011}),\
  \Eprint{http://arxiv.org/abs/1110.4510}{arXiv:1110.4510 [hep-lat]}%
  \bibAnnoteFile{NoStop}{McNeile:2011ng}%
\bibitem{Gamiz:2009ku}%
  \BibitemOpen
  \bibfield{author}{%
  \bibinfo {author} {\bibfnamefont{E.}~\bibnamefont{{G\'amiz}}}, \bibinfo
  {author} {\bibfnamefont{C.~T.~H.}\ \bibnamefont{Davies}}, \bibinfo {author}
  {\bibfnamefont{G.~P.}\ \bibnamefont{Lepage}}, \bibinfo {author}
  {\bibfnamefont{J.}~\bibnamefont{Shigemitsu}},\ and\ \bibinfo {author}
  {\bibfnamefont{M.}~\bibnamefont{Wingate}} (\bibinfo {collaboration}
  {HPQCD}),\ }%
  \bibfield{journal}{%
  \Doi{10.1103/PhysRevD.80.014503}{\bibinfo {journal} {Phys. Rev.}}\ }%
  \textbf{\bibinfo {volume} {D80}},\ \bibinfo {pages} {014503} (\bibinfo {year}
  {2009}),\ \Eprint{http://arxiv.org/abs/0902.1815}{arXiv:0902.1815 [hep-lat]}%
  \bibAnnoteFile{NoStop}{Gamiz:2009ku}%
\bibitem{Albertus:2010nm}%
  \BibitemOpen
  \bibfield{author}{%
  \bibinfo {author} {\bibfnamefont{C.}~\bibnamefont{Albertus}} \emph{et~al.}
  (\bibinfo {collaboration} {RBC and UKQCD}),\ }%
  \bibfield{journal}{%
  \Doi{10.1103/PhysRevD.82.014505}{\bibinfo {journal} {Phys. Rev.}}\ }%
  \textbf{\bibinfo {volume} {D82}},\ \bibinfo {pages} {014505} (\bibinfo {year}
  {2010}),\ \Eprint{http://arxiv.org/abs/1001.2023}{arXiv:1001.2023 [hep-lat]}%
  \bibAnnoteFile{NoStop}{Albertus:2010nm}%
\bibitem{Bazavov:2010pi}%
  \BibitemOpen
  \bibfield{author}{%
  \bibinfo {author} {\bibfnamefont{A.}~\bibnamefont{Bazavov}} \emph{et~al.}
  (\bibinfo {collaboration} {MILC}),\ }%
  \bibfield{journal}{%
  \bibinfo {journal} {PoS}\ }%
  \textbf{\bibinfo {volume} {LATTICE2010}},\ \bibinfo {pages} {320} (\bibinfo
  {year} {2010}),\ \Eprint{http://arxiv.org/abs/1012.1265}{arXiv:1012.1265
  [hep-lat]}%
  \bibAnnoteFile{NoStop}{Bazavov:2010pi}%
\bibitem{Follana:2007uv}%
  \BibitemOpen
  \bibfield{author}{%
  \bibinfo {author} {\bibfnamefont{E.}~\bibnamefont{Follana}}, \bibinfo
  {author} {\bibfnamefont{C.~T.~H.}\ \bibnamefont{Davies}}, \bibinfo {author}
  {\bibfnamefont{G.~P.}\ \bibnamefont{Lepage}},\ and\ \bibinfo {author}
  {\bibfnamefont{J.}~\bibnamefont{Shigemitsu}} (\bibinfo {collaboration}
  {HPQCD}),\ }%
  \bibfield{journal}{%
  \Doi{10.1103/PhysRevLett.100.062002}{\bibinfo {journal} {Phys. Rev. Lett.}}\
  }%
  \textbf{\bibinfo {volume} {100}},\ \bibinfo {pages} {062002} (\bibinfo {year}
  {2008}),\ \Eprint{http://arxiv.org/abs/0706.1726}{arXiv:0706.1726 [hep-lat]}%
  \bibAnnoteFile{NoStop}{Follana:2007uv}%
\bibitem{Mertens:1997wx}%
  \BibitemOpen
  \bibfield{author}{%
  \bibinfo {author} {\bibfnamefont{B.~P.~G.}\ \bibnamefont{Mertens}}, \bibinfo
  {author} {\bibfnamefont{A.~S.}\ \bibnamefont{Kronfeld}},\ and\ \bibinfo
  {author} {\bibfnamefont{A.~X.}\ \bibnamefont{El-Khadra}},\ }%
  \bibfield{journal}{%
  \Doi{10.1103/PhysRevD.58.034505}{\bibinfo {journal} {Phys. Rev.}}\ }%
  \textbf{\bibinfo {volume} {D58}},\ \bibinfo {pages} {034505} (\bibinfo {year}
  {1998}),\
  \Eprint{http://arxiv.org/abs/hep-lat/9712024}{arXiv:hep-lat/9712024}%
  \bibAnnoteFile{NoStop}{Mertens:1997wx}%
\end{thebibliography}%
\bibliographystyle{apsrev4-1} 

\end{document}